\documentclass{nature}
\usepackage{graphicx}
\makeatletter
\let\saved@includegraphics\includegraphics
\AtBeginDocument{\let\includegraphics\saved@includegraphics}
\renewenvironment*{figure}{\@float{figure}}{\end@float}
\makeatother
\usepackage[dvipsnames]{xcolor}
\usepackage{caption}
\usepackage{amsmath}
\usepackage{amssymb}

\title{Galaxies lacking dark matter produced by close encounters in a cosmological simulation}

\author{Jorge Moreno$^{*,1,2,3}$, Shany Danieli$^{4,5}$, James S.~Bullock$^2$, Robert Feldmann$^6$, Philip F. Hopkins$^3$, Onur \c{C}atmabacak$^6$, Alexander Gurvich$^7$, Alexandres Lazar$^2$, Courtney Klein$^2$, Cameron B.~Hummels$^{3}$, Zachary Hafen$^2$, Francisco  J. Mercado$^2$, Sijie Yu$^2$, Fangzhou Jiang$^{3,8}$, Coral Wheeler$^9$, Andrew Wetzel$^{10}$, Daniel Angl\'es-Alc\'azar$^{11,12}$, Michael Boylan-Kolchin$^{13}$, Eliot Quataert$^{4}$, Claude-Andr\'{e} Faucher-Gigu\`{e}re$^7$ \& Du\v{s}an Kere\v{s}$^{15}$}

\begin{document}

\maketitle

\begin{affiliations}

\item Department of Physics and Astronomy, Pomona College, Claremont, CA 91711, USA
\item Department of Physics and Astronomy, 4129 Reines Hall, University of California, Irvine, CA 92697, USA
\item TAPIR, California Institute of Technology, Pasadena, CA 91125, USA
\item Department of Astrophysical Sciences, 4 Ivy Lane, Princeton University, Princeton, NJ 08544, USA
\item Institute for Advanced Study, 1 Einstein Drive, Princeton, NJ 08540, USA
\item Institute for Computational Science, University of Zurich, Winterthurerstrasse 190, Zurich CH-8057, Switzerland
\item Department of Physics \& Astronomy and CIERA, Northwestern University, 1800 Sherman Ave, Evanston, IL 60201, USA
\item Carnegie Observatories, 813 Santa Barbara Street, Pasadena, CA 91101, USA
\item Department of Physics and Astronomy, California State Polytechnic University, Pomona, Pomona, CA 91768, USA
\item Department of Physics \& Astronomy, University of California, Davis, One Shields Ave, Davis, CA 95616
\item Department of Physics, University of Connecticut, 196 Auditorium Road, U-3046, Storrs, CT 06269-3046, USA 
\item Center for Computational Astrophysics, Flatiron Institute, 162 5th Ave, New York, NY 10010, USA
\item Department of Astronomy, The University of Texas at Austin, 2515 Speedway, Stop C1400, Austin, TX 78712-1205, USA
\item Department of Astrophysical Sciences, Princeton University, Princeton, NJ 08544, USA
\item Department of Physics, Center for Astrophysics and Space Sciences, University of California San Diego, 9500 Gilman Drive, La Jolla, CA 92093, USA

\end{affiliations}

\section*{Abstract}

\begin{abstract} 
The standard cold dark matter plus cosmological constant model predicts that galaxies form within dark-matter haloes, and that low-mass galaxies are more dark-matter dominated than massive ones. The unexpected discovery of two low-mass galaxies lacking dark matter immediately provoked concerns about the standard cosmology and ignited explorations of alternatives, including self-interacting dark matter and modified gravity. Apprehension grew after several cosmological simulations using the conventional model failed to form adequate numerical analogues with comparable internal characteristics (stellar masses, sizes, velocity dispersions and morphologies). Here we show that the standard paradigm naturally produces galaxies lacking dark matter with internal characteristics in agreement with observations. Using a state-of-the-art cosmological simulation and a meticulous galaxy-identification technique, we find that extreme close encounters with massive neighbours can be responsible for this. We predict that $\sim$30\% of massive central galaxies (with at least $10^{11}$ solar masses in stars) harbour at least one dark- matter-deficient satellite (with $10^{8}–10^{9}$ solar masses in stars). This distinctive class of galaxies provides an additional layer in our understanding of the role of interactions in shaping galactic properties. Future observations surveying galaxies in the aforementioned regime will provide a crucial test of this scenario.

\end{abstract}

\section*{Editor's Summary}
{\bf A cosmological simulation shows that low-mass galaxies can form with far less dark matter than expected, with results matching some observed characteristics. Roughly one-third of massive central galaxies may host at least one such dark-matter- deficient satellite.}

\begin{figure*}
\includegraphics[width=\columnwidth]{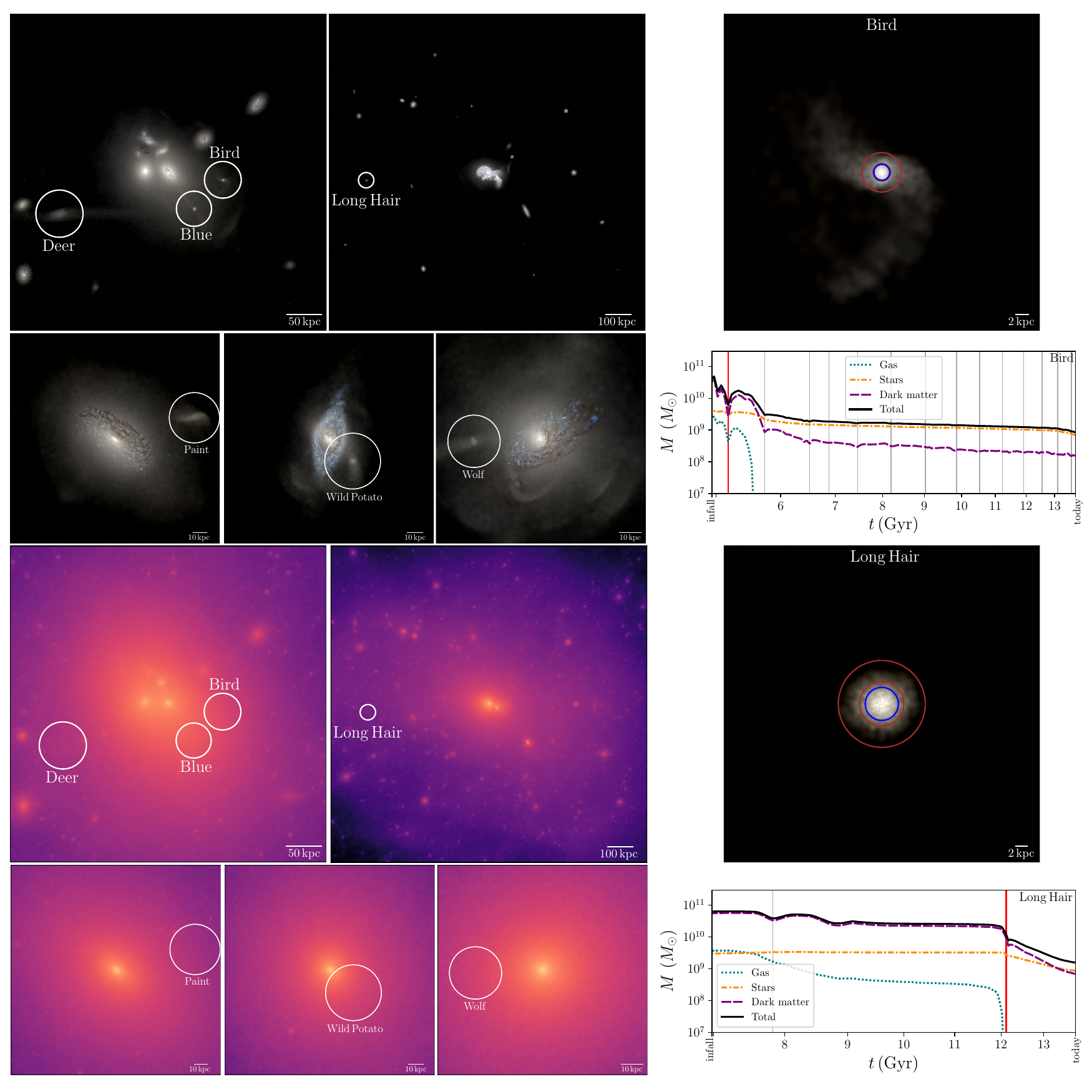}
\vspace{-1.5cm}
    \caption{{\bf Galaxies lacking dark matter}. Top left: mock stellar Hubble Space Telescope $u/g/r$ composite images  of the environments around our seven dark-matter deficient galaxies at present time ($z=0$), with a cut-off of 28.9 mag arcsec$^{-2}$ (corresponding to $10^5 L_{\odot}$ kpc$^{-2}$). Bottom left: dark matter mass surface density maps (down to $10^5 M_{\odot}$ kpc$^{-2}$). The satellites are clearly visible in the stellar images but not in the dark matter maps. Right: magnified mock stellar images (top) and masses inside the subhalo radius versus time ($t$, bottom) of Bird (top) and Long Hair (bottom). The red circles indicate the radii containing 50 and 80\% of the stellar mass, and the blue circle refers to the effective radius in the g-band. The grey vertical lines mark pericentric passages, red vertical lines mark the closest pericentric passage. } 
 \label{fig:fig1}
\end{figure*}

\section*{Main}

For over half a century, cold dark matter has been a key ingredient in our understanding of galaxy formation\cite{Zwicky1933,Rubin1970}. In the low-mass regime, the standard cold dark matter paradigm\cite{White1978} predicts that galaxies should be more dark-matter dominated\cite{Behroozi2013}. This is supported by observations in the nearby Universe\cite{Tollerud2011}. For this reason, the detection of DF2 and DF4, two low-mass galaxies devoid of dark matter, was not anticipated\cite{vanDokkum2018,vanDokkum2019}. This intriguing discovery immediately sparked several searches of numerical analogues in cosmological simulations\cite{Yu2018,Jing2019,Haslbauer2019,Haslbauer2019b,Carleton2019,Sales2020,Saulder2020,Shin2020,Applebaum2021,Jackson2021,Wright2021}, with limited success. Indeed, although idealised numerical experiments\cite{Ogiya2018,Huo2020,Nusser2020,Yang2020,Maccio2021} had hinted at the possibility of the existence of such analogues within the standard model -- no cosmological simulation had successfully generated numerical galaxies whose internal properties (stellar masses, velocity dispersions and morphologies) simultaneously matched those of their observed counterparts. Unsurprisingly, the absence of dark-matter deficient low-mass galaxies in some of the earlier cosmological simulations also raised doubts on the validity of the standard paradigm itself\cite{Haslbauer2019b,Yang2020} -- although, these studies did not conduct an exhaustive comparison to previous works (see Supplementary Information). In this paper we demonstrate that a novel cosmological hydrodynamical simulation (which presupposes this paradigm) naturally creates adequate numerical versions of the observed dark matter-deficient galaxies. This simulation utilises the `Feedback In Realistic Environments' (\texttt{FIRE-2}) physics model\cite{Hopkins2018}, which successfully reproduces an array of galaxy properties\cite{CAFG2018} -- and is state-of-the-art in its ability to resolve the internal structure of individual galaxies within a large cosmological environment (see Methods Section for details). By directly comparing with observations, we confirm that one of our simulated galaxies resembles DF2 and DF4 in arresting ways. This finding alleviates the aforementioned concerns -- and mitigates the exigency for alternative explanations invoking new (non-standard) physics, such as self-interacting dark matter\cite{Yang2020} and modified gravity\cite{Famaey2018,Haghi2019,Haslbauer2019b,Kroupa2018,Moffat2019,Kalifeh2021}.

We identify seven galaxies lacking dark matter within a volume of $10^4$ Mpc$^3$. Fig.~\ref{fig:fig1} shows present-time images of the environments around these simulated dark-matter deficient galaxies (white circles), which we name in honour of the seven Cherokee clans: Bird, Blue, Deer, Long Hair, Paint, Wild Potato and Wolf. The top-left panels show mock Hubble Space Telescope composite $u/g/r$ stellar images (down to 28.9 mag arcsec$^{-2}$, corresponding to $10^5L_{\odot}$ kpc$^{-2}$) and the bottom-left panels display corresponding dark matter mass surface density maps (down to $10^5M_{\odot}$ kpc$^{-2}$). The clear presence of these galaxies in the stellar images and not in the dark matter maps is remarkable. To illustrate, the right-hand panels display two special cases: Bird and Long Hair, the members with the most and least active interaction histories in our set of seven.

At the present time (redshift $z=0$), Bird, Blue and Deer inhabit the same galaxy group (virial mass $M_{\rm vir}=1.8 \times 10^{13} M_{\odot}$) and are at 80 kpc, 66 kpc and 174 kpc away from its center, respectively. Long Hair also lives in a group ($M_{\rm vir}=1.4 \times 10^{13} M_{\odot}$) and has the largest halo-centric distance (484 kpc) amongst the seven. Paint, Wild Potato and Wolf exist in haloes slightly more massive than that of the Milky Way ($M_{\rm vir}=3.6, \, 2.5,$ {\rm and} $\, 3.3 \times 10^{12}  M_{\odot}$) and have the following halo-centric distances: 72 kpc, 57 kpc and 36 kpc. The two observed galaxies lacking dark matter, DF2\cite{vanDokkum2018} and DF4\cite{vanDokkum2019}, are at close projected separations (80 kpc and 165 kpc, which set a lower limit to the 3D separations) from NGC 1052\cite{Shen2021b, Danieli2020} (stellar mass $M_{\star}=1.05 \times 10^{11}M_{\odot}$\cite{Forbes2017}). Following Ref.\cite{Zahid2018}, we estimate that NGC 1052 has $M_{\rm vir}=6.2^{+25.8}_{-4.2} \times 10^{12} M_{\odot}$ at 90\% confidence, which is consistent with the virial masses of our hosts. Although halo-centric separations vary with time, their present-day values can provide information about their current membership status as satellites (see below).

\begin{figure}
\includegraphics[width=\columnwidth]{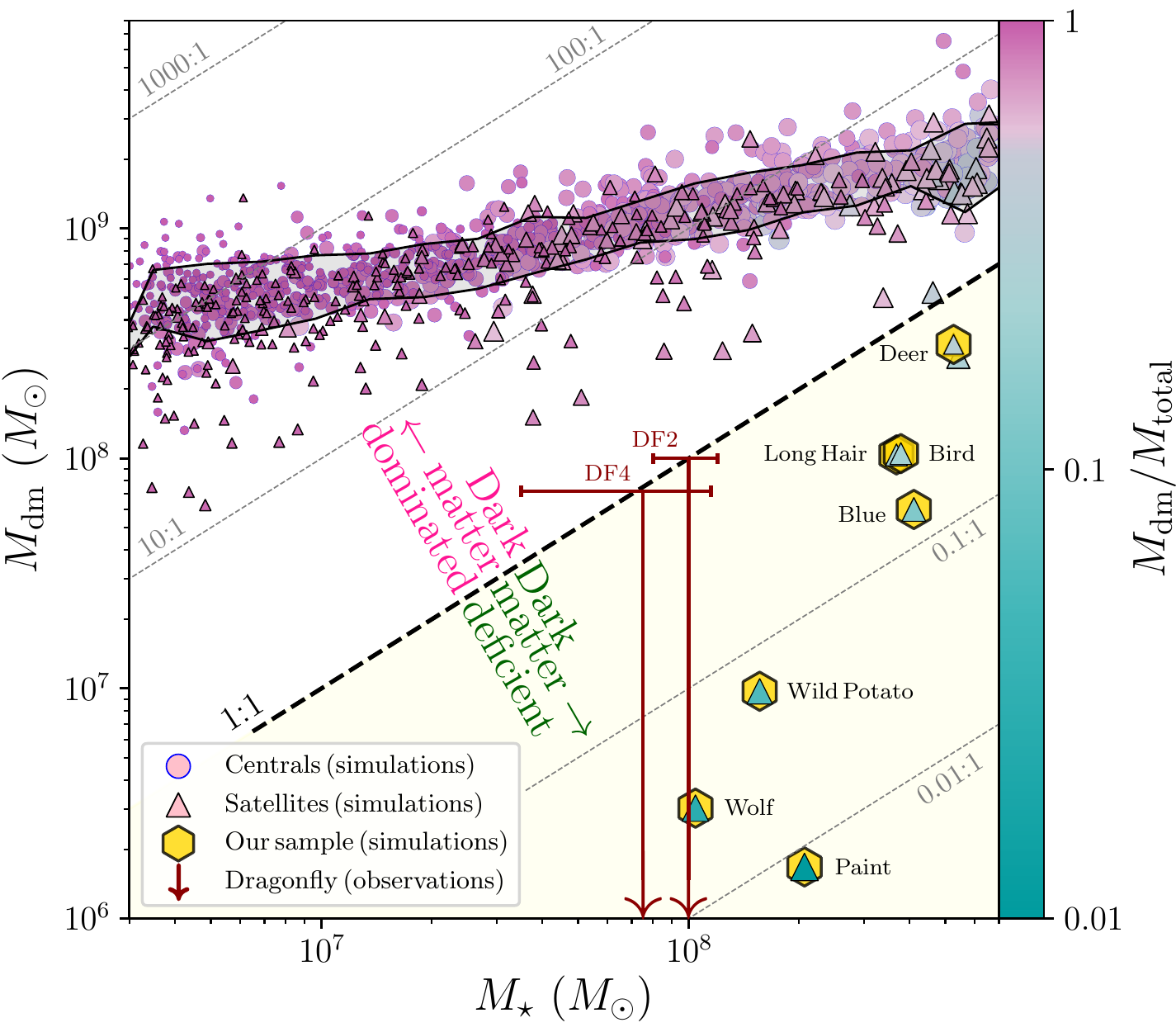}
\vspace{-1.5cm}
\caption{{\bf Fraction of mass in dark matter in simulated and observed galaxies.} Dark matter mass versus stellar mass, colour-coded by fraction of total mass in dark matter ($M_{\rm dm}/M_{\rm total}$). Masses are determined within $r^{\star}_{\rm 50}$, the radius containing 50\% of the stellar mass. The solid circles and triangles denote central and satellites galaxies (simulations). Symbol dimensions scale inversely with $M_{\rm dm}/M_{\rm total}$. The light-gray band represents the median and standard deviation (simulated centrals only). The light-yellow region denotes the dark-matter deficient regime. The yellow hexagons highlight our sample of simulated galaxies lacking dark matter. The red arrows with error bars (1 s.d.) represent observations\cite{vanDokkum2018,vanDokkum2019} conducted with the Dragonfly Telephoto Array.}
\label{fig:fig2}
\end{figure}

Fig.~\ref{fig:fig2} shows dark matter mass ($M_{\rm dm}$) versus stellar mass ($M_{\star}$). Hereafter we quote quantities within $r^{\star}_{\rm 50}$, the radius containing 50\% of the stellar mass, to compare directly with observations\cite{Danieli2019} (dark-red arrows with error bars). We note, however, that our seven galaxies are also dark-matter deficient out to larger galactocentric radii (Supplementary Table~2 and Supplementary Fig.~9). The diagonal lines represent constant dark-matter-to-stellar mass ratios. We select our sample of seven dark-matter deficient galaxies (yellow hexagons) by requiring $M_{\rm dm}<M_{\star}$ (light-yellow region below the $1:1$ line). Wolf's stellar mass is consistent with both DF2 and DF4. We do not include galaxies with $M_{\star}<M_{\rm dm}<M_{\rm baryon}$ in this sample. These are likely simulated analogues of the recently-discovered baryon-rich galaxies\cite{Mancera2019,Guo2020,Mancera2021}, which we plan to study separately. (We also discard one object -- the triangle in close proximity to Deer -- because it is participating in an on-going gas-rich merger, unlike DF2 and DF4.)

\begin{figure}
\includegraphics[width=\columnwidth]{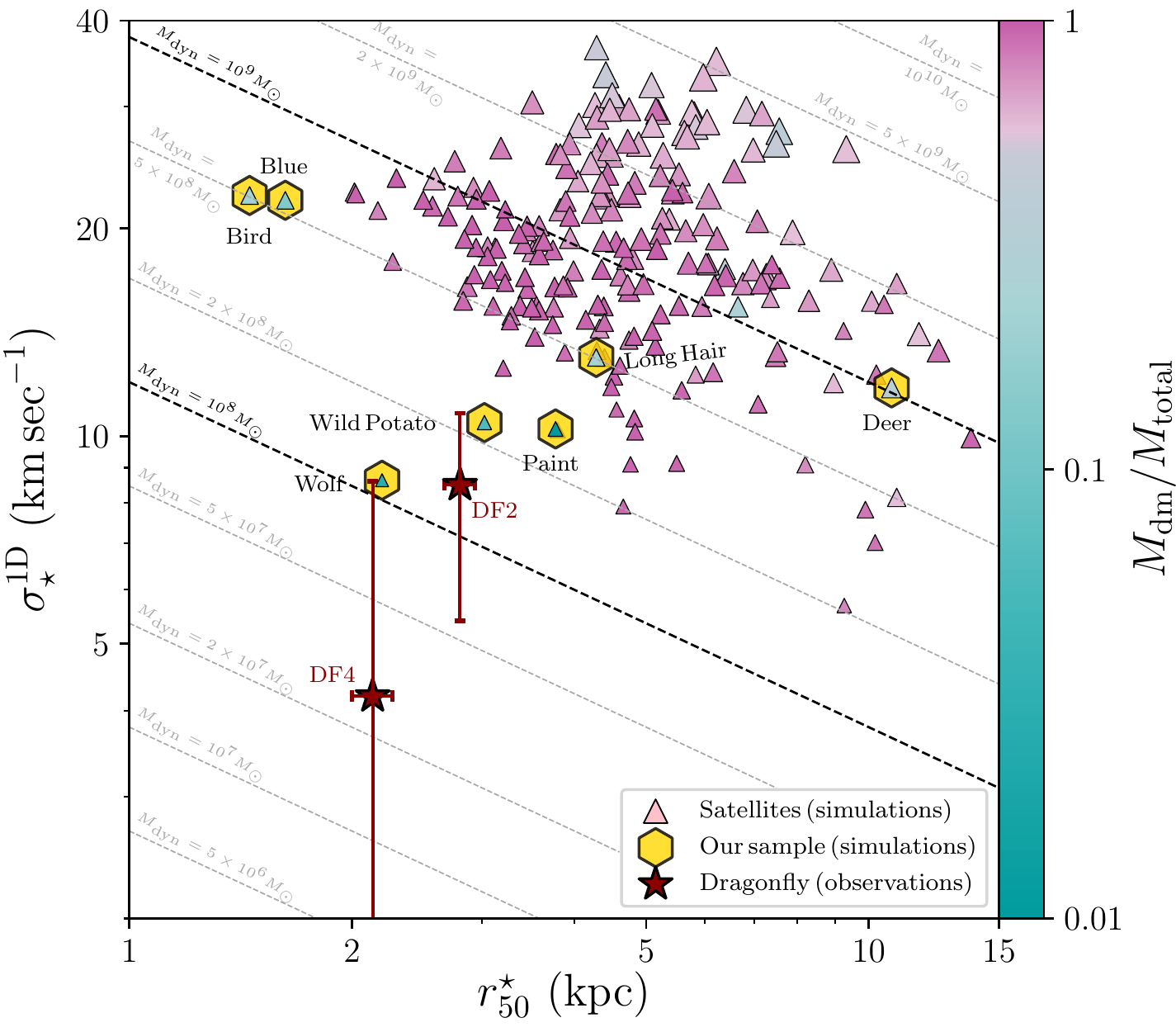}
\vspace{-1.5cm}
\caption{{\bf Comparison of internal properties between simulated and observed galaxies.} Quantities are measured within $r^{\star}_{\rm 50}$. The triangles denote simulated satellites with $M_{\star}=10^{8-9}M_{\odot}$, their dimensions scale with $M_{\rm total}$ and are colour-coded by $M_{\rm dm}/M_{\rm total}$. The red stars with error bars (1 s.d.) represent Dragonfly observations\cite{vanDokkum2019,Danieli2019}. The diagonal dashed lines show various fixed dynamical masses according to an analytic mass estimator\cite{Wolf2010}: $M_{\rm dyn}(<r^{\star}_{\rm 50})\propto (\sigma^{\rm 1D}_\star)^2 \, r^{\star}_{\rm 50}$.}
\label{fig:fig3}
\end{figure}

\begin{figure}
\includegraphics[width=\columnwidth]{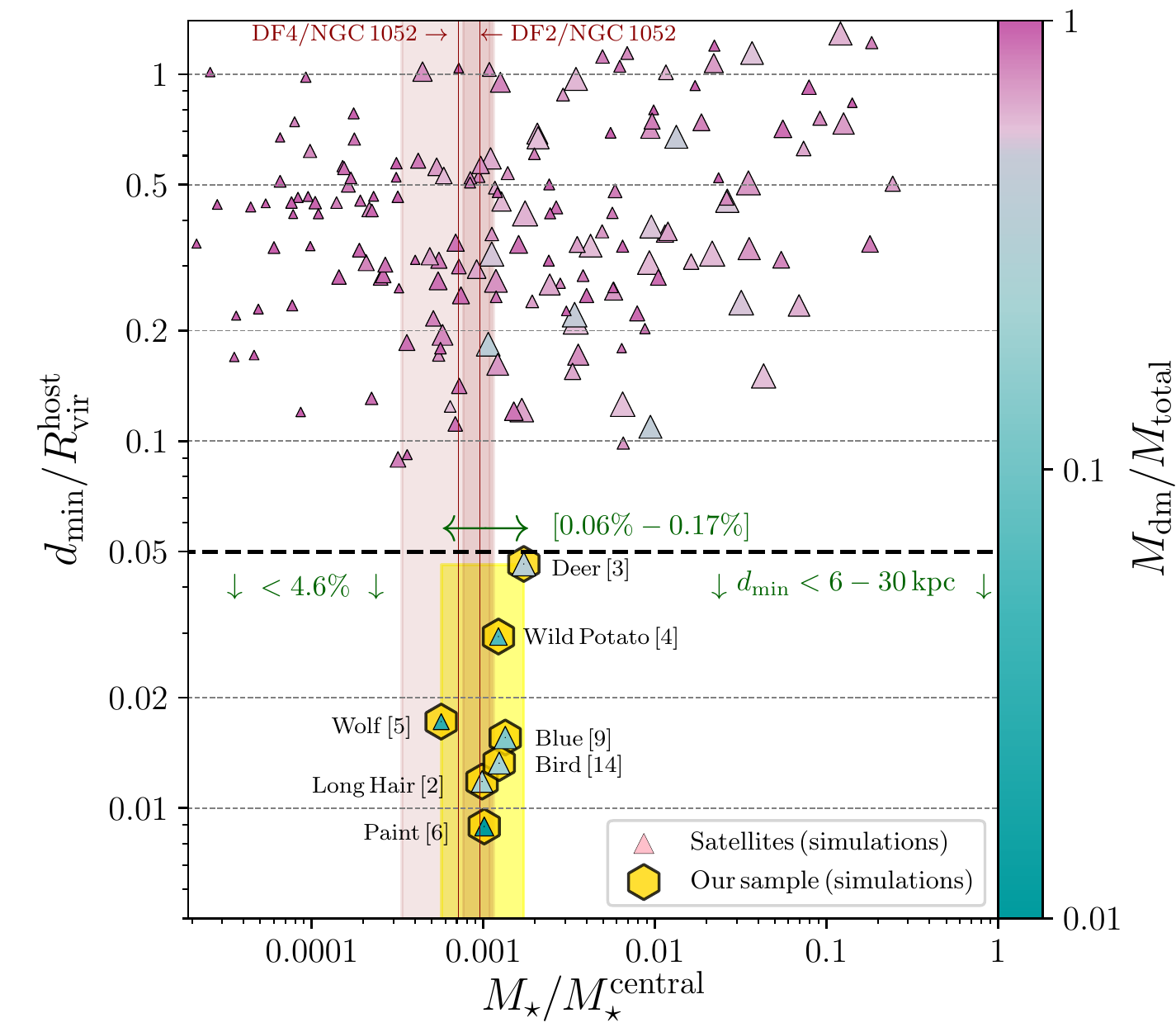}
\vspace{-1.5cm}
\caption{{\bf Conditions for creating a galaxy lacking dark matter.} The minimum halo-centric distance ever achieved (in units of the hosts' virial radius) versus the satellite-central stellar mass ratio. The triangles denote simulated satellites with $M_{\star}=10^{8-9}M_{\odot}$, colour coded by $M_{\rm dm}/M_{\rm total}$, with dimensions scaled by $M_{\star}$ (simulations). Quantities are within $r^{\star}_{\rm 50}$ for satellites and $r^{\star}_{\rm 80}$ for central companions. The yellow rectangle, percentages within parenthesis and values between downward arrows highlight the region of parameter space our seven galaxies occupy (yellow hexagons) and the numbers in parentheses indicate number of pericentric passages. The vertical red lines (and bands) denote the stellar mass ratios (and uncertainty) between DF2/DF4 and NGC 1052 (observations).} 
\label{fig:fig4}
\end{figure}

Fig.~\ref{fig:fig3} shows the 1D line-of-sight velocity dispersion ($\sigma^{\rm 1D}_{\star}$) versus the galactic size ($r^{\star}_{\rm 50}$).
For observations of DF2 and DF4, we use deprojected measurements\cite{Wolf2010} by Ref.\cite{Danieli2019} and Ref.\cite{vanDokkum2019}, respectively. For comparison, we also include simulated galaxies with $M_{\star}=10^{8-9}M_{\odot}$ (within $r^{\star}_{\rm 50}$), in line with our sample of seven. We note that although this stellar-mass regime contains DF2 and DF4, it is ``too generous". Namely, only galaxies with $M_{\star}$ (within $r^{\star}_{\rm 50}$) close to $10^8 M_{\odot}$ may qualify as realistic numerical analogues of DF2 and DF4 (the rest should thus be regarded as slightly more massive predictions.) To guide the eye, the dashed diagonal lines track fixed dynamical masses within $r^{\star}_{\rm 50}$ according to an analytic mass estimator\cite{Wolf2010}. Additionally, to quantify the morphology of our seven galaxies, we performed S\'ersic\cite{Sersic1963} profile fittings in the $g$-band. We measure effective radii ($R^{\rm 2D}_{\rm e}$), which are in line with our $r^{\star}_{\rm 50}$ values after deprojection (see the Supplementary Figs.~2 and 3, plus Supplementary Table 1). With this procedure, we also obtain S\'ersic indices ($n_{\rm S\acute{e}rsic}$), which range from $0.60-0.77$, in excellent agreement with observations ($n_{\rm S\acute{e}rsic}=0.60$ and $0.79$ for DF2 and DF4, respectively). Our results thus far suggest that Wolf is the only dark-matter deficient galaxy in the cosmological-simulation literature that simultaneously matches observations on the following fronts: stellar mass (within $r^{\star}_{\rm 50}$), 1D  line-of-sight velocity dispersion, size and morphology (Wild Potato has $\sigma^{\rm 1D}_\star = 7.0$ km sec$^{-1}$ along one of the three orthogonal directions we report, but is has marginally higher stellar mass than DF2 and DF4 -- see Supplementary Fig.~4 and Supplementary Table 1). 

Previous cosmological simulations have not been able to match the aforementioned internal properties of the two observed dark-matter deficient galaxies. The most promising object in Ref.\cite{Jackson2021} has $\sim$66\% more mass in dark matter relative to stars -- whilst the stellar mass and 1D velocity dispersion in the simulations reported by Ref.\cite{Jing2019} exceed observational measurements by more than 1-$\sigma$. See Supplementary Fig.~9 for a direct comparison against those two works. Furthermore, no other cosmological simulation has been able to produce low-mass galaxies with dark-matter mass fractions as low as ours: three members of our sample have dark-matter mass fractions below 10\% and one has only 1\% of its mass in dark matter -- rendering them almost dark-matter free galaxies. One possible explanation is that our simulation has the distinct advantage of being able to model the small-scale ($\sim 20$ pc) interstellar medium and the high densities ($>300$ cm$^{-3}$) of star-forming gas of individual galaxies within a cosmological region ($\sim 20$ Mpc on the side) large enough to contain several massive groups of galaxies. See the Supplementary Fig.~9 and associated text for a detailed comparison with results by other groups.

Below we show that close encounters are responsible for the creation of dark-matter deficient galaxies in our simulation. Observationally, it is challenging to infer if a galaxy has experienced close interactions in its history. For instance, Bird exhibits `S-shaped' low surface brightness tails, whilst Long Hair appears undisturbed (see magnified examples in the right-hand side of Fig.~\ref{fig:fig1} -- and the Supplementary Fig.~5 for a more detailed analysis of Long Hair). However, our simulation reveals that close encounters with a massive neighbour are responsible for dark-matter deficiency in all of our seven galaxies. They all started out more massive, gas rich and with fairly typical stellar-to-dark-matter mass ratios for galaxies of their mass -- but subsequently experienced multiple close interactions with their host galaxies (see the mass-evolution insets in Fig.~\ref{fig:fig1} and Supplementary Fig.~7). With the exception of Deer, these galaxies lost all of their gas during this process, in agreement with DF2\cite{Sardone2019}. These satellites also lost between $97.9-99.99\%$ of their dark matter mass, whilst losing only $45-97\%$ of their stellar mass (within the subhalo radius) -- see Supplementary Table~2 and Supplementary Fig.~8 for details). We suspect that stars are more resilient, as they move on more circular orbits in contrast to the eccentric orbits of the dark matter particles. Differences in elongation and dynamical times of the orbits followed by dark matter particles relative to those followed by stars thus leave them more susceptible to tidal stripping\cite{Penarrubia2010,Bryan2012,Valluri2013,Zhu2017}.

To quantify the role of interactions, we evaluate how close satellite galaxies come relative to their hosts and their stellar mass ratios relative to these massive central companions. Fig.~\ref{fig:fig4} shows $d_{\rm min}/R^{\rm host}_{\rm vir}$, the minimum halo-centric distance ever attained by each satellite (in units of the final virial radius of the host) versus $M_{\star}/M^{\rm central}_{\star}$, the present-day satellite-central stellar-mass ratio. For satellites, we continue to employ $r^{\star}_{\rm 50}$, whilst for centrals, we switch to $r^{\star}_{\rm 80}$, which better captures their spatial extent\cite{Mowla2019}. Our simulation suggests that, to become a dark-matter deficient galaxy, a satellite must pierce within $\sim$5\% (below horizontal black dashed line) of the host virial radius (within $\sim 6-30$ kpc, depending on the host). In other words, the satellite must transit through the host's stellar body. We also predict that their satellite-central stellar-mass ratio must be $\sim0.1\%$. The light-yellow region indicates these two conditions. The stellar mass ratios of DF2 and DF4 relative to NGC 1052 are consistent with our simulation; however, no orbital information is available to determine their minimum halo-centric distance. 

Whether or not the galaxies in our simulation were formed in the same way as DF2 and DF4 is still unclear. Our results support a scenario where DF2 and DF4 became dark-matter deficient due to past close encounters with NGC\,1052, the only massive galaxy in their neighbourhood (note that our simulation allows hosts to have multiple dark-matter deficient satellites). Their projected (80 kpc and 165 kpc) and line-of-sight separations ($2.1 \pm 0.5$ Mpc)\cite{Shen2021b} -- as well as our estimates of the virial diameter of the NGC\,1052 halo -- suggest that DF2 and DF4 could be satellites of the NGC\,1052 group. Concretely, using Ref.\cite{Zahid2018}, we estimate this galaxy group to have a virial diameter of $0.7-1.7$ Mpc at 90\% confidence, allowing for the possibility that NGC\,1052 could encompass both objects. This possibility is amplified if either DF2 or DF4 are a `backsplash'\cite{Wetzel2014} galaxy -- i.e., a galaxy on an extremely radial orbit that carried it beyond the virial volume of NGC\,1052 after a close encounter (we note that our simulated sample includes two recent backsplash galaxies: Long Hair and Deer). Future investment in characterizing accurate 3D separations between DF2 and DF4 relative to NGC 1052 will provide an important test of this interaction-based scenario.

Although we have successfully created a galaxy (Wolf) in a cosmological simulation that matches  DF2 and DF4 remarkably (in terms of $M_{\star}$, $r^{\star}_{\rm 50}$, $\sigma^{\rm 1D}_{\star}$ and $n_{\rm S\acute{e}rsic}$), a few outstanding issues remain. In addition to being dark-matter deficient, DF2 and DF4 are intriguing because they (1) inhabit the same group and (2) are extremely similar to each other. Namely, they resemble each other in stellar-mass, size, (extremely-low) velocity dispersion, S\'ersic index, colour, and the presence of anomalous globular cluster populations\cite{Shen2021a}. Our cosmological simulation has insufficient resolution to explore the globular-cluster problem, but zoom-in simulations with our model can potentially address this\cite{Kim2018,Ma2020}. However, we are able to measure stellar-population properties in our seven dark-matter deficient galaxies (see Supplementary Table 3). In particular, Blue and Bird have stellar ages of 10 and 8.7 Gyr, in line with DF2\cite{vanDokkum2018} ($8.9\pm1.5$ Gyr). Indeed, just like DF2 and DF4, these two simulated galaxies also inhabit the same group (Fig.~\ref{fig:fig1}). However, although they have some of the lowest metallicities in our set of seven (${\rm \langle [Fe/H] \rangle} = -0.61$ and $-0.34$), they are not as metal-poor as DF2\cite{Fensch2019} (${\rm \langle [Fe/H] \rangle} =-1.35\pm0.12$). One possibility is that DF2 and DF4 had infall stellar-masses similar to Blue and Bird, but sat initially on the lower envelop of their mass-metallicity relation. Confirming the existence of such objects would require a larger simulation box or multiple distinct realsations of this simulation (with the same physics, but different initial conditions). 

In summary, our results demonstrate that galaxies resembling DF2 and DF4 can arise naturally within the standard cold-dark-matter based cosmological paradigm. We note that we did not expect this to occur a priori (i.e., our simulation was not originally designed for this purpose). Although there certainly is still room for new physics beyond the standard paradigm, discriminating between alternative models will now rely on predicted differences between the expected properties of dark-matter deficient galaxies, rather than their mere existence\cite{Haslbauer2019b}. 

To efficiently uncover more galaxies devoid of dark matter, observers should focus on satellites with $M_{\star}=10^{8-9}M_{\odot}$, near massive companions ($M_{\star} \geq 10^{11}M_{\odot}$). Concretely, we predict 30\% of such massive galaxies to have at least one dark-matter deficient satellite. Interestingly, we also predict that only 30\% of haloes with virial masses $\geq 2 \times 10^{12}$ (the lower bound for NGC 1052) harbour such peculiar objects. This mass range is slightly above the virial masses of the Milky Way and Andromeda -- which might explain why no analogueues of DF2 and DF4 have been detected in the Local Group. However, we cannot exclude a situation where a central-satellite pair outside this mass regime could have a sufficiently unorthodox orbital configuration to produce the same effect. Verifying how strict this regime is will require a larger volume (or more realisations). More generally, to truly constrain the varied processes and underlying physics governing these galaxies, ambitious observational campaigns aimed to enumerate and characterise large samples of galaxies with unusual dark-matter properties -- will be desirable in the years ahead.

\begin{methods}
\label{sec:methods}

\subsection{Our cosmological simulation.}
\label{subsec:oursim}

We ran a high-resolution cosmological simulation of galaxy formation with box size of $21$ comoving Mpc and the following number of baryonic and dark matter particles: $N_{\rm b}=1,024^3$ and $N_{\rm dm}=1,024^3$. Initially the baryonic particle masses are $m_{\rm baryon}=6.3\times10^4\,M_{\odot}$ for gas and star particles, whilst the dark matter particles have $m_{\rm dm}=3.3\times10^5\, M_{\odot}$. The force resolution is set at a fixed $h_{\rm star}=12$ pc (physical) for star particles and 80 pc for dark matter particles. 
For gas, the force resolution is set equal to the adaptive smoothing length down to a minimum of 1.5 pc, which occurs only in the densest regions of galaxies.
We identify galaxies (and merger history trees) with the \texttt{AMIGA Halo Finder} (\texttt{AHF}\cite{Knollmann2009}), which uses an iterative unbinding procedure to identify gravitationally-bound objects\cite{Knebe2011}. This method assumes the virial-mass definition of Ref.\cite{Bryan1998} by construction. We used \texttt{yt}\cite{Turk2011} to interface with the particle data. The cosmological parameters are set to $\Omega_{\rm m} = 0.3089$, $\Omega_{\Lambda} = 0.6911$, $\Omega_{\rm b} = 0.0486$, $\sigma_{8} = 0.8159$ and $h = 0.6774$.

We employ the `Feedback In Realistic Environments' (\texttt{FIRE-2}) model\cite{Hopkins2018} for baryonic physics, which includes radiative cooling, star formation in dense self-gravitating gas and accounts for stellar feedback in the form of supernovae, stellar mass-loss and radiation interacting with the surrounding gas (as described therein). We follow eleven separately-tracked species: H, He, C, N, O, Ne, Mg, Si, S, Ca, and F. This model has been extensively validated in a number of publications analyzing properties of galaxies across a range in stellar masses and numerical resolutions\cite{CAFG2018}.  In our run, star formation occurs only in gas denser than $n=300$ cm$^{-3}$.  With this condition, the inter-particle separation in star-forming gas is $\sim20$ pc (or less) and we are able to resolve giant-molecular-cloud complexes within the interstellar medium of individual galaxies. 

\subsection{Our simulated galaxy samples.}
\label{subsec:parent_sample}

In this work we focus on galaxies with at least 100 stellar particles at redshift $z=0$. The actual number of particles per object is typically much larger owing to an additional contribution of gas and dark matter particles for each object, either at the present time or at some point in its history. To avoid discarding galaxies lacking dark matter, our application of \texttt{AHF} utilises all types of particles, not just dark matter. This comes at a price: the halo finder falsely identifies large numbers of gas clumps within galaxies as individual `galactic units'. To address this, first we discard objects with baryon-to-total mass ratios greater than 0.5. This may naturally discard the very objects we seek (and keep false clumps below this threshold). To ameliorate this, we create surface density maps of every halo with substructures and visually recover objects that might have been discarded erroneously by our 50\% cut. The images shown in Fig.~\ref{fig:fig1} belong to this set. We focus on manually deleting clumps embedded inside massive galaxies and on recovering galaxies that are either clearly disjoint from their neighbours -- or in tidal tails, to avoid rejecting tidally-formed candidates\cite{Elmegreen1993}. We highlight that every galaxy lacking dark matter studied in this paper is recovered during this step. Lastly, because we are interested in the low-mass regime, we constrain our sample to have stellar masses under $10^{9} M_{\odot}$ (within $r^{\star}_{\rm 80}$, the radius containing 80\% of the stellar mass). This produces our final parent sample, which contains 1,218 resolved galaxies: 886 centrals and 332 satellites. See Supplementary Fig.~1 for a visual description of these two sets.

Our dark-matter deficient set consists of seven galaxies with $M_{\rm dm}<M_{\star}$ that are not participating in a major merger (we discard an eighth galaxy meeting the former condition, but not the latter). We adopt this extra condition because the two observed Dragonfly galaxies are not currently merging with a companion of similar mass. We also exclude 19 low-mass galaxies (15 centrals and 4 satellites) with $M_{\rm dm}<M_{\rm baryon}$ (but $M_{\rm dm}>M_{\star}$) -- which are possibly numerical analogues of the recently-observed baryon-rich galaxies\cite{Mancera2019,Guo2020,Mancera2021}. The final set of seven are satellites with $M_{\star} = 10^{8-9}M_{\odot}$; they orbit five different massive central galaxies with $M_{\star}>10^{11}M_{\odot}$. In total, our simulation contains fifteen massive centrals that host 47 satellite galaxies in the target stellar mass range ($10^{8-9}M_{\odot}$). Of these satellites, $\sim15\%$ belong to our set of seven, which orbit $\sim30\%$ of the available massive centrals. 

The absence of dark-matter deficient satellite galaxies at stellar masses below $10^8 M_\odot$ could arise from the fact that, at lower masses, galaxies are born as centrals with very high dark matter mass fractions, such that environmental effects cannot act to remove dark matter without completely destroying the galaxy. The absence of dark-matter deficient galaxies at lower stellar masses is likely not driven by lack of resolution because our simulation produces a large population of galaxies with at least 1,000 stellar particles that are less massive than Wolf (which has 3,280 stellar particles at redshift zero and had a factor of $\sim$35 higher before becoming a satellite). For a discussion about potential numerical effects on tidal disruption, see Supplementary Fig.~8 and associated text.

\subsection{Galactic sizes.}
\label{subsec:sizes}

We calculate galactic sizes as follows. For each object, we record the distance to the nearest major companion (with stellar mass of at least a tenth of that of our object of interest). If no such companion exists, we record the virial radius (for centrals) or subhalo radius (for satellites). This step is necessary because \texttt{AHF} considers substructures as part of their host. Once this region is identified, we calculate the stellar center-of-mass. Next we redefine the radius of this region by subtracting the distance from the origin to this center-of-mass. Within this region, we then calculate cumulative mass profiles (using 200 identical linear radial bins) and extract the radius containing 80\%  of the stellar mass within this region by linearly interpolating between the two radii below and above this threshold. The sizes of our seven dark-matter deficient galaxies are listed in Supplementary Table~1 and displayed in Supplementary Figs.~2 and 3.

\subsection{1D line-of-sight velocity dispersions.}
\label{subsec:vdisp}

We also calculate 1D line-of-sight velocity dispersions ($\sigma^{\rm 1D}_{\star}$) within $r^{\star}_{\rm 50}$. We perform this calculation along three random directions (in the frame of each galaxy), corresponding to the $x$, $y$ and $z$ directions in the frame of the box. We create histograms of line-of-sight velocity minus the average line-of-sight velocity within the $r^{\star}_{\rm 50}$. These histograms have bin-widths of $2$ km sec$^{-1}$ and extend from $-70$ to $70$ km sec$^{-1}$. We then fit a Gaussian curve and record its $\sigma$-value. In the main text we report the simple average of these three numbers. The $\sigma^{\rm 1D}_{\star}$-values for our seven dark-matter deficient galaxies are listed Supplementary Table~1 and displayed (along the aforementioned three directions) in Supplementary Fig.~4

\section*{Peer review}

\textit{\textbf{Peer review information.}} Nature Astronomy thanks the anonymous reviewers for their contribution to the peer review of this work.

\subsection{Data availability.}
The datasets generated during and/or analyzed during the current study are available from the corresponding author on reasonable request, contingent on approval by the FIRE Collaboration on a case-by-case basis.

\subsection{Code availability.}
Below we list the URLs of the codes we use.
\begin{itemize}
    \item \texttt{GIZMO}: https://bitbucket.org/phopkins/gizmo-public/src/master/.
    \item \texttt{yt}: https://yt-project.org/.
    \item \texttt{Amiga Halo Finder}: http://popia.ft.uam.es/AHF/Download.html.
    \item \texttt{FIRE Studio}: https://github.com/agurvich/FIRE\_studio.
    \item \texttt{WebPlotDigitizer}: https://apps.automeris.io/wpd/.
\end{itemize}
The Python scripts used to create the figures in the current study are available from the corresponding author on reasonable request, contingent on approval by the FIRE Collaboration on a case-by-case basis.

\end{methods}

\begin{addendum}
 
\item Sabbatical leave support from Pomona College and the Harry and Grace Steele Foundation (J.M.). NASA Hubble Fellowship grant number HST-HF2-51454.001-A awarded by the Space Telescope Science Institute, which is operated by the Association of Universities for Research in Astronomy, Incorporated, under NASA contract number NAS5-26555 (S.D.). NSF grant number AST- 1910346 (J.S.B., F.J.M. and S.Y.). Swiss National Science Foundation grant numbers PP00P2\_157591, PP00P2\_194814 and 200021\_188552; Swiss National Supercomputing (CSCS) project IDs s697 and s698 (R.F.). NSF Research Grant numbers 1911233 and 20009234, NSF CAREER grant number 1455342, NASA grant numbers 80NSSC18K0562 and HST-AR-15800.001-A (P.F.H.). NASA grant number 80NSSSC20K1469 (A.L.). Gary McCue postdoctoral fellowship through the Center for Cosmology at UC Irvine (Z.H.). National Science Foundation Graduate Research Fellowship under grant number DGE-1839285 (C.K.). NSF CAREER grant number 2045928; NASA ATP grant numbers 80NSSC18K1097 and 80NSSC20K0513; HST grant numbers AR-15809 and GO-15902 from STScI; Scialog Award from the Heising-Simons Foundation; Hellman Fellowship (A.W.). NSF grant number AST-2009687 and by the Flatiron Institute, which is supported by the Simons Foundation (D.A.-A). NSF CAREER award number AST-1752913, NSF grant number AST-1910346, NASA grant numbers NNX17AG29G and HST-AR-15006, HST-AR- 15809, HST-GO-15658, HST-GO-15901, HST-GO-15902, HST-AR-16159 and HST-GO-16226 from STScI (M.B.-K.). Simons Investigator Award from the Simons Foundation; NSF grant number AST-1715070 (E.Q.). NSF through grant numbers AST- 1715216 and CAREER award AST-1652522; by NASA through grant number 17-ATP17-0067; by STScI through grant number HST-AR-16124.001-A; and by the Research Corporation for Science Advancement through a Cottrell Scholar Award and a Scialog Award (C.-A.F.-G.). NSF grant number AST-1715101 (D.K.). Numerical calculations were run on the Caltech compute cluster “Wheeler,” allocations FTA-HopkinsAST20016 supported by the NSF and TACC and NASA HEC SMD-16-7592. We acknowledge PRACE for awarding us access to MareNostrum at the Barcelona Supercomputing Center (BSC), Spain. This research was partly carried out via the Frontera computing project at the Texas Advanced Computing Center. Frontera is made possible by NSF award number OAC-1818253. We acknowledge access to Piz Daint at the Swiss National Supercomputing Centre, Switzerland under the University of Zurich’s share with the project ID uzh18. Additional computing support was provided by S3IT resources at the University of Zurich. We thank Y. Jing for patiently answering every question we had regarding their 2019 paper, and for sharing the data we requested. We also thank P. van Dokkum, J. Hudgings, D. Whitaker, D. Tanenbaum and R. Gaines for comments on an earlier draft, and C. Hayward for data-transfer support. J.M. thanks J.S.B., P.F.H., P. Torrey and L. Hernquist for sabbatical-leave hospitality. J.M. (an astronomer of Indigenous ancestry, non-Cherokee) thanks D. Ingram (a Cherokee physicist) for sharing knowledge about the Cherokee Nation and cultural appropriation. This work was conducted on Tongva-Gabrielino land.

\newpage
\item[Author Contributions.] J.M. conducted the analysis and designed the manuscript, with substantial input from S.D. and J.S.B. S.D. collected observational data from the literature and verified the theory–observations comparisons. R.F. ran the cosmological simulation used in this work. O.Ç. generated the halo catalogues and merger trees. A.G. provided support for the creation of mock images. A.L. and C.B.H. provided support with yt. C.K. calculated g-band effective radii, S\'ersic indices and surface brightness contours. C.K. and F.J.M. calculated stellar ages, metallicities and colours; and S.D. conducted the observational comparison. J.S.B, R.F., O.Ç. and F.J.M. worked on the discussion of numerical stripping, with substantial input from F.J. Z.H. contributed to the literature comparison. All authors contributed to the final creation of the manuscript and figures.

\item[Competing Interests.] The authors declare that they have no competing financial interests.

\item[Correspondence.] Correspondence and requests for materials
should be addressed to \newline J.M.~(email: jorge.moreno@pomona.edu).

Published online version available at https://www.nature.com/articles/s41550-021-01598-4.

Supplementary material available at https://doi.org/10.1038/s41550-021-01598-4.

DOI number: 10.1038/s41550-021-01598-4.

Scheduled for publication in Nature Astronomy on 14 February 2022 at 16:00 (London time), 14 February 2022 at 11:00 (US Eastern Time).

\end{addendum}

\newpage



\section*{Supplementary Information}

Supplementary Figs. 1–9, Tables 1–4 and Discussion.

\begin{figure}
   
\includegraphics[width=\columnwidth]{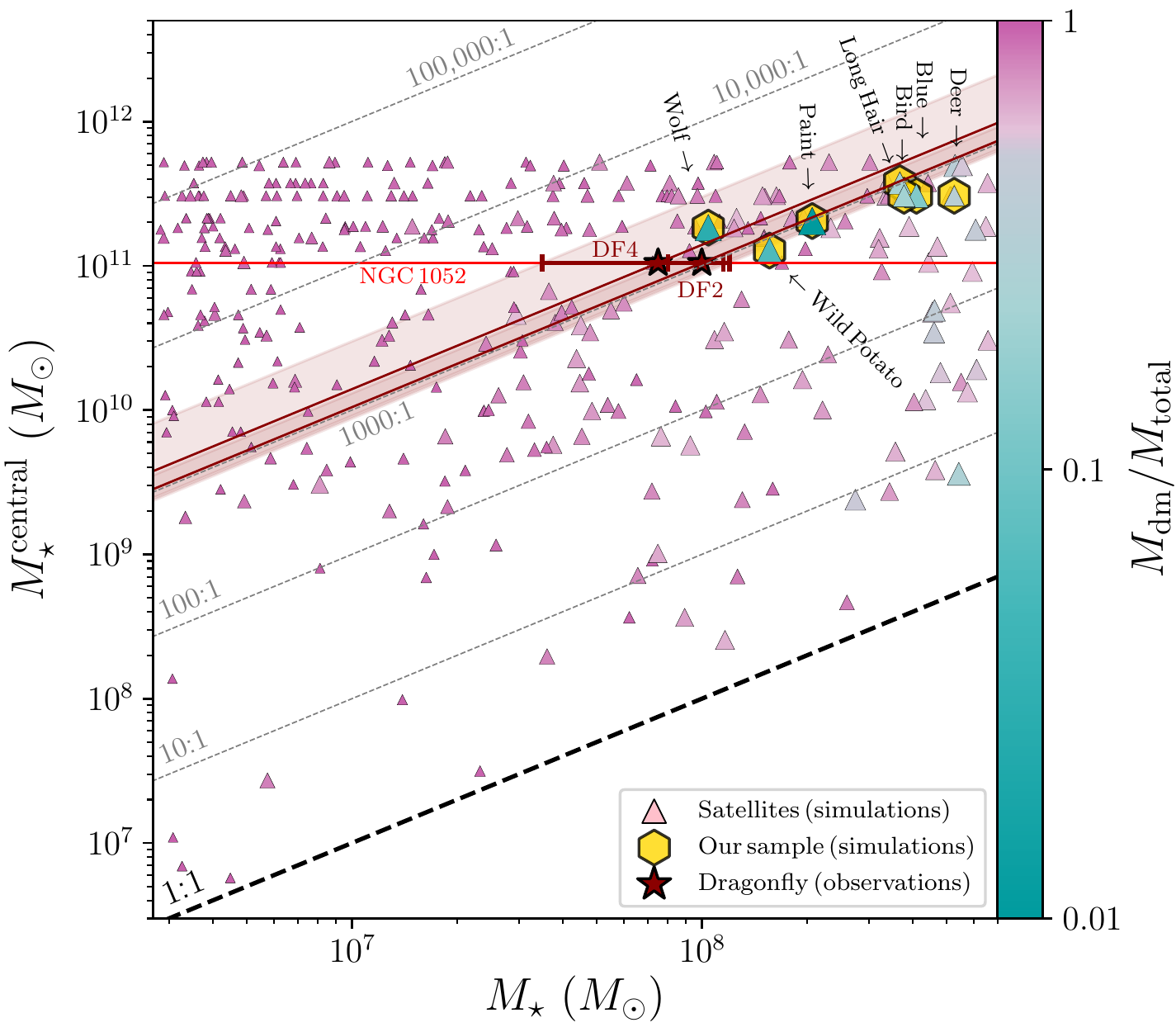}
\def\figurename{Supplementary Figure}
\vspace{-.7cm}
\caption{{\bf Stellar masses of satellite and central galaxies.}  We display the stellar mass of the central galaxy (within $r^{\star}_{\rm 80}$) versus that of each satellite (within $r^{\star}_{\rm 50}$). We adopt the same symbol and colour scheme as in Figure~4. The horizontal red line refers to NGC 1052's stellar mass and the dark-red stars represent DF2 and DF4. The diagonal light-gray dashed lines indicate constant central-to-satellite stellar mass ratios. The brown diagonal lines (and bands) indicate the stellar-mass ratios between DF2 or DF4 and NGC 1052 (and their uncertainties).}
\label{fig:figsi1}
\end{figure}

\subsection{Central and satellite stellar masses.}
\label{subsec:host}

Supplementary Fig.~\ref{fig:figsi1} shows the stellar mass of the central galaxies (hosting galaxies in our low-mass sample) versus that of their satellites. We adopt $r^{\star}_{\rm 80}$ for the former (to better capture the spatial extent of the galaxy\cite{Mowla2019}) and $r^{\star}_{\rm 50}$ for the latter (for easy comparisons with observations). The diagonal lines denote constant stellar-mass ratios. Our seven dark-matter deficient galaxies (with stellar masses between $10^{8-9}M_{\odot}$) accompany massive central galaxies in a narrow stellar-mass regime between $(1.3-3.7)\times 10^{11}M_{\odot}$. The stellar mass ratios of our seven galaxies are similar to that of DF2 and DF4 relative to NGC 1052 (diagonal brown lines and bands).

\begin{figure}
\def\figurename{Supplementary Figure}
    \vspace{-.5cm}
    \hbox{
    \hspace{0.cm}
	\includegraphics[width=1.57in]{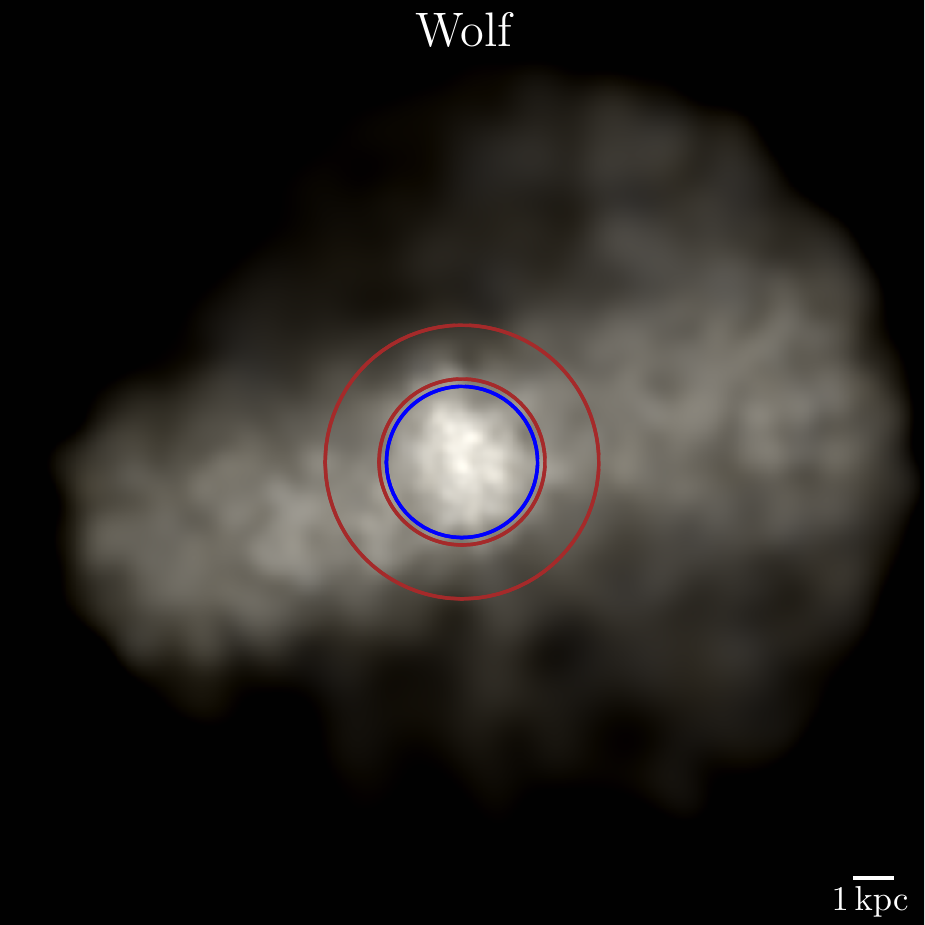}
	\includegraphics[width=1.57in]{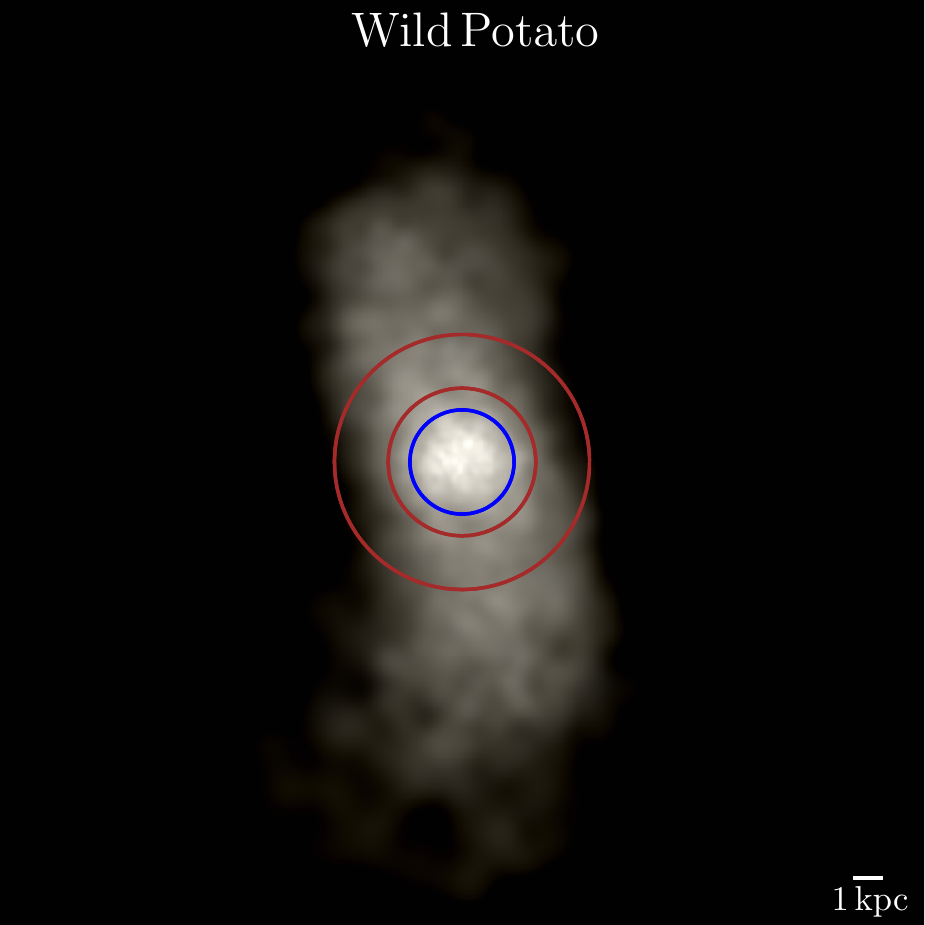}
	\includegraphics[width=1.57in]{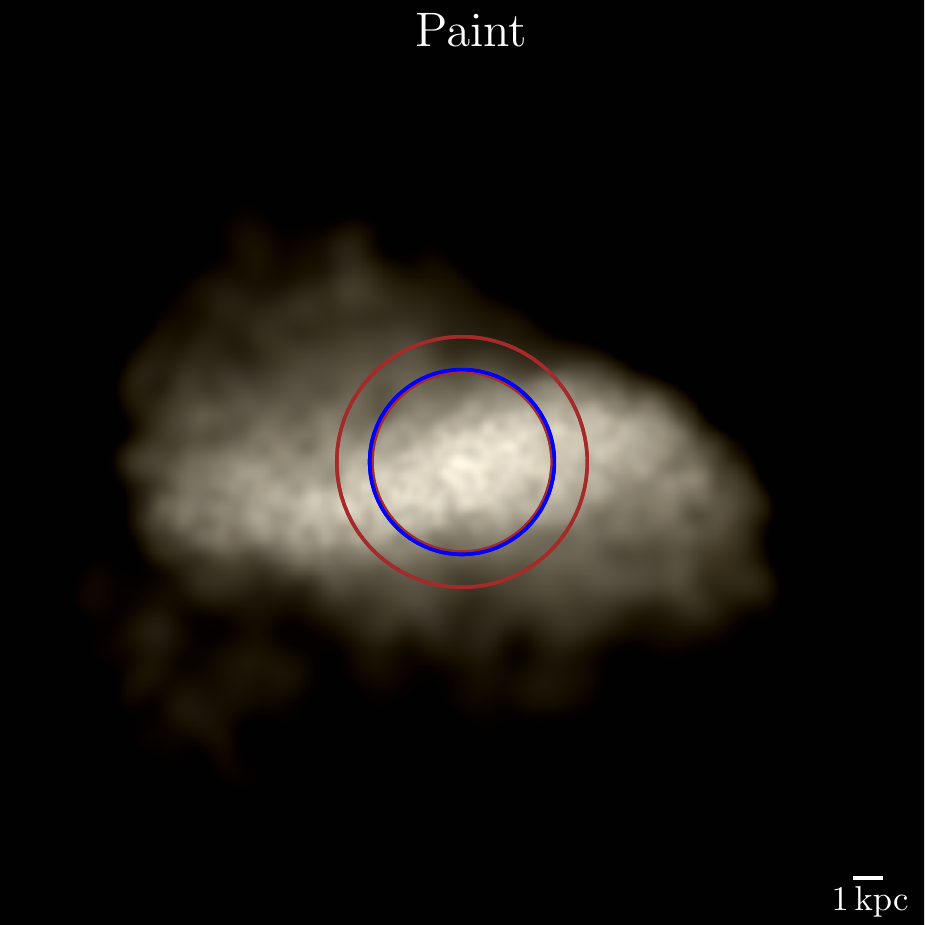}
	\includegraphics[width=1.57in]{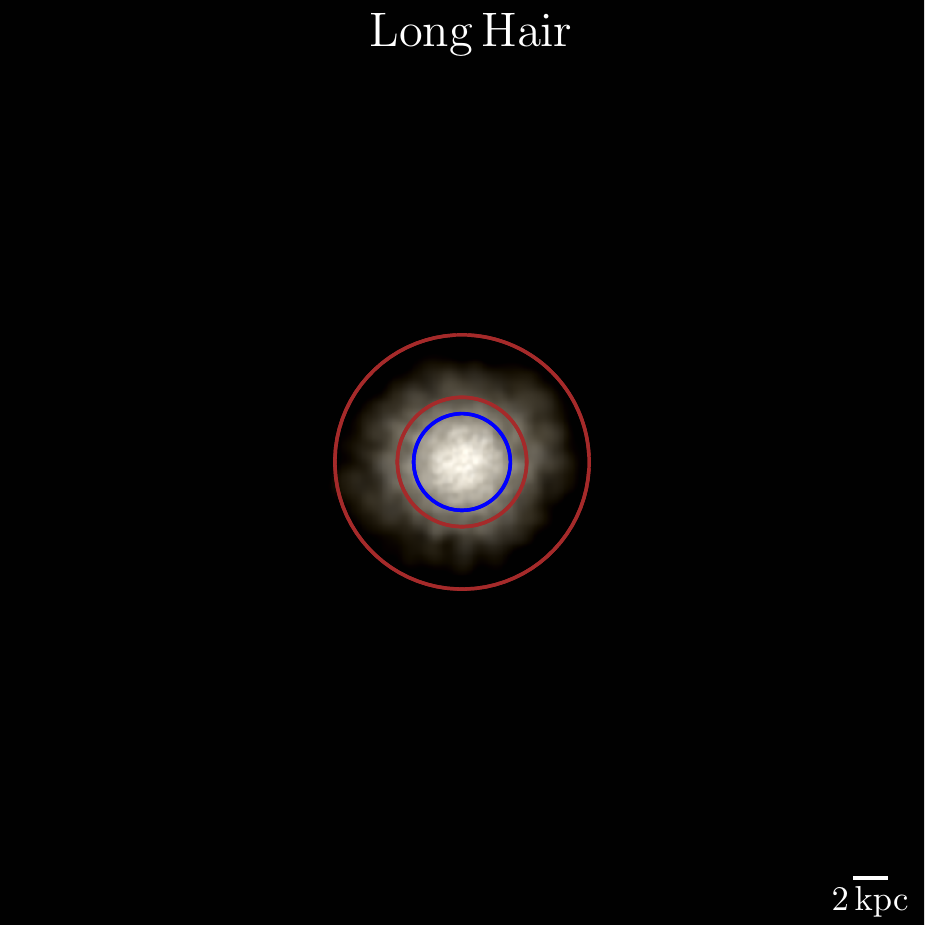}
    } 
    \hbox{
    \hspace{0cm}
	\includegraphics[width=1.57in]{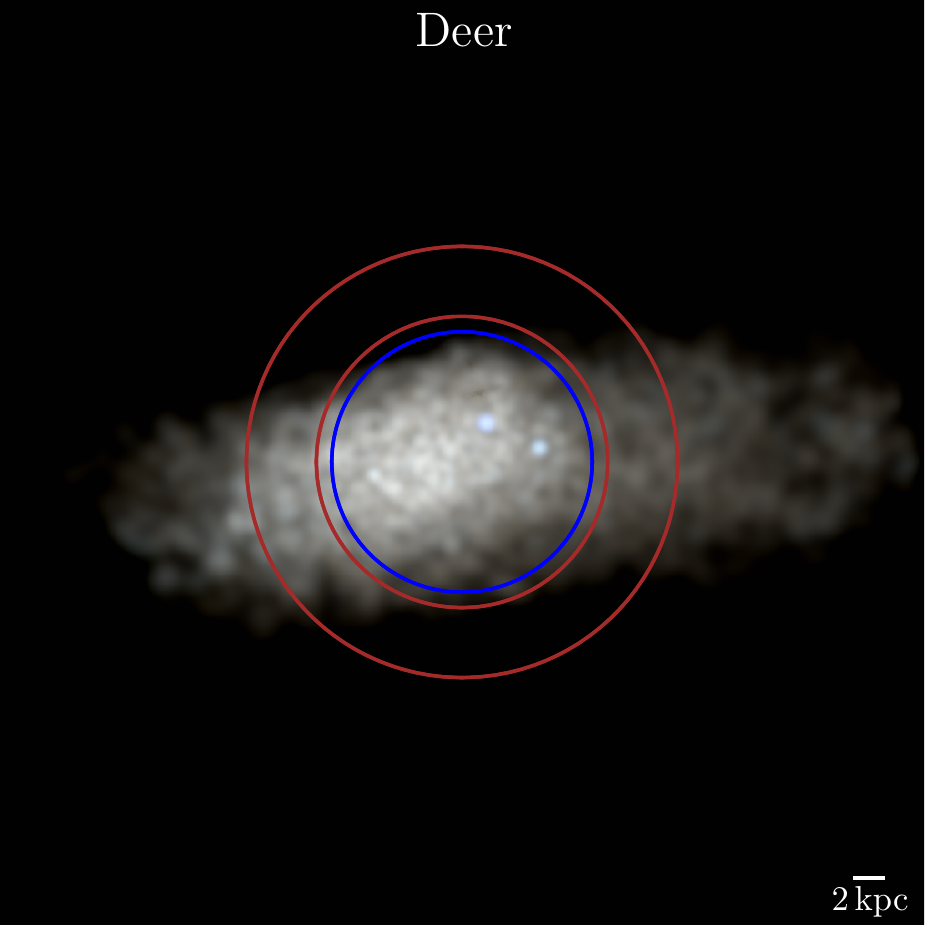}
	\includegraphics[width=1.57in]{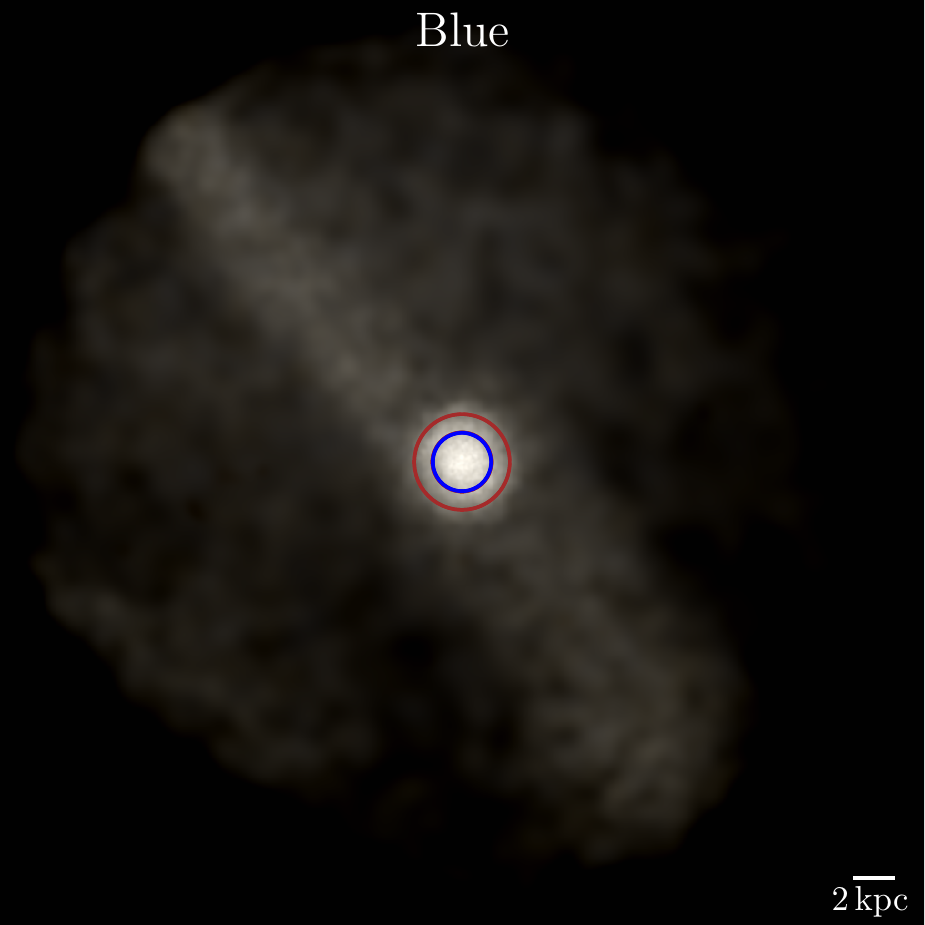}
        \includegraphics[width=1.57in]{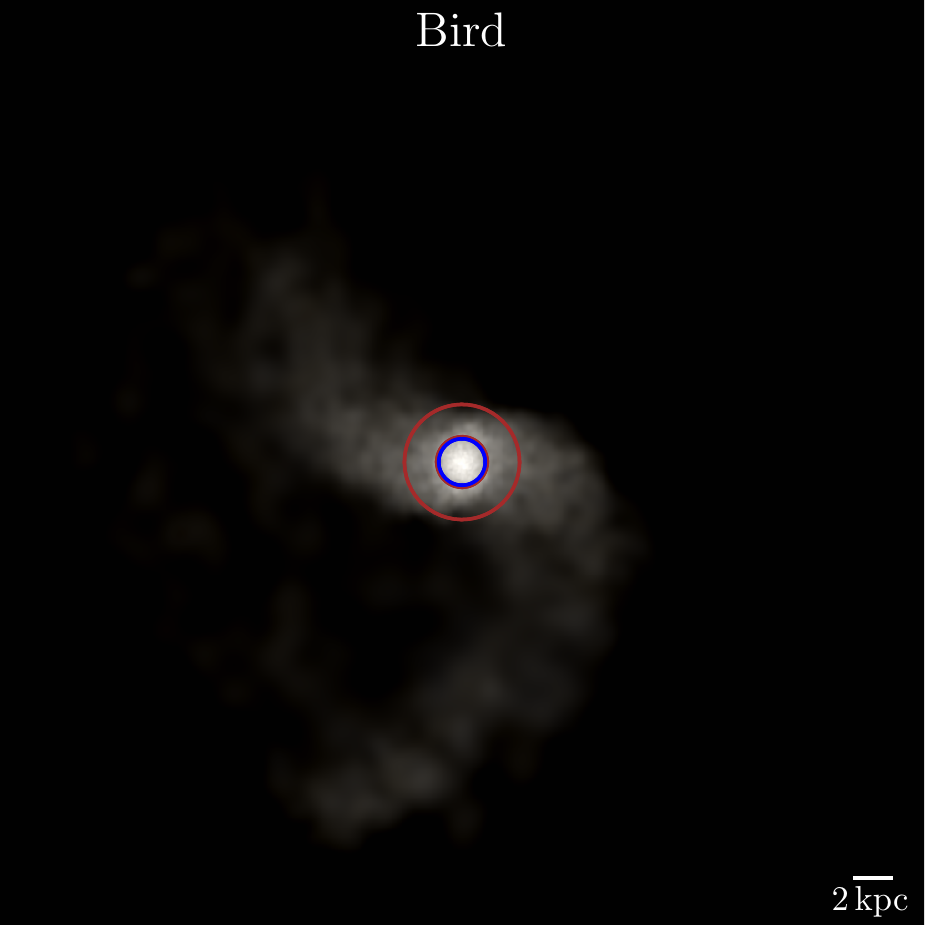}
    } 
    \vspace{-.2cm}
    \caption{{\bf Images of the seven dark-matter deficient galaxies}. Mock Hubble Space Telescope $u/g/r$ composite stellar images (inner and outer red circles denote $r^{\star}_{\rm 50}$ and $r^{\star}_{\rm 80}$, blue circle represents $(4/3)R^{\rm 2D}_{\rm e}$ in the $g$-band; field-of-view equals the subhalo radius).}
    \label{fig:figsi2}
\end{figure}

\begin{figure}
\def\figurename{Supplementary Figure}
    \vspace{-.5cm}
    \hbox{
    \hspace{0.cm}
	\includegraphics[width=1.57in]{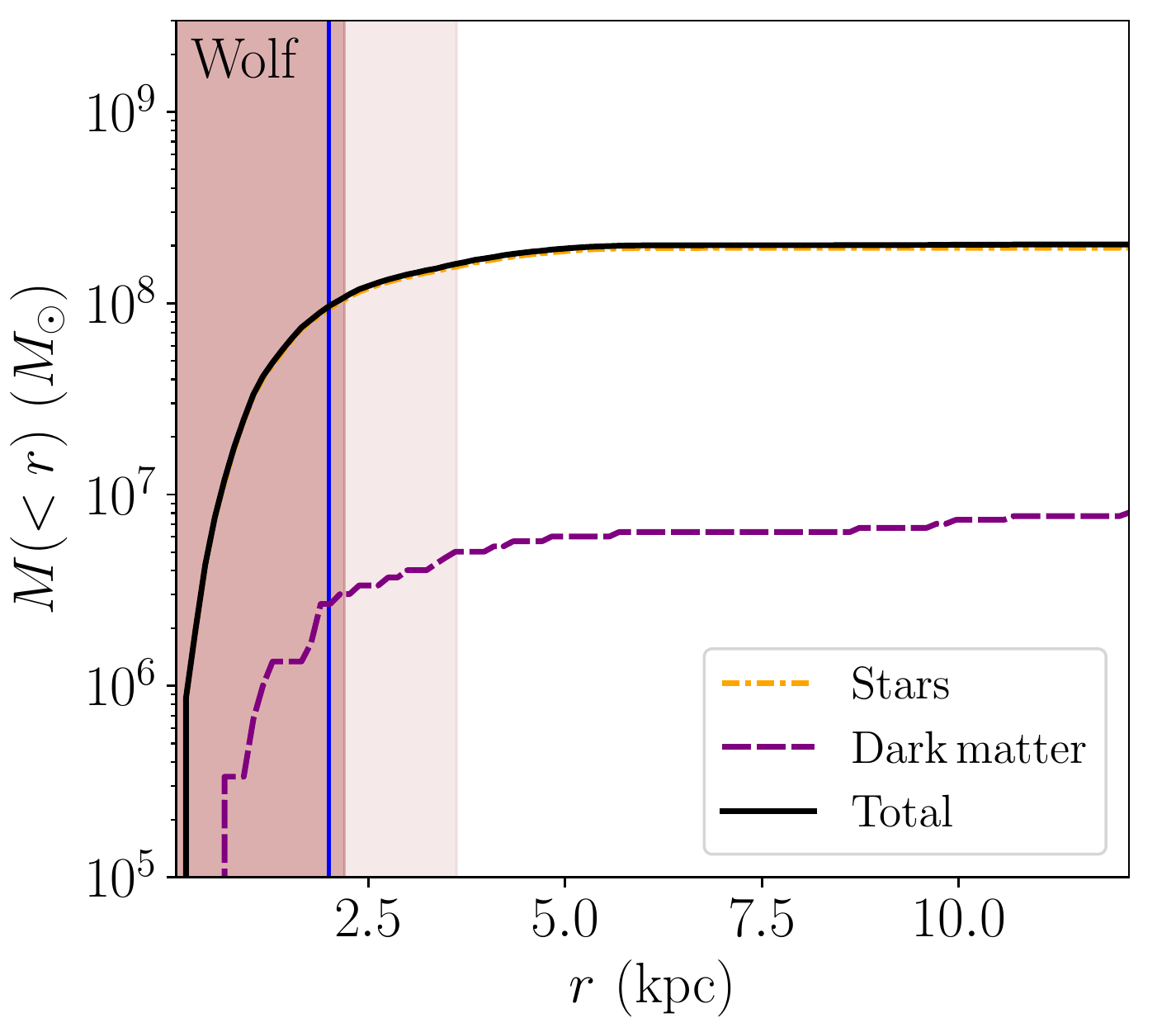}
	\includegraphics[width=1.57in]{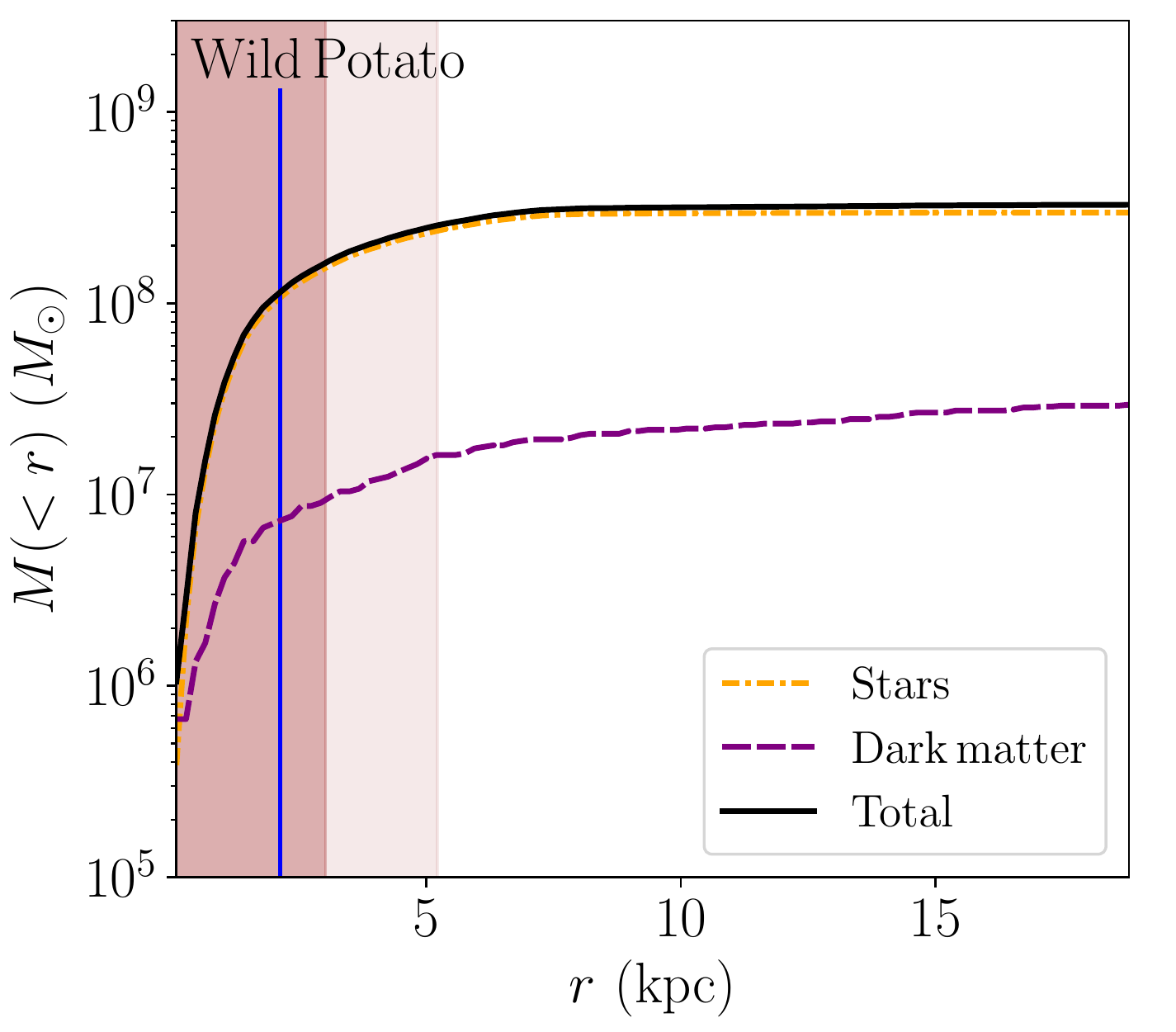}
	\includegraphics[width=1.57in]{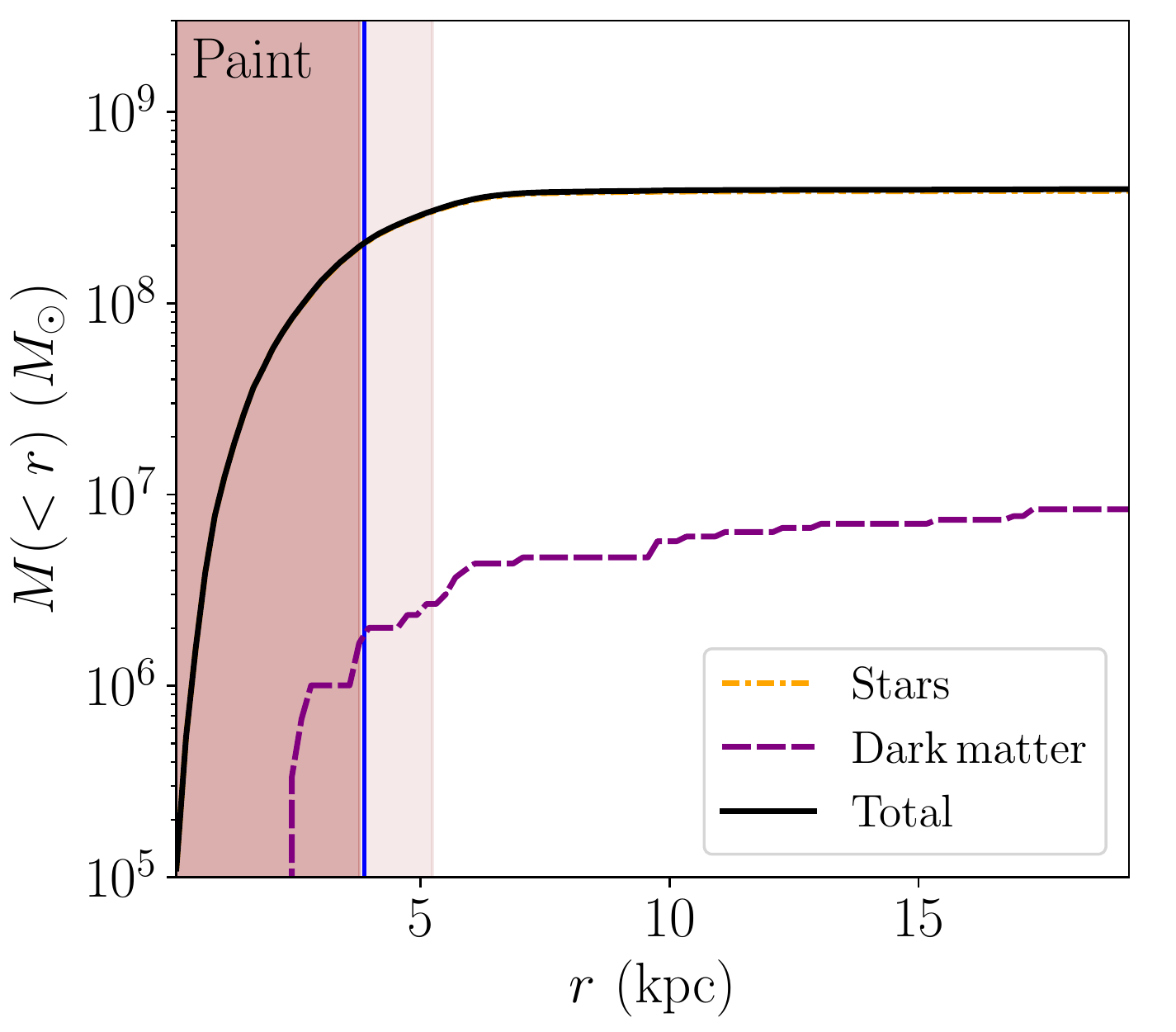}
	\includegraphics[width=1.57in]{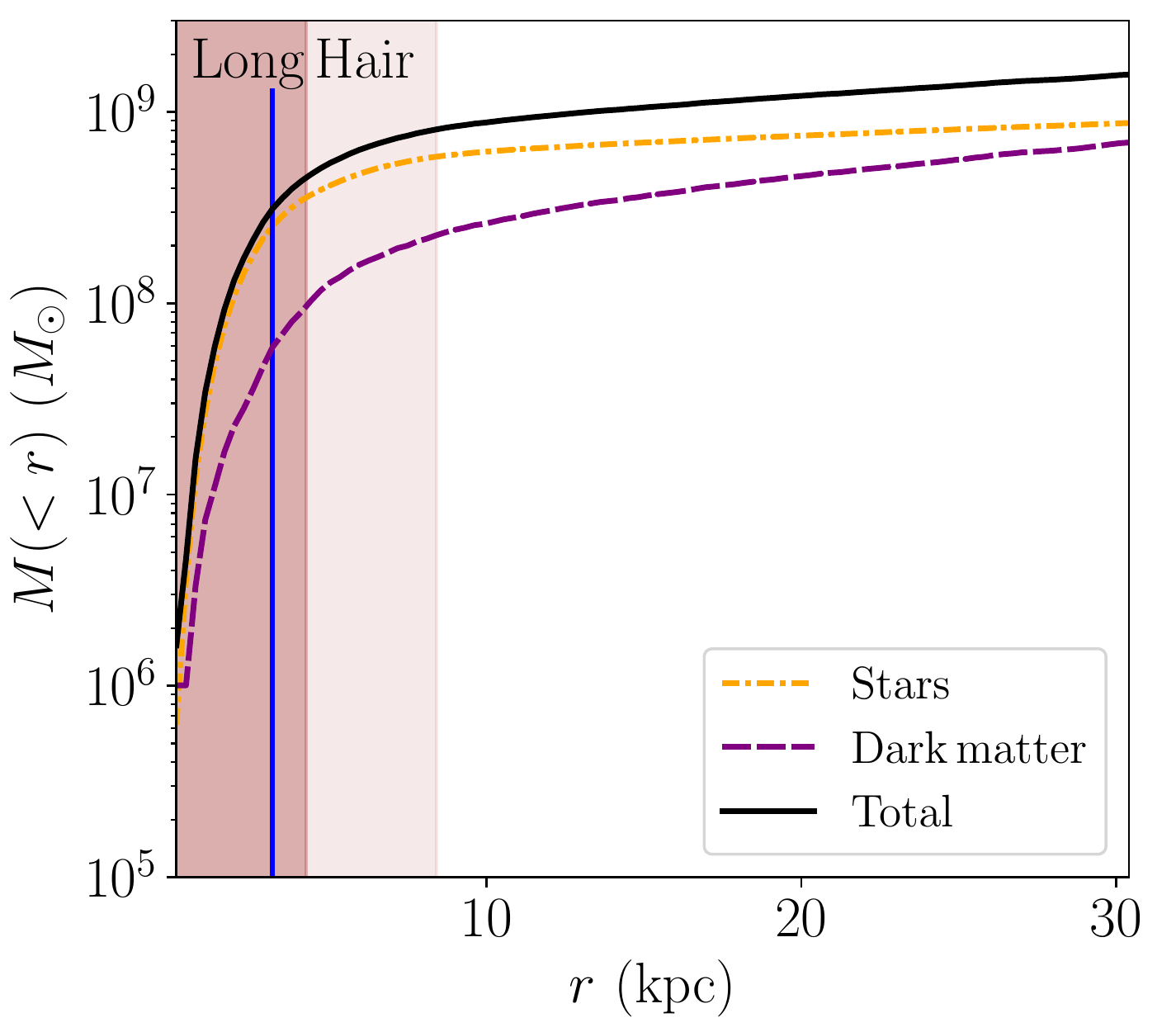}
    } 
    \hbox{
    \hspace{0cm}
	\includegraphics[width=1.57in]{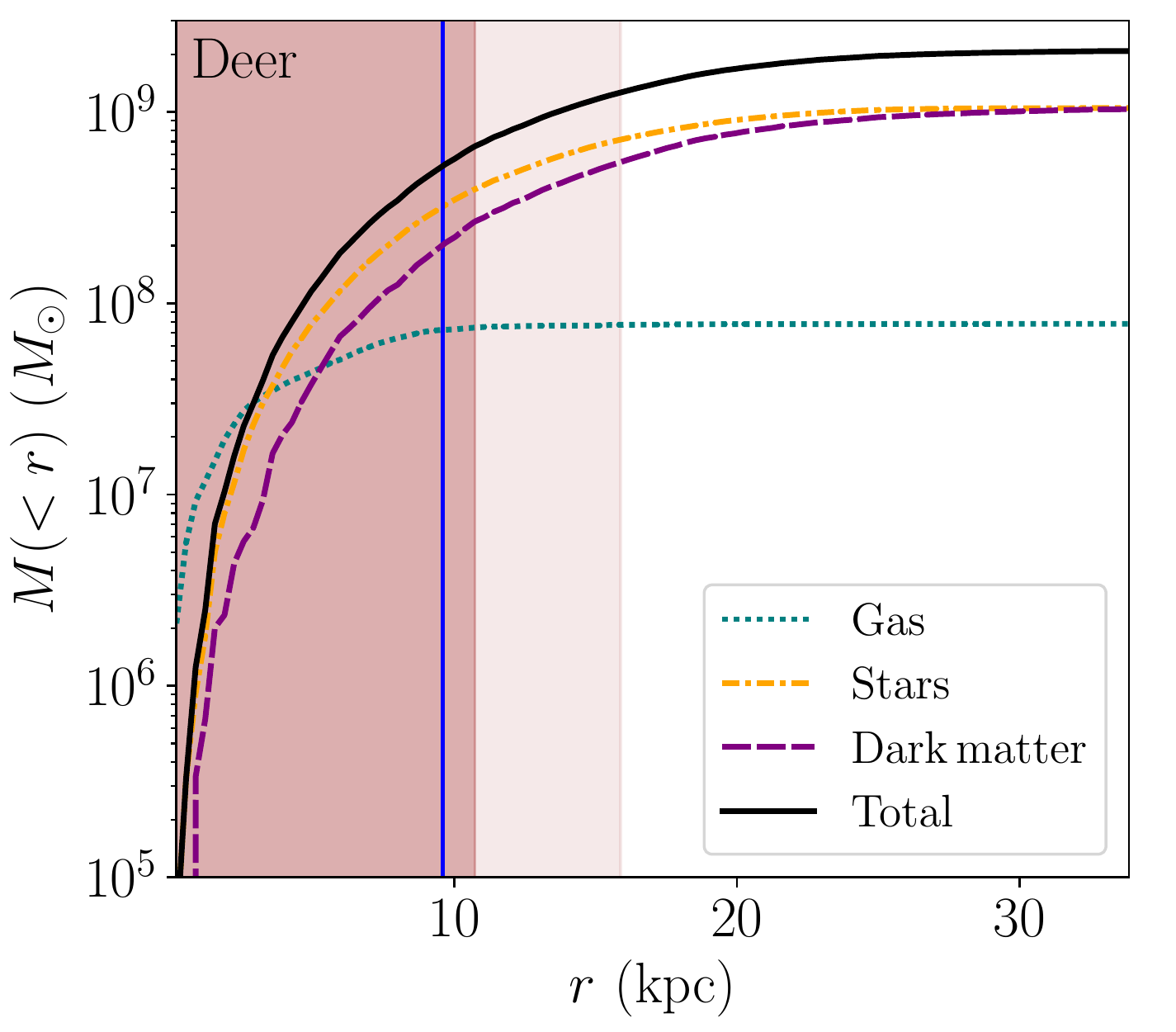}
	\includegraphics[width=1.57in]{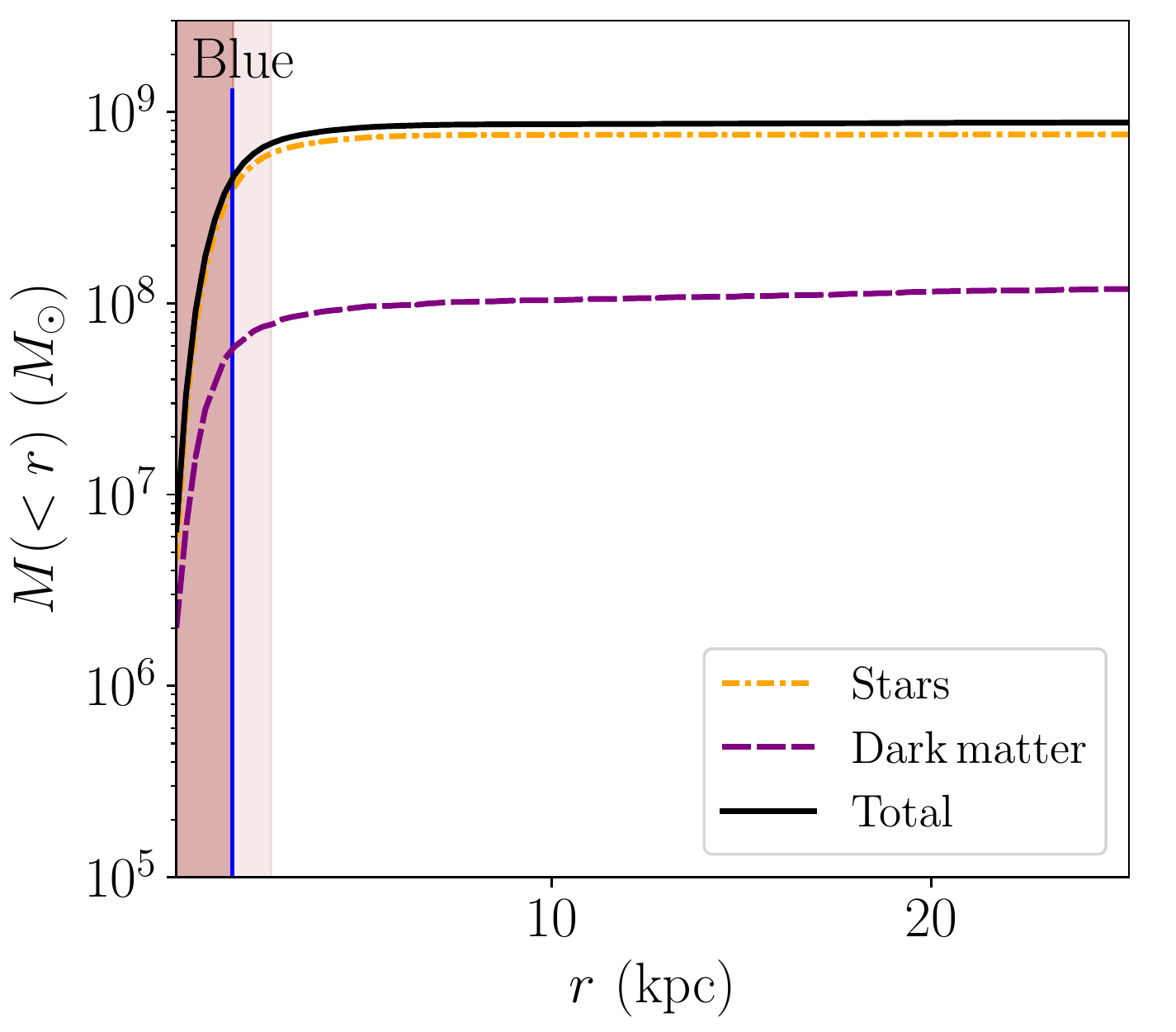}
        \includegraphics[width=1.57in]{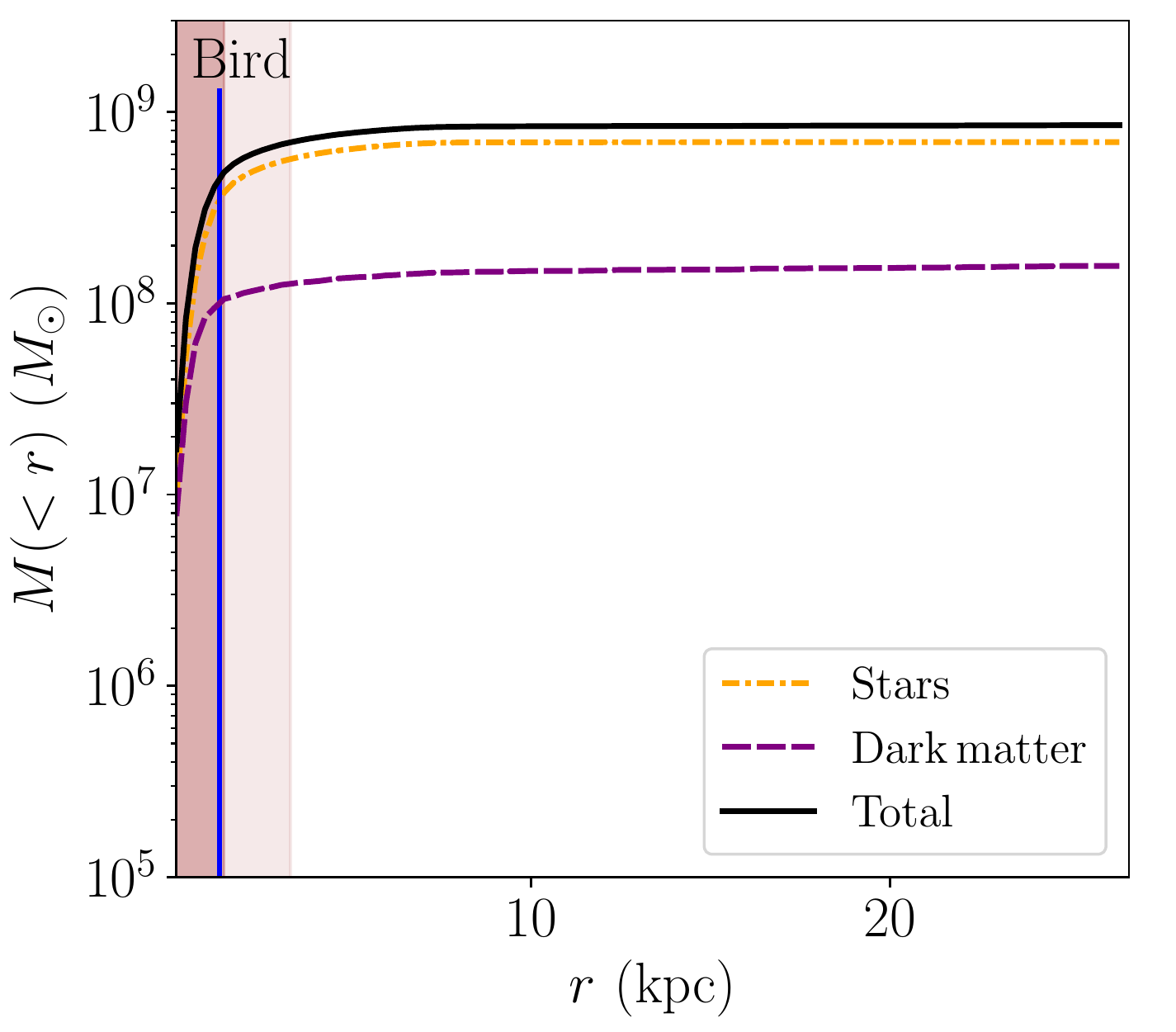}
    } 
    \vspace{-.2cm}
    \caption{{\bf Cumulative mass profiles of the seven dark-matter deficient galaxies}. Dotted-teal, dot-dashed-orange and dashed-purple curves denote gas, stars and dark matter. Solid black represents their sum. The dark and light rectangles denote $r^{\star}_{\rm 50}$ and $r^{\star}_{\rm 80}$; the blue vertical line denotes $(4/3)R^{\rm 2D}_{\rm e}$ ($g$-band).}
    \label{fig:figsi3}
\end{figure}

\begin{figure}
\def\figurename{Supplementary Figure}
    \vspace{-.5cm}
    \hbox{
    \hspace{0.cm}
	\includegraphics[width=1.57in]{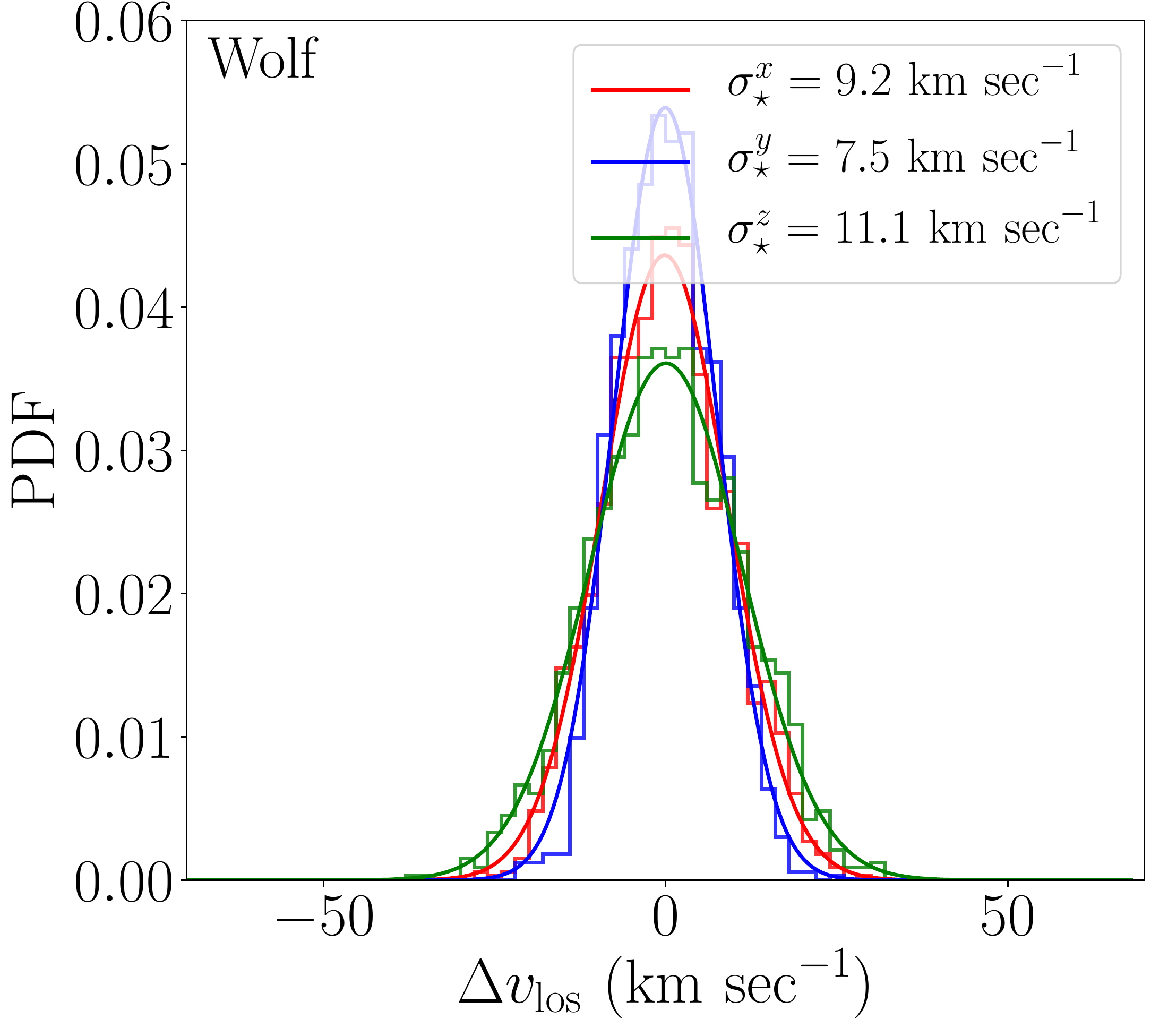}
	\includegraphics[width=1.57in]{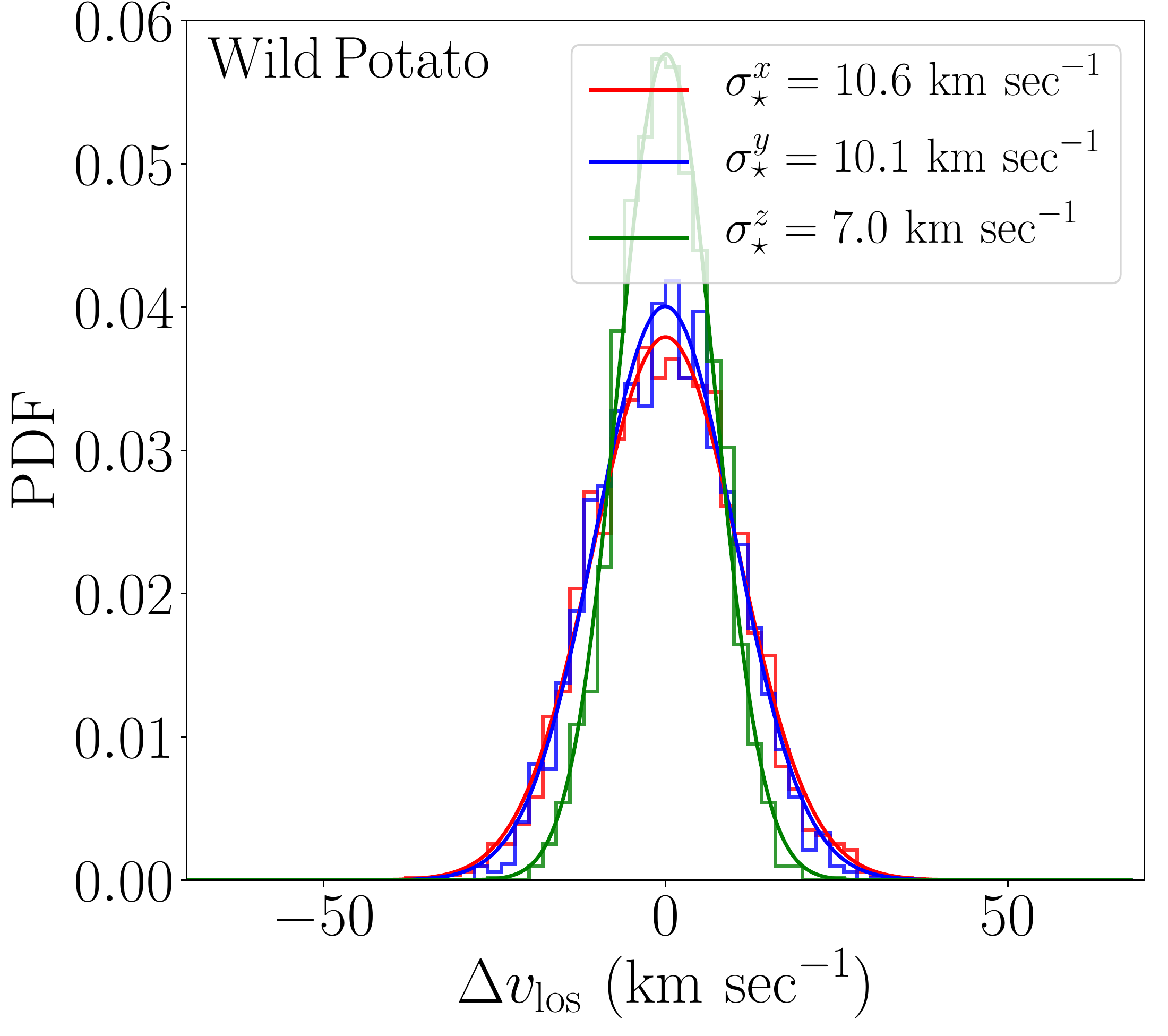}
	\includegraphics[width=1.57in]{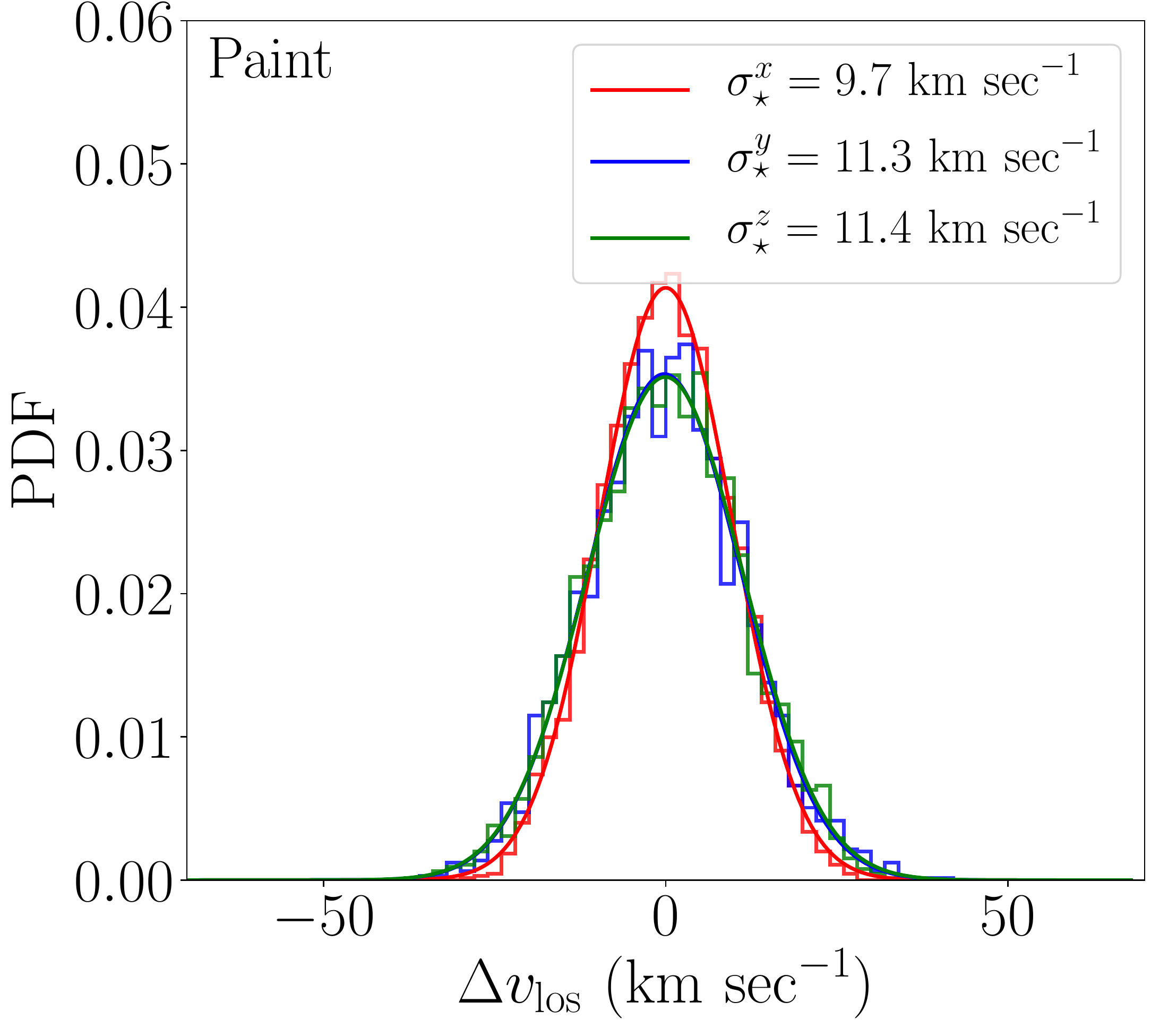}
	\includegraphics[width=1.57in]{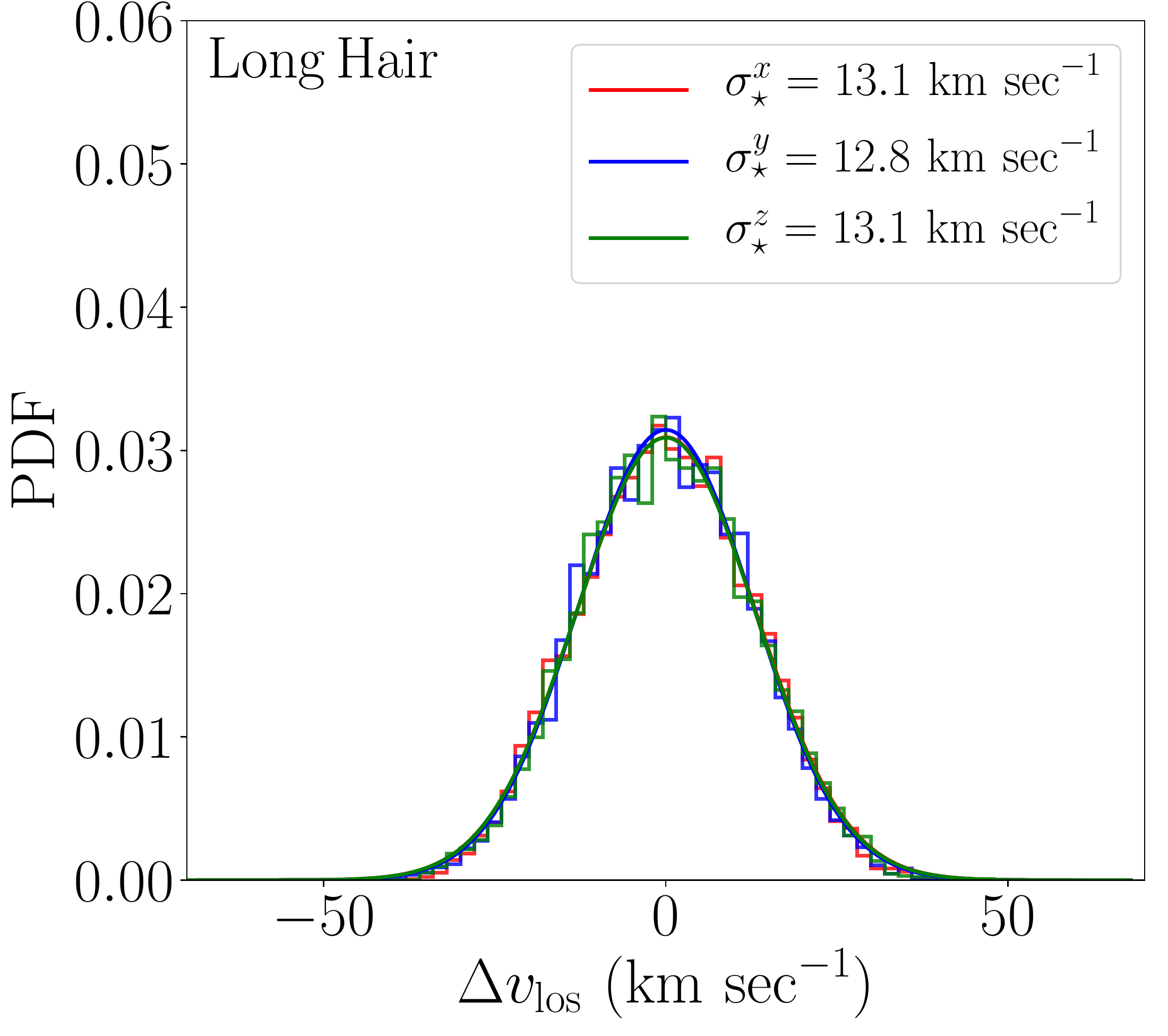}
    } 
    \hbox{
    \hspace{0cm}
	\includegraphics[width=1.57in]{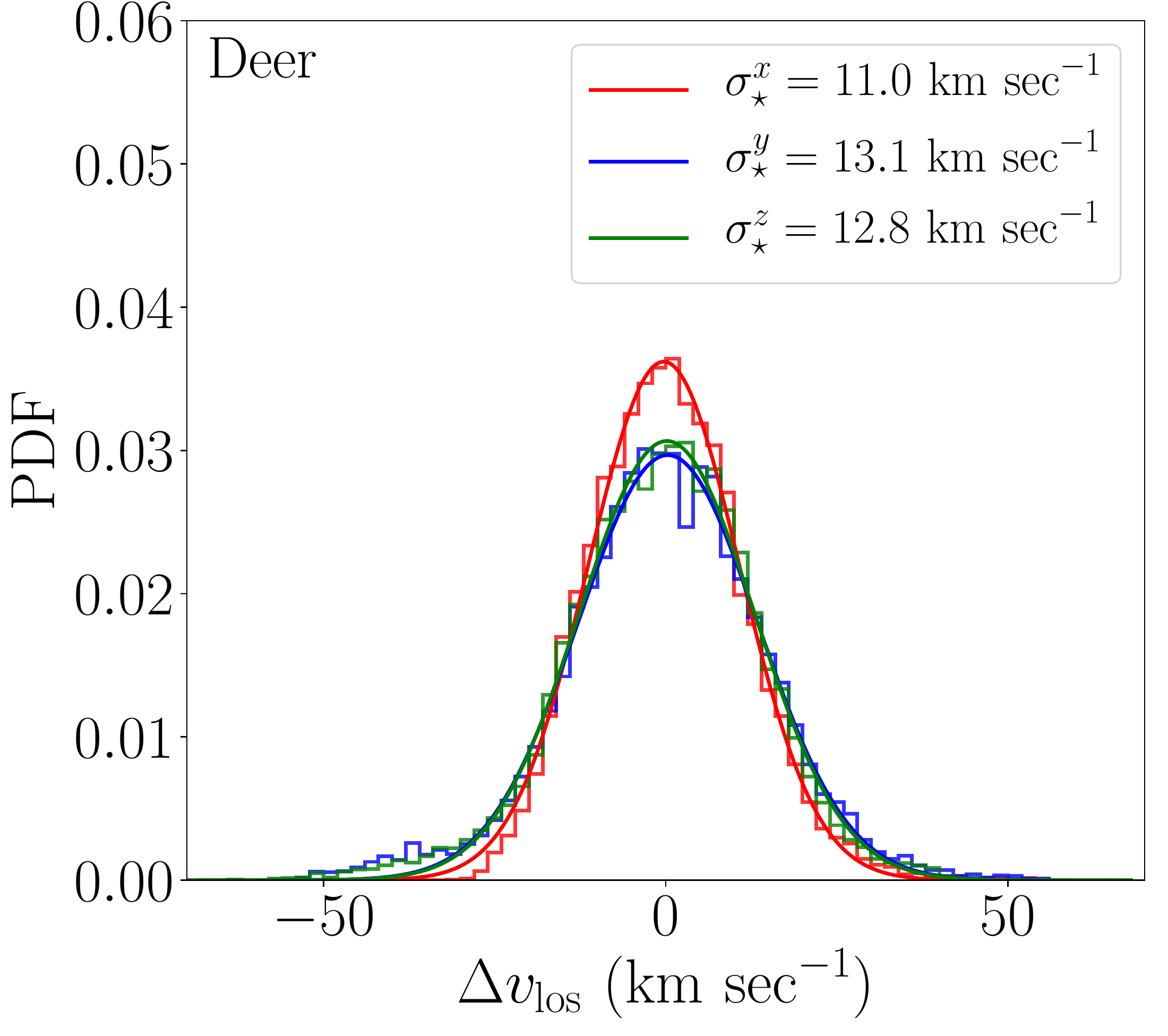}
	\includegraphics[width=1.57in]{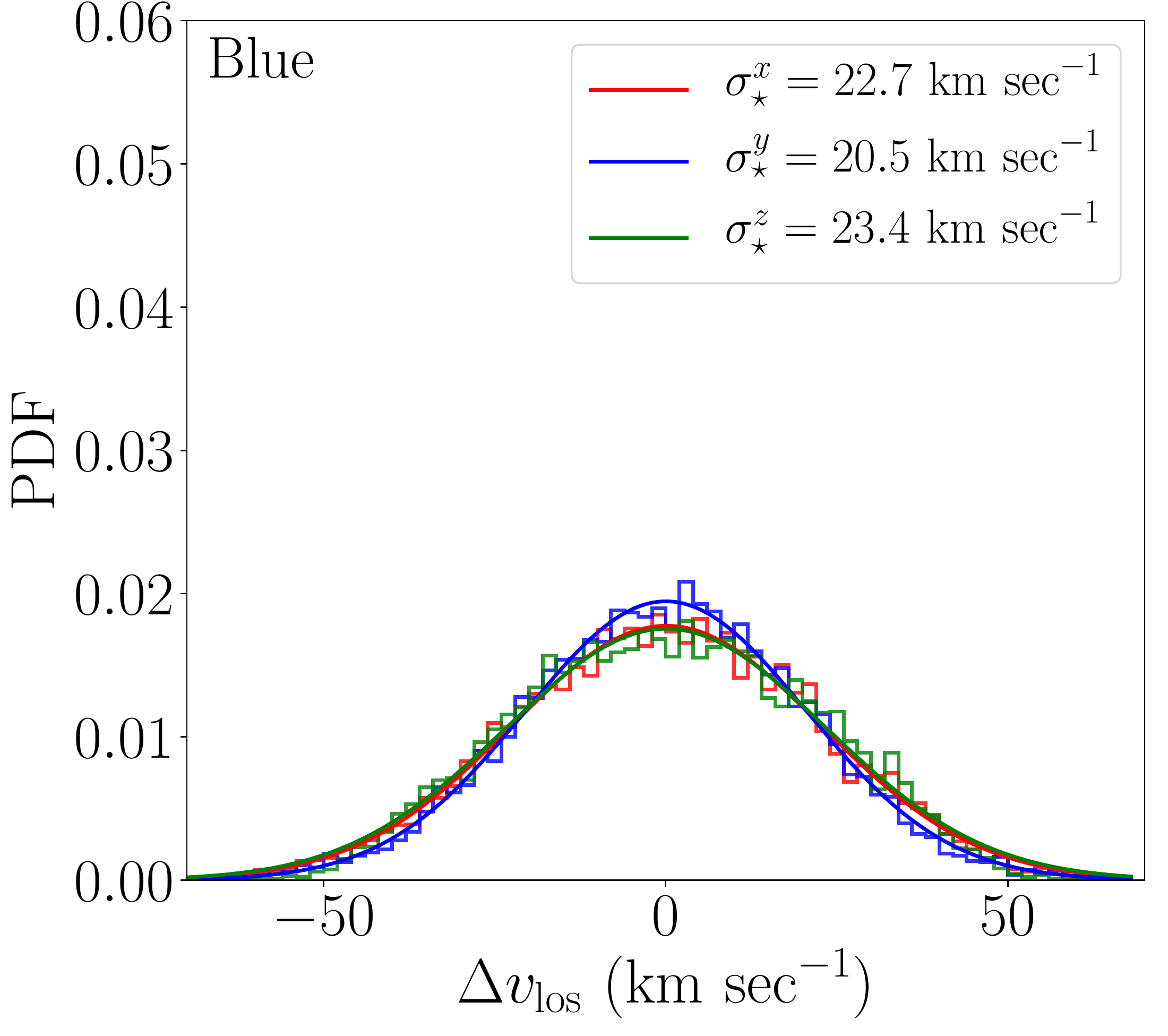}
        \includegraphics[width=1.57in]{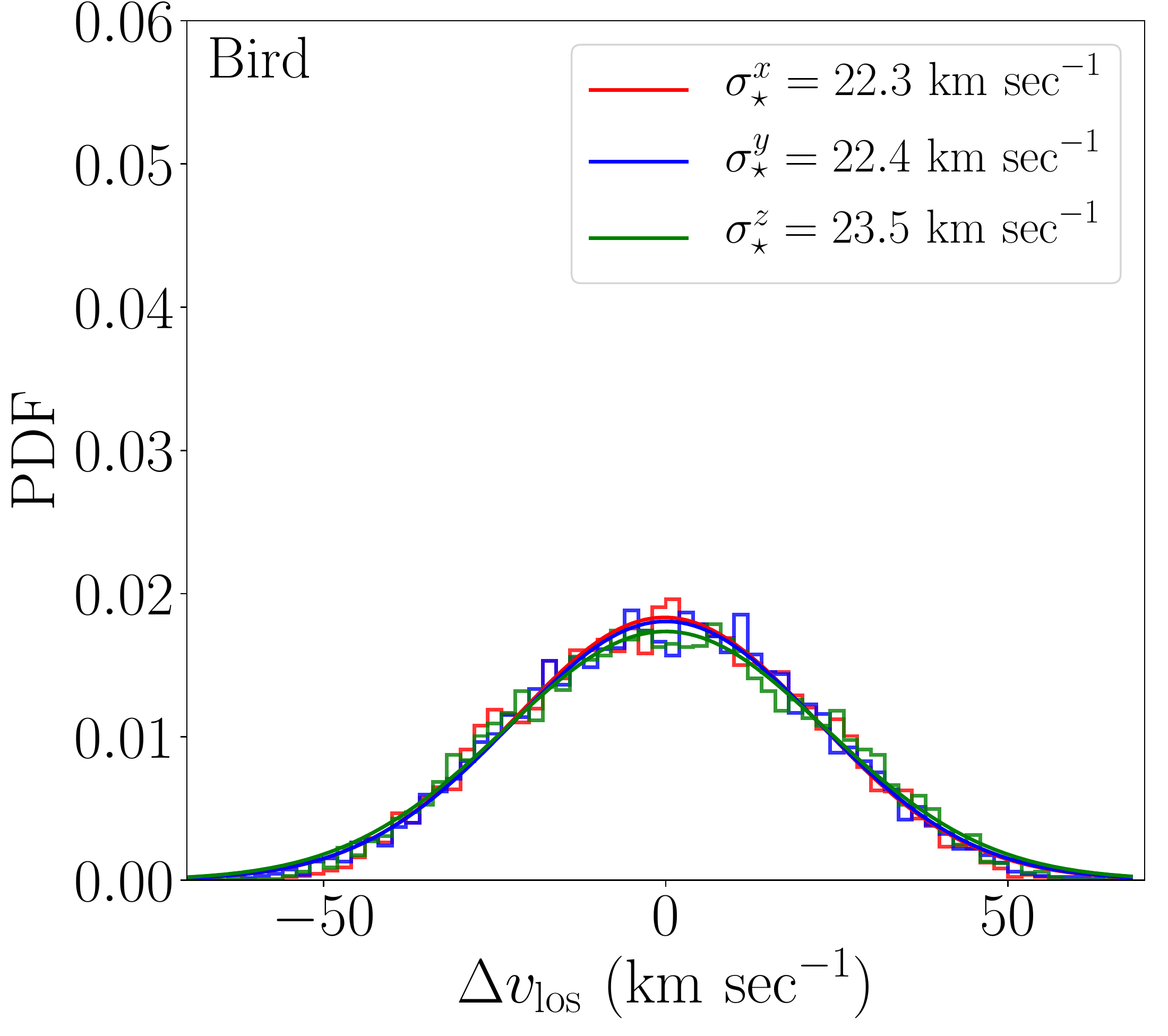}
    } 
    \vspace{-.2cm}
    \caption{{\bf Velocity dispersions for the seven dark-matter deficient galaxies}. Line-of-sight quantities within $r^{\star}_{\rm 50}$ (stellar component) along three random orthogonal directions ($x$, $y$ and $z$ directions in the frame of the box).}
    \label{fig:figsi4}
\end{figure}

\begin{table}
\begin{center}
\def\tablename{Supplementary Table}
 \begin{tabular}{c  c c c c c c c  c c }
 \hline
  & ${\rm Bird}$ & ${\rm Blue}$ & ${\rm Deer}$ & ${\rm LH}$ & ${\rm Paint}$ & ${\rm WP}$ & ${\rm Wolf}$ & ${\rm DF2}$ & ${\rm DF4}$\\ [0.5ex] 
 \hline\hline
 $M_{\star} \, (10^{8}M_{\odot})$ & ${\rm 3.8}$ & ${\rm 4.1}$ & ${\rm 5.3}$ & ${\rm 3.7}$ & ${\rm 2.1}$ & ${\rm 1.6}$ & ${\rm 1.0}$ & ${\rm 1.0}\pm{\rm 0.2}$ & ${\rm 0.75}\pm{\rm 0.4}$ \\ 
 \hline
  $M_{\rm gas} \, (10^{8}M_{\odot})$ & ${\rm 0.0}$ & ${\rm 0.0}$ & ${\rm 0.5}$ & ${\rm 0.0}$ & ${\rm 0.0}$ & ${\rm 0.0}$ & ${\rm 0.0}$ & $-$ & $-$ \\ 
 \hline
 $M_{\rm dm } \, (10^{8}M_{\odot})$ & ${\rm 1.0}$ & ${\rm 0.6}$ & ${\rm 3.1}$ & ${\rm 1.0}$ & ${\rm 0.02}$ & ${\rm 0.1}$ &  ${\rm 0.03}$ & $-$ & $-$ \\ 
 \hline
 $M_{\rm total} \, (10^{8}M_{\odot})$ & ${\rm 4.8}$ & ${\rm 0.6}$ & ${\rm 8.9}$ & ${\rm 4.7}$ & ${\rm 2.1}$ & ${\rm 1.7}$ &  ${\rm 1.1}$ & $1.3 \pm 0.8$ & $-$ \\ 
 \hline
  $\sigma^{\rm 1D}_{\star} \, ({\rm km}\,{\rm sec}^{-1})$ & ${\rm 22.3}$ & ${\rm 22.0}$ & ${\rm 11.7}$ & ${\rm 13.0}$ & ${\rm 10.2}$ & ${\rm 10.5}$ &  ${\rm 8.6}$ & $8.5^{+2.3}_{-3.1}$ & $4.2^{+4.4}_{-2.2}$ \\ 
 \hline
$r^{\star}_{\rm 50} \,  ({\rm kpc})$ & ${\rm 1.5}$ & ${\rm 1.6}$ & ${\rm 10.7}$ & ${\rm 4.3}$ & ${\rm 3.8}$ & ${\rm 3.0}$ &  ${\rm 2.2}$ & $-$ & $-$ \\ 
 \hline
$R^{\rm 2D}_{\rm e} \,  ({\rm kpc})$ & ${\rm 1.0}$ & ${\rm 1.2}$ & ${\rm 7.2}$ & ${\rm 2.4}$ & ${\rm 2.9}$ & ${\rm 1.6}$ &  ${\rm 1.5}$ & $2.2 \pm 0.1$ & $1.6 \pm 0.1$ \\ 
\hline
$n_{\rm S\acute{e}rsic}$ & ${\rm 0.77}$ & ${\rm 0.60}$ & ${\rm 0.70}$ & ${\rm 0.67}$ & ${\rm 0.58}$ & ${\rm 0.73}$ &  ${\rm 0.60}$ & $0.6$ & $0.79$ \\
\hline
$M^{\rm central}_{\star} \, (10^{11}M_{\odot})$ & ${\rm 3.1}$ & ${\rm 3.1}$ & ${\rm 3.1}$ & ${\rm 3.7}$ & ${\rm 2.0}$ & ${\rm 1.3}$ &  ${\rm 1.8}$ & $1.0$ & $1.0$ \\ 
\hline
$M^{\rm host}_{\rm vir} \, (10^{13}M_{\odot})$ & ${\rm 1.8}$ & ${\rm 1.8}$ & ${\rm 1.8}$ & ${\rm 1.4}$ & ${\rm 0.4}$ & ${\rm 0.3}$ &  ${\rm 3.3}$ & $0.6^{+2.6}_{-0.4}$ & $0.6^{+2.6}_{-0.4}$ \\ 
\hline
$d_{\rm host} \, ({\rm kpc})$ & ${\rm 80.2}$ & ${\rm 66.3}$ & ${\rm 174.1}$ & ${\rm 483.8}$ & ${\rm 72.3}$ & ${\rm 56.9}$ &  ${\rm 36.2}$ & $>80$ & $>165$ \\ 
\hline
$d_{\rm min} \, ({\rm kpc})$ & ${\rm 9.1}$ & ${\rm 10.7}$ & ${\rm 31.7}$ & ${\rm 7.4}$ & ${\rm 3.6}$ & ${\rm 10.4}$ &  ${\rm 6.7}$ & $-$ & $-$ \\ 
 [1ex] 
 \hline
\end{tabular}

\caption{{\bf Characteristics of our dark-matter deficient galaxies.} `LH' and `WP' denote Long Hair and Wild Potato. Measurements are taken within $r^{\star}_{\rm 50}$, the radius containing 50\% of the stellar mass, unless stated otherwise. (i) Stellar mass; (ii) gas mass; (iii) dark matter mass; (iv) total mass; (v) 1D line-of-sight velocity dispersion; (vi) galactic size; (vii) effective 2D radius ($g$-band stellar light); (viii) S\'{e}rsic\cite{Sersic1963} index; (ix) stellar mass of central within $r^{\star}_{\rm 80}$; (x) host's virial mass (90\% confidence\cite{Zahid2018}); 
(xii) present-time halo-centric distance (projected distance is used for observations); (xiii) minimum halo-centric distance.}

\label{table:tablesi1}
\end{center}
\end{table}

\subsection{Characteristics of our dark-matter deficient sample.}
\label{subsec:sample}

Supplementary Table~1 lists various properties for our seven dark-matter deficient galaxies. Supplementary Fig.~\ref{fig:figsi2} shows mock Hubble space telescope $u/g/r$ composite stellar images for this set. These images were created with \texttt{Fire Studio}, an open source Python visual package. We employ the same surface brightness limit as in Figure~1: 28.9 mag arcsec$^{-2}$. The field of view corresponds to the subhalo radius, defined as the truncation radius\cite{Knollmann2009} where the density profile experiences an upward ``turn over." The inner and outer red circles denote $r^{\star}_{\rm 50}$ and $r^{\star}_{\rm 80}$, respectively. The blue circles denote $R_{\rm e}$, the $g$-band effective radius (more on how this is measured below). Supplementary Fig.~\ref{fig:figsi3} shows cumulative 3D-radial mass profiles within the subhalo radius. The dark and light vertical rectangles represent $r^{\star}_{\rm 50}$ and $r^{\star}_{\rm 80}$, respectively. The vertical blue line represents $R_{\rm e}$. The dotted-teal, dot-dashed-orange, dashed-purple and solid-black lines represent gas (when available), stars, dark matter and the sum of these (the total mass). Supplementary Fig.~\ref{fig:figsi4} shows histograms and Gaussian best-fitting lines for 1D line-of-sight velocity dispersions with $r^{\star}_{\rm 50}$ along three random orthogonal directions ($x$, $y$ and $z$ in the frame of the simulation box). We report values using particles within a sphere with this radius. Using a projected circular region yields very similar results. 

With the exception of Long Hair, all of  our simulated dark-matter deficient galaxies exhibit low surface brightness tidal features (down to 28.9 mag arcsec$^{-2}$ -- but see Supplementary Fig.~\ref{fig:figsi5} below for an analysis of this object at deeper surface-brightness levels). Deer is the only member of our set with a detectable (albeit subdominant) gaseous component. For Wolf, Wild Potato and Paint, the stellar cumulative mass profiles dominate at all galactocentric radii (the orange and black curves are almost indistinguishable). This is particularly evident for Paint, the galaxy with the lowest dark-matter-to-total mass fraction in our simulation (the purple curve is $\sim$2 dex below the orange curve at all radii). Wolf, Wild Potato and Paint are the only galaxies that achieve 1D line-of-sight velocity dispersions below 10 km sec$^{-1}$ -- which has been a challenge to produce in cosmological simulations\cite{Carleton2019,Sales2020}. In addition to lacking low surface brightness features, Long Hair is also the galaxy with the largest halo-centric distance today. Note that its line-of-sight velocity dispersion histograms and curves along the three orthogonal directions are almost indistinguishable, suggesting that this galaxy has had enough time at distances far away from its host to become dynamically relaxed. 

We recognize that it is observationally impossible to directly measure stellar-mass-based 3D sizes. To emulate what is often done, we also measure the effective radii ($R^{\rm 2D}_{\rm e}$) of the seven dark-matter deficient galaxies by fitting a S\'{e}rsic\cite{Sersic1963} profile to the average surface brightness in the $g$-band along each principal axis. In Supplementary Fig.~\ref{fig:figsi2}, we report the simple average of $R^{\rm 2D}_{\rm e}$ calculated along three (random) orthogonal directions (blue circles, using a deprojection factor\cite{Wolf2010} of $4/3$) -- which agrees well with our 3D method. Supplementary Table~\ref{table:tablesi1} includes these values, along with the associated S\'ersic indices ($n_{\rm S\acute{e}rsic}\simeq0.60-0.77$), which are also in line with observations ($n_{\rm S\acute{e}rsic}=0.60$ and $0.79$ for DF2 and DF4, respectively).

\begin{figure}
\def\figurename{Supplementary Figure}
\includegraphics[width=\columnwidth]{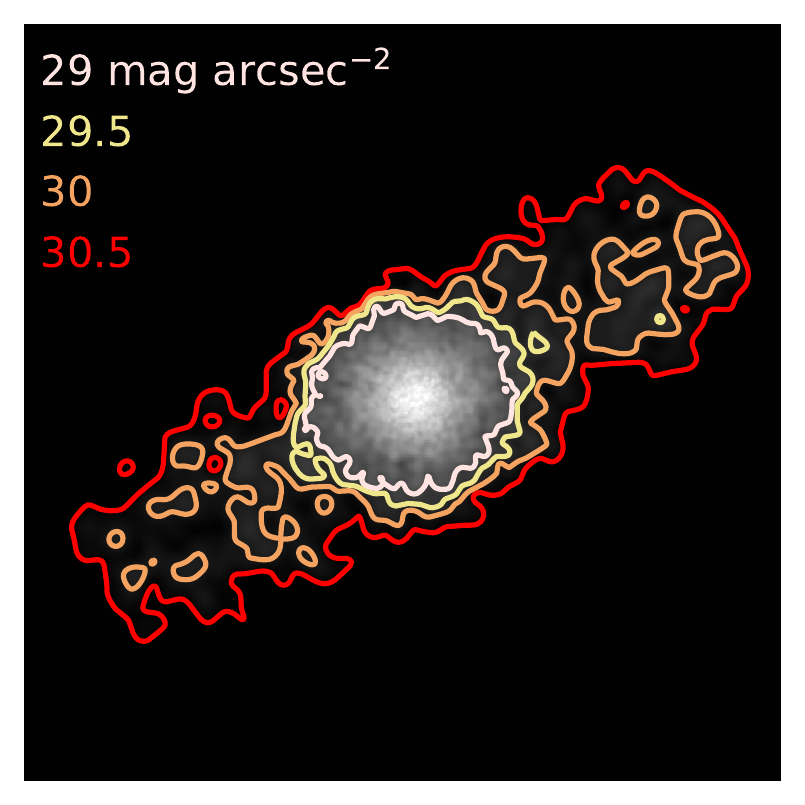}
\vspace{-2cm}
\caption{ {\bf Tidal features.} Mock $g$-band image of Long Hair, limited to 30.5 mag arcsec$^{-2}$. The contours represent $g$-band surface-brightness levels at 29, 28.5, 30 and 30.5 mag arcsec$^{-2}$. Tidal features for this object are evident at surface-brightness levels deeper than 30 mag arcsec$^{-2}$.}
\label{fig:figsi5}
\end{figure}

\subsection{Tidal features.}
\label{subsec:tidal}

Past interactions may leave a detectable imprint on galaxies. This has instigated a debate on whether or not DF2 and DF4 do exhibit tidal features\cite{Montes2020,Montes2021,Keim2021}. The majority of our dark-matter deficient simulated galaxies exhibit tidal features down to 28.9 mag arcsec$^{-2}$. One exception is Long Hair. This object also has the largest halo-centric distance amongst our set of seven, and is dynamically relaxed (see the upper-right panel of Supplementary Fig.~\ref{fig:figsi2}). Supplementary Fig.~\ref{fig:figsi5} explores this object in more detail. The background $g$-band image is now limited to 30.5 mag arcsec$^{-2}$ and the contours correspond to 29, 29.5, 30 and 30.5 mag arcsec$^{-2}$ surface-brightness levels in the $g$-band. We find that, at 30 and 30.5 mag arcsec$^{-2}$, Long Hair does display evident tidal features. We defer further studies on this issue to future work because a careful analysis may require more realistic synthetic data products\cite{Bottrell2019}.

\subsection{The evolution of our dark-matter deficient sample.}
\label{subsec:evolution}

\begin{figure}
\def\figurename{Supplementary Figure}
    \vspace{-.5cm}
    \hbox{
    \hspace{0.cm}
	\includegraphics[width=3.3in]{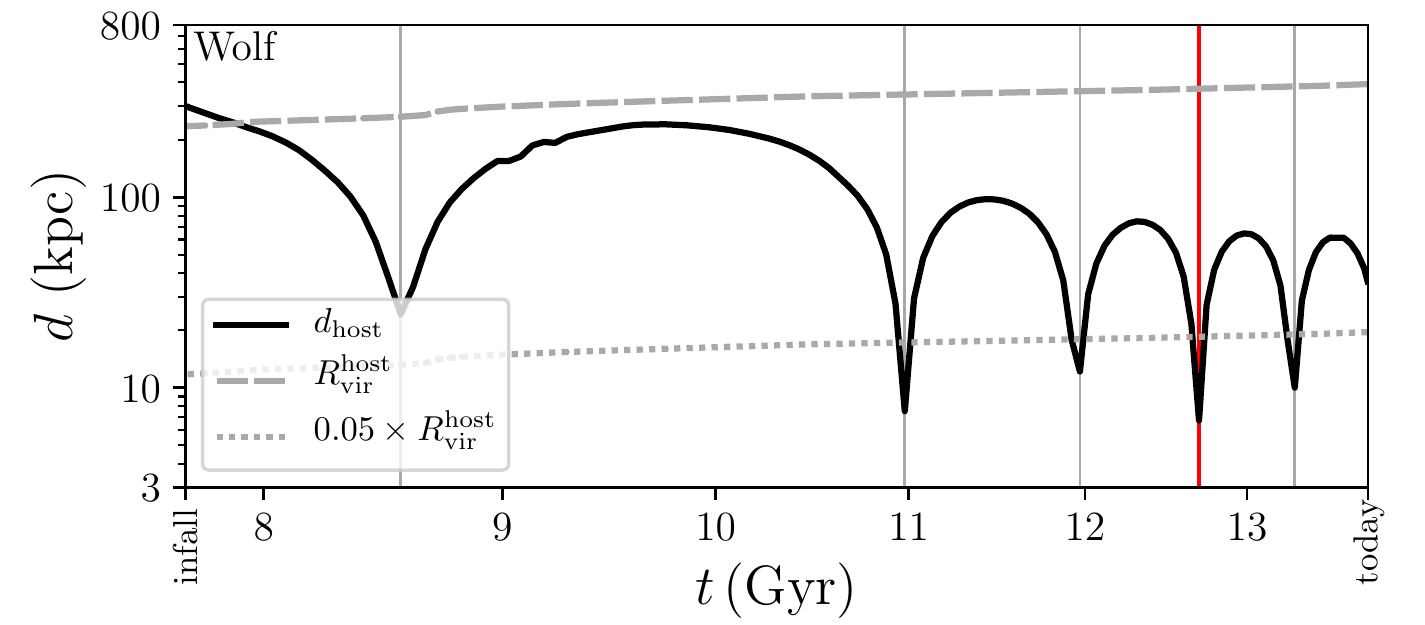}
	\includegraphics[width=3.3in]{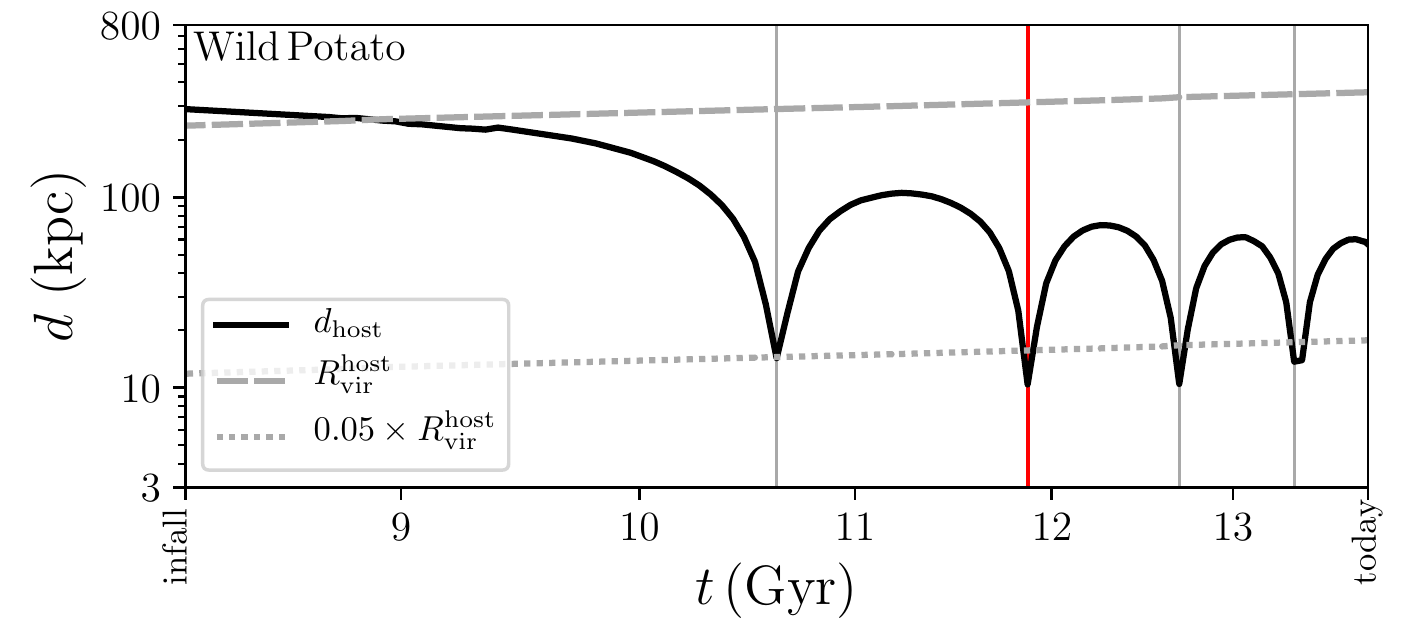}
    } 
    \hbox{
    \hspace{0cm}
        \includegraphics[width=3.3in]{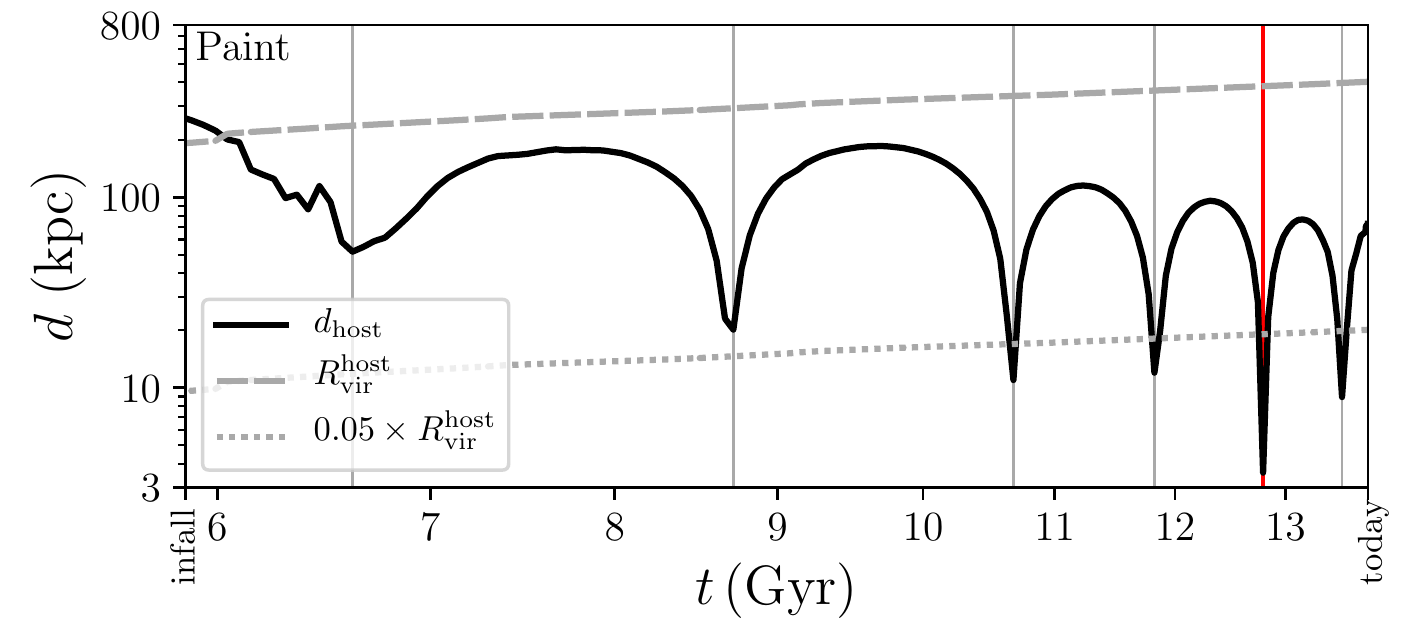}
	\includegraphics[width=3.3in]{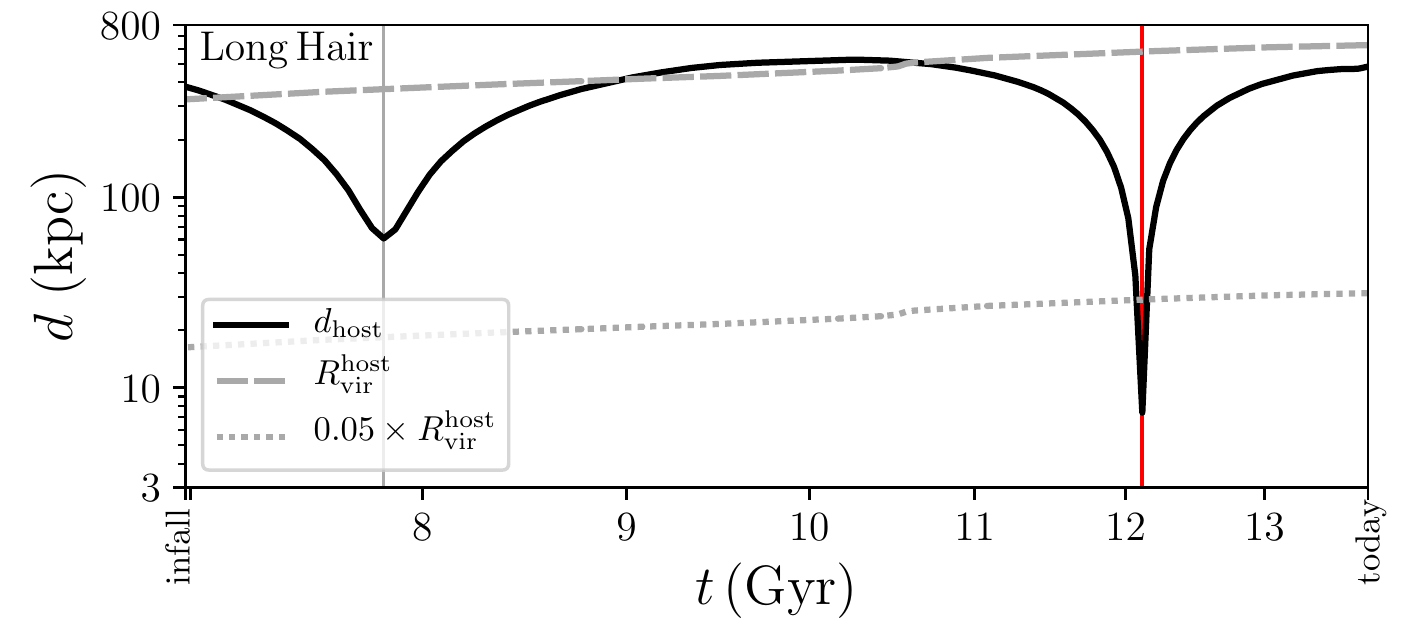}
    } 
    \hbox{
    \hspace{0cm}
        \includegraphics[width=3.3in]{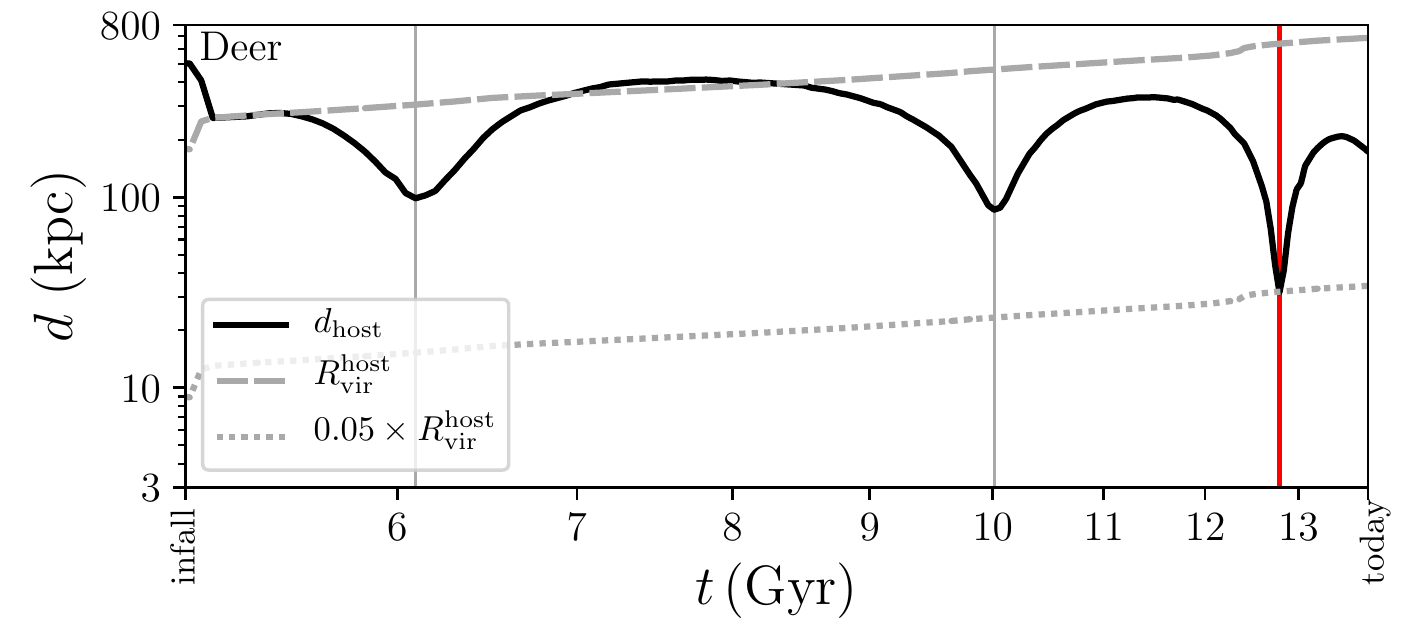}
	\includegraphics[width=3.3in]{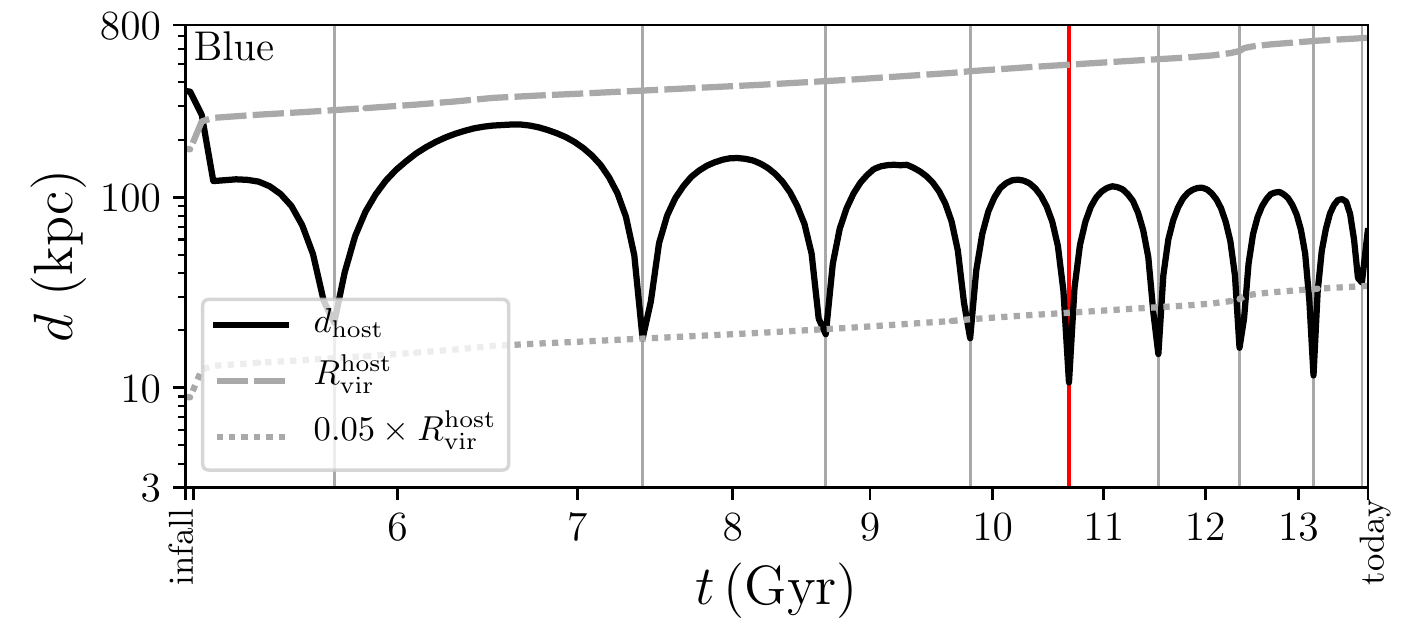}
    } 
    \hbox{
    \hspace{0cm}
        \includegraphics[width=3.3in]{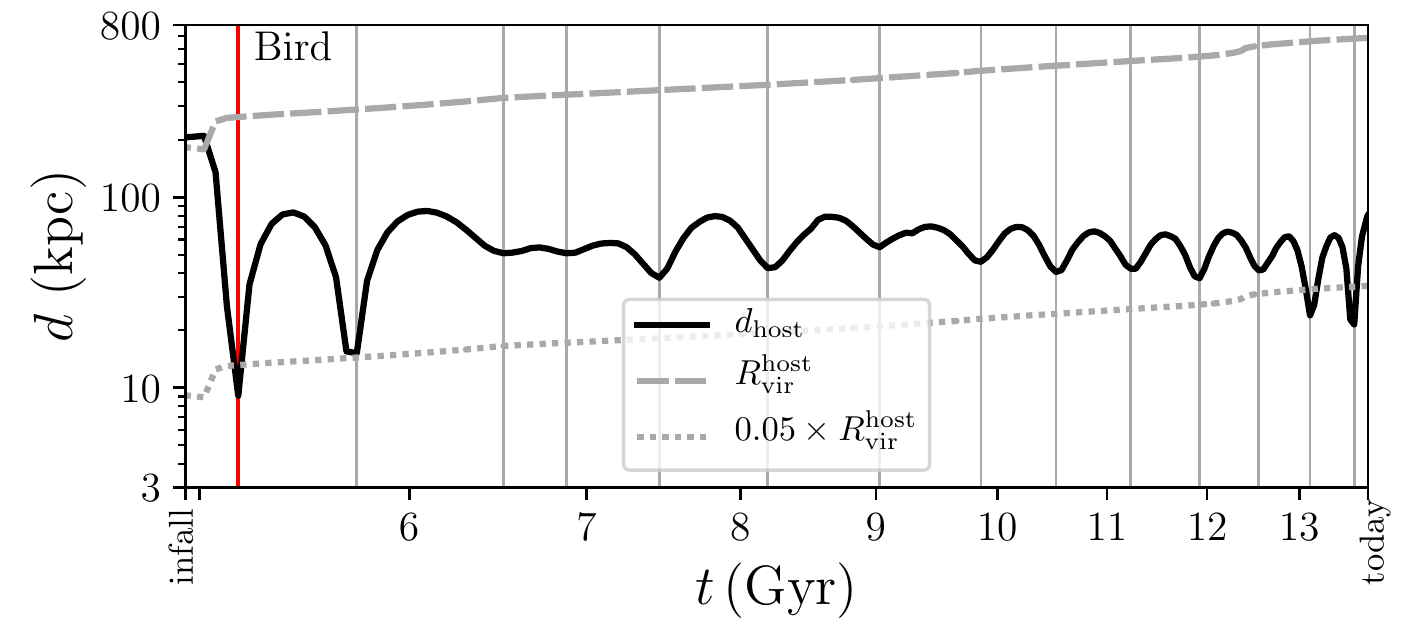}
    } 
    \vspace{-.2cm}
    \caption{{\bf Halo-centric distance (in physical units) versus cosmic time.} Each panel corresponds to one of our seven dark-matter deficient galaxies. The dashed and dotted gray curves represents the virial radius of the host and 5\% thereof. The time range extends from $t_{\rm infall}$, the infall time when the galaxy became a satellite for the first time, to the present time (13.8 Gyr). The vertical thin gray lines indicate pericentric passages. The vertical thick red line highlights the time of the minimum halo-centric distance.}
    \label{fig:figsi6}
\end{figure}

\begin{figure}
\def\figurename{Supplementary Figure}
    \vspace{-.5cm}
    \hbox{
    \hspace{0.cm}
	\includegraphics[width=3.3in]{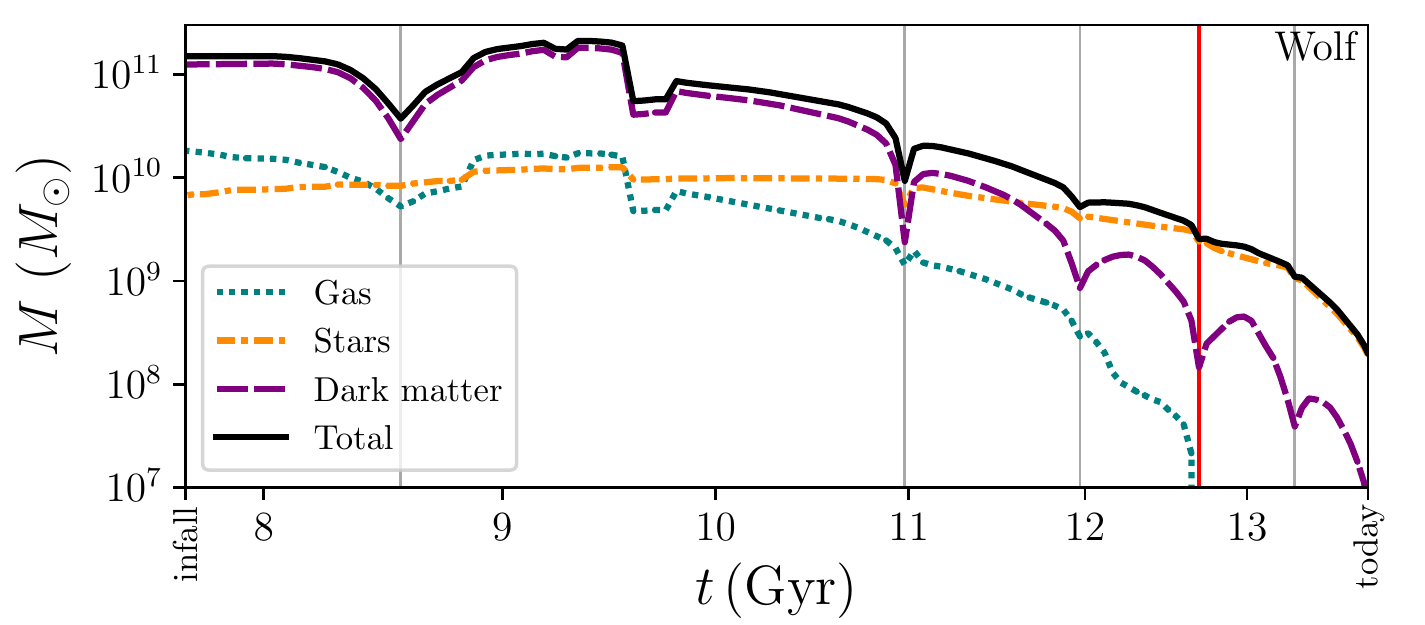}
	\includegraphics[width=3.3in]{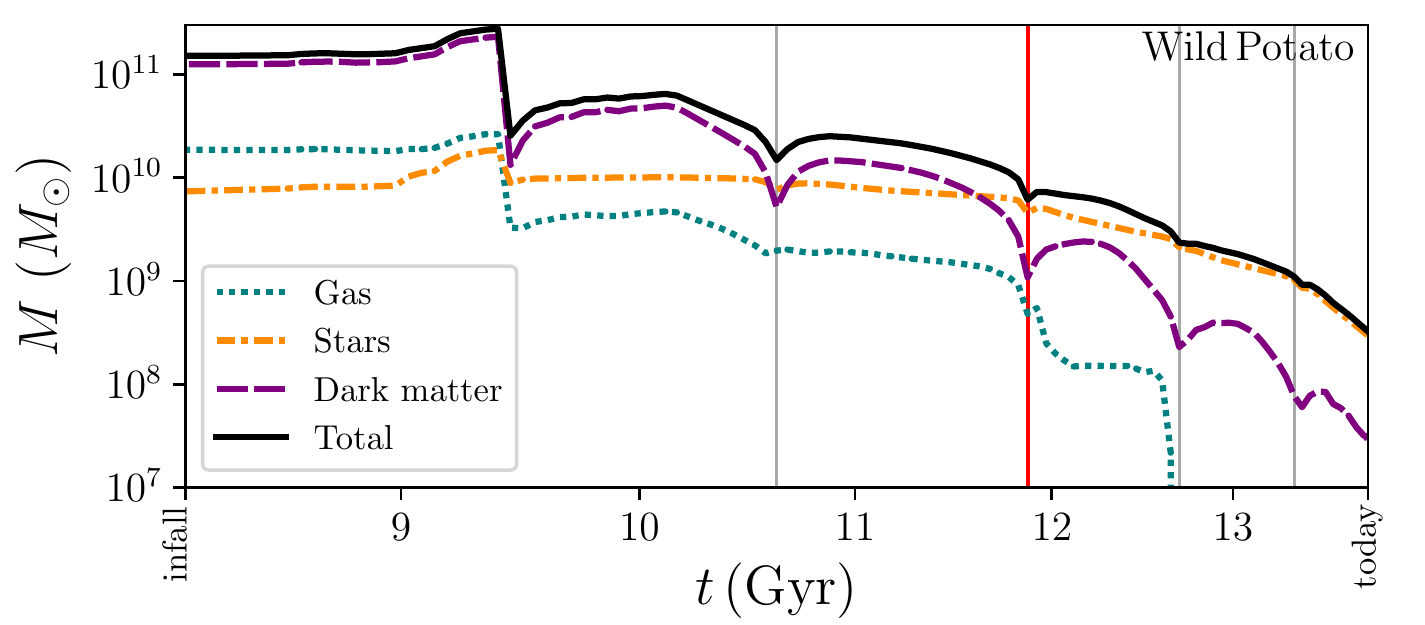}
    } 
    \hbox{
    \hspace{0cm}
        \includegraphics[width=3.3in]{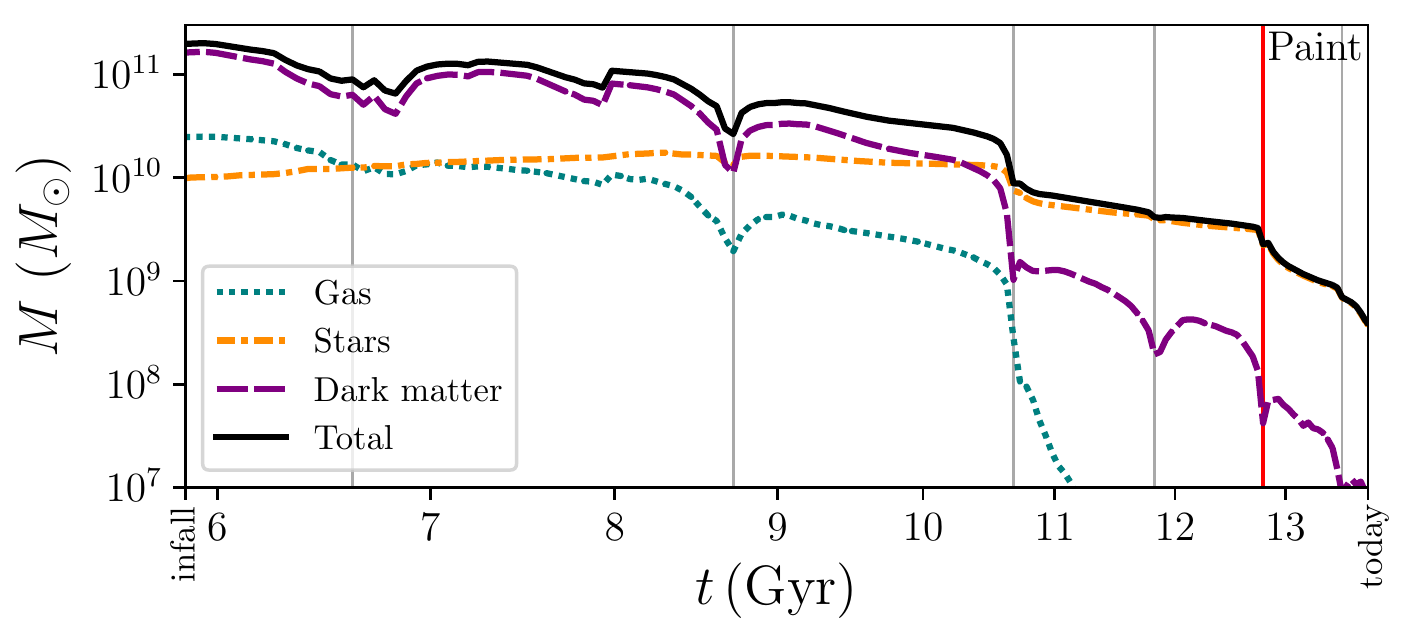}
	\includegraphics[width=3.3in]{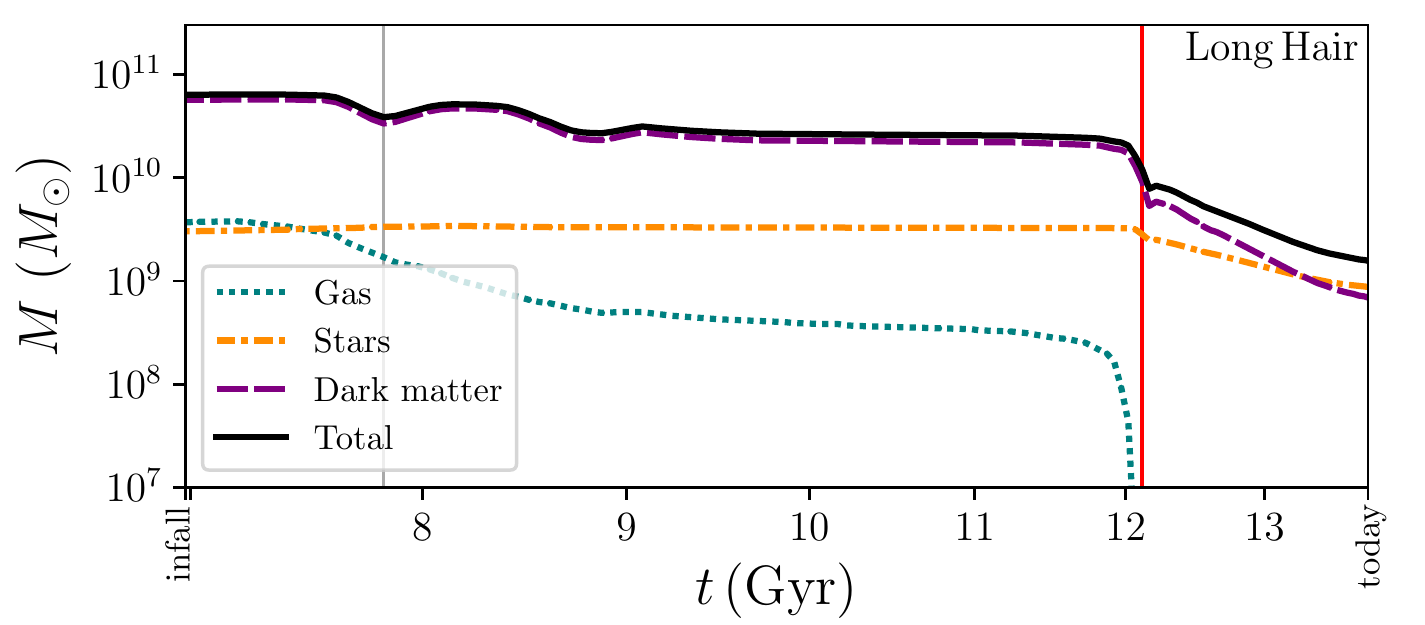}
    } 
    \hbox{
    \hspace{0cm}
        \includegraphics[width=3.3in]{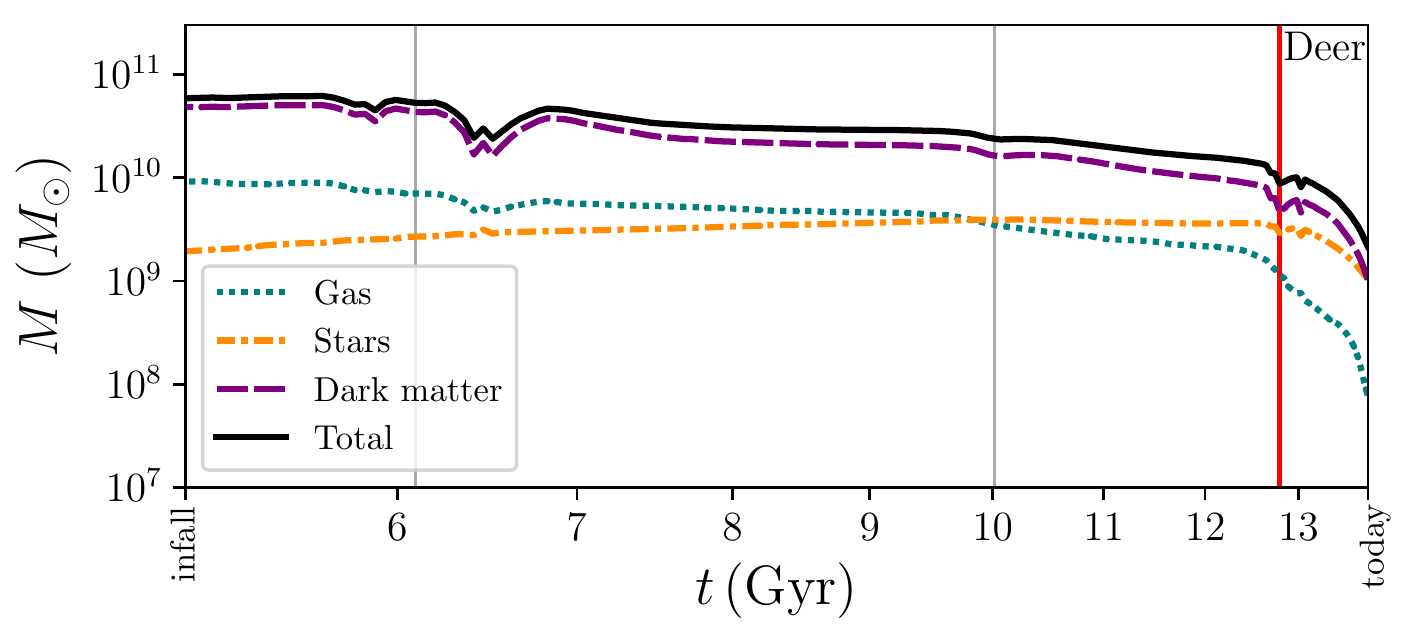}
	\includegraphics[width=3.3in]{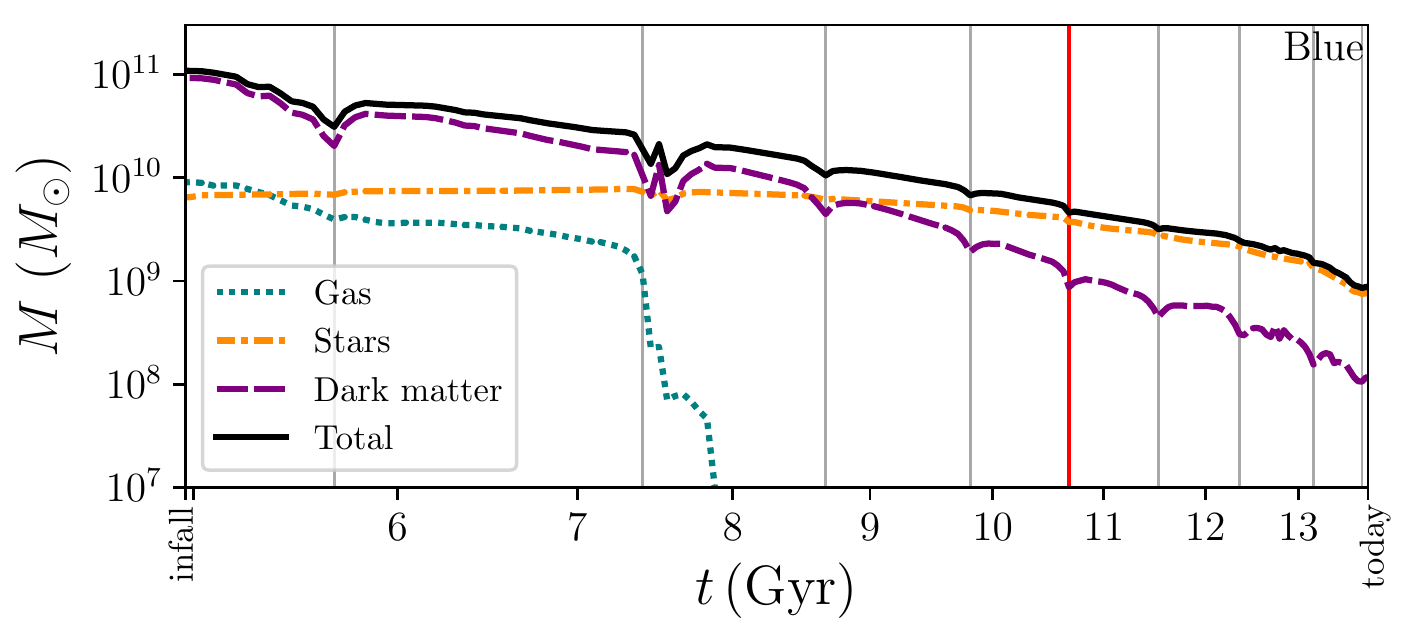}
    } 
    \hbox{
    \hspace{0cm}
        \includegraphics[width=3.3in]{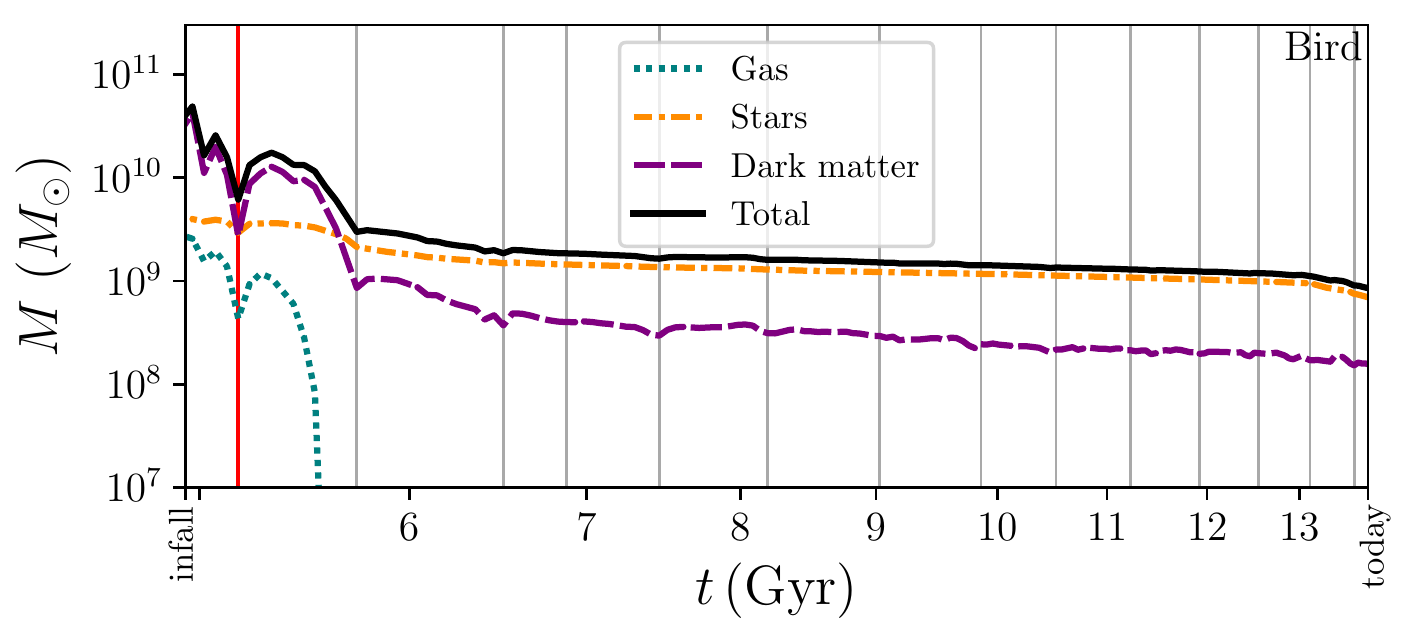}
    } 
    \vspace{-.2cm}
    \caption{{\bf Mass evolution versus cosmic time.} Each panel corresponds to one of our seven dark-matter deficient galaxies. The dotted-teal, dot-dashed-orange and dashed-purple curves indicate mass in gas, stars and dark matter. The solid black curve represents total mass. The time range extends from $t_{\rm infall}$, the infall time when the galaxy became a satellite for the first time, to the present time (13.8 Gyr). The vertical thin gray lines indicate pericentric passages. The vertical thick red line highlights the time of the minimum halo-centric distance.}
    \label{fig:figsi7}
\end{figure}

Supplementary Fig.~\ref{fig:figsi6} shows halo-centric distance (in physical units) versus cosmic time, between infall time (defined as the first time the galaxy became a satellite) and today. Each panel displays one of our seven dark-matter deficient galaxies. The dashed and dotted gray lines represent the growth of the virial radius of the host and 5\% thereof. The vertical thin-gray lines indicate pericentric passages and the vertical thick-red line corresponds to the minimum halo-centric distance. We confirmed the aformentioned number of close passages by inspecting 3D representations of these orbits (not included). Supplementary Fig.~\ref{fig:figsi7} show mass (within the subhalo radius) versus time. The dotted-teal, dot-dashed-orange, dashed-purple and solid-black curves represent mass in gas, stars, dark matter and the sum of these, respectively. Before becoming satellites, our seven galaxies were more massive ($M_{\star}\sim2-10\times 10^{9} M_{\odot}$), gas-rich ($M_{\rm gas}\sim3-25\times 10^{9} M_{\odot}$) and dark-matter dominated ($M_{\rm dm}\sim4-16\times 10^{10} M_{\odot}$). All of them lost over $99\%$ of their dark-matter content between infall and the present time. See Supplementary Table~\ref{table:tablesi2} for numerical values.

\begin{table}
\def\tablename{Supplementary Table}
\begin{center}
 \begin{tabular}{c c c c c c c c } 
 \hline
  & ${\rm Bird}$ & ${\rm Blue}$ & ${\rm Deer}$ & ${\rm LH}$ & ${\rm Paint}$ & ${\rm WP}$ & ${\rm Wolf}$ \\ [0.5ex] 
 \hline\hline
 $M^{z=0}_{\star}$ & ${\rm 7.0}$ & ${\rm 7.6}$ & ${\rm 10.0}$ & ${\rm 8.8}$ & ${\rm 3.9}$ & ${\rm 3.0}$ & ${\rm 2.0}$  \\ 
  \hline
$M^{z=z_{\rm infall}}_{\star}$ & ${\rm 39.4}$ & ${\rm 64.5}$ & ${\rm 19.5}$ & ${\rm 30.2}$ & ${\rm 98.5}$ & ${\rm 73.4}$ & ${\rm 68.8}$  \\ 
 \hline
  $M^{z=0}_{\rm gas}$ & ${\rm 0.0}$ & ${\rm 0.0}$ & ${\rm 0.8}$ & ${\rm 0.0}$ & ${\rm 0.0}$ & ${\rm 0.0}$ & ${\rm 0.0}$  \\ 
 \hline
$M^{z=z_{\rm infall}}_{\rm gas}$ & ${\rm 26.7}$ & ${\rm 89.7}$ & ${\rm 91.8}$ & ${\rm 36.8}$ & ${\rm 247.2}$ & ${\rm 187.3}$ & ${\rm 176.5}$  \\ 

 \hline
  $M^{z=0}_{\rm dm}$ & ${\rm 1.6}$ & ${\rm 1.2}$ & ${\rm 10.0}$ & ${\rm 6.9}$ & ${\rm 0.08}$ & ${\rm 0.3}$ & ${\rm 0.08}$  \\
\hline
$M^{z=z_{\rm infall}}_{\rm dm}$ & ${\rm 353.2}$ & ${\rm 920.0}$ & ${\rm 480.8}$ & ${\rm 565.9}$ & ${\rm 1,622.6}$ & ${\rm 1,252.6}$ & ${\rm 1,252.9}$  \\ 
 \hline
$f^{\rm lost}_{\star}$ & ${\rm 82.36\%}$ & ${\rm 88.21\%}$ & ${\rm 45.40\%}$ & ${\rm 70.99\%}$ & ${\rm 96.08\%}$ & ${\rm 95.94\%}$ & ${\rm 97.17\%}$ \\ 
 \hline
$f^{\rm lost}_{\rm gas}$ & ${\rm 100\%}$ & ${\rm 100\%}$ & ${\rm 99.15\%}$ & ${\rm 100\%}$ & ${\rm 100\%}$ & ${\rm 100\%}$ & ${\rm 100\%}$ \\ 
\hline
  $f^{\rm lost}_{\rm dm}$ & ${\rm 99.56\%}$ & ${\rm 99.87\%}$ & ${\rm 97.85\%}$ & ${\rm 98.78\%}$ & ${\rm 99.99\%}$ & ${\rm 99.98\%}$ & ${\rm 99.64\%}$ \\ 
 \hline
  $z_{\rm infall}$ & ${\rm 1.26}$ & ${\rm 1.23}$ & ${\rm 1.23}$ & ${\rm 0.76}$ & ${\rm 1.00}$ & ${\rm 0.56}$ & ${\rm 0.63}$ \\ 
 [1ex] 
 \hline
\end{tabular}
\caption{{\bf Mass evolution of our seven dark-matter deficient galaxies.} `LH' and `WP' denote for Long Hair and Wild Potato and $z_{\rm infall}$ denotes the infall redshift. We display masses in units of $10^{8} M_{\odot}$ to emphasize changes. Quantities are within $R_{\rm vir}$ (or $R_{\rm subhalo}$ after infall) unless stated otherwise. (i) Stellar mass at $z=0$; (ii) stellar mass at infall; (iii) gas mass at $z=0$; (iv) gas mass at infall; (v) dark matter mass at $z=0$; (vi) dark matter mass at infall; (vii) fraction of stellar mass lost since infall; (viii) fraction of gas mass lost since infall; (ix) fraction of dark matter mass lost since infall; (x) infall redshift. Note: Strictly speaking, the final $z=0$ values are lower limits, given that our galaxies might be impacted by artificial disruption. However, the final $M^{z=0}_{\rm dm}/M^{z=0}_{\star}$ ratio is unlikely to be affected by this effect (see below for a discussion).}
\label{table:tablesi2}
\end{center}
\end{table}

\subsection{Stellar populations.}
\label{subsec:stars}

Supplementary Table~\ref{table:tablesi3} compares properties of the stellar populations in our seven dark-matter deficient galaxies against observations. For simulations, we report quantities within $r^{\star}_{\rm 50}$. There are no measurements available for the age or metallicity of DF4, but given its similarity in colour to DF2, it is reasonable to expect that these two objects have similar stellar populations. Bird and Blue have similar stellar ages to DF2. Our other dark-matter deficient galaxies are younger (7.3 Gyr at most). In addition to being the oldest, Bird and Blue also have below-average $\rm \langle [Fe/H] \rangle$ values relative to the set of seven. However, we do not achieve metallicities as low as DF2 (or $V-I$ colours as blue as this object). Comparing to Local Group data\cite{Kirby2013}, Bird, Blue and DF2 lie marginally within the scatter of the local stellar mass-metallicity relation -- with Blue and Bird sitting on the upper envelop, and DF2 on the lower one. One possibility is that DF2 and DF4 had lower than usual metallicities than Bird and Blue at infall (within the scatter, which is also broader in the massive regime, where the progenitors of these galaxies existed at infall).

\begin{table}
\def\tablename{Supplementary Table}
\begin{center}
 \begin{tabular}{c c c c c c c c c c} 
 \hline
  & ${\rm Bird}$ & ${\rm Blue}$ & ${\rm Deer}$ & ${\rm LH}$ & ${\rm Paint}$ & ${\rm WP}$ & ${\rm Wolf}$ & ${\rm DF2}$ & ${\rm DF4}$\\ [0.5ex] 
 \hline\hline
 ${\rm Age_{50}}$ & ${\rm 10.0}$ & ${\rm 8.7}$ & ${\rm 5.5}$ & ${\rm 7.3}$ & ${\rm 5.5}$ & ${\rm 4.3}$ &  ${\rm 4.2}$ & $ 8.9 \pm  1.5 $ & $ - $ \\
\hline
$ \langle [{\rm Fe/H}] \rangle $ & ${\rm -0.61}$ & ${\rm -0.34}$ & ${\rm -0.46}$ & ${\rm -0.35}$ & ${\rm -0.08}$ & ${\rm -0.08}$ &  ${\rm -0.13}$ & $ -1.35 \pm 0.12 $ & $ - $ \\
\hline
${\rm V-I}$ & ${\rm 0.70}$ & ${\rm 0.72}$ & ${\rm 0.60}$ & ${\rm 0.70}$ & ${\rm 0.75}$ & ${\rm 0.67}$ &  ${\rm 0.67}$ & $ 0.37 \pm 0.05$ & $ 0.32 \pm 0.1 $ \\
 [1ex] 
 \hline
\end{tabular}

\caption{{\bf The stellar populations of our seven dark-matter deficient galaxies.} `LH' and `WP' denote Long Hair and Wild Potato. Measurements in simulations are taken within $r^{\star}_{\rm 50}$. (i) stellar age (median for simulated galaxies in Gyr; see Ref.\cite{Fensch2019} for DF2's stellar light); (ii) stellar iron-to-hydrogen abundance, dex (see Ref.\cite{vanDokkum2018b} for DF2, who report this value only for the globular cluster population); (iii) V$-$I colour (mag, see Ref.\cite{vanDokkum2018} for DF2 and Ref.\cite{Cohen2018} for DF4, who use $V_{\rm 606}-I_{\rm 814}$ from the Hubble Space Telescope Advanced Cameras for Surveys in the AB system).}

\label{table:tablesi3}
\end{center}
\end{table}

\subsection{Resolution and tidal disruption.}
\label{subsec:numerics}

Our dark-matter deficient galaxies are all tidal remnants of satellites which experienced multiple pericentric passages. To resolve such tidal remnants requires excellent numerical resolution. It has been shown that subhaloes with insufficient number of particles simulated with non-optimal softening length suffer from artificial tidal mass loss and even complete numerical disruption\cite{vdB2018}. Therefore, it is natural to be cautious about the robustness of their dark-matter deficient status. Although it is quite possible that our dark-matter deficient galaxies experienced some artificial tidal dark-matter mass loss by $z=0$, most likely they were already in the dark-matter deficient territory before artificial tidal mass loss became significant. Below we gauge the severity of numerical noise and artificial tidal evolution in two ways. 

First, we use the study of artificial disruption in Ref.\cite{vdB2018} -- especially the results tabulated in their Figure 10, which report the timescales for numerical disruption in terms of the initial particle number of the satellite and the softening length. These authors find that, for subhaloes orbiting within 10\% of the virial radius of the host, if the initial particle number is $\sim3\times10^{5}$ and the softening length is $\sim0.03$ times the scale radius of the subhalo at infall ($r_{\rm s, \, sub}$), then artificial tidal stripping kicks in after $\sim5$ Gyr of evolution (and disrupts the subhalo completely after $\sim10$ Gyr). To illustrate, we focus our attention on two of our dark-matter deficient galaxies: Wolf (the one with the lowest final stellar mass) and Paint (the one with the lowest final dark-matter mass fraction). Both of these galaxies started with dark-matter masses of $\sim1.6$ and $\sim 1.3 \times 10^{11} M_{\odot}$ (corresponding to $\sim5$ and $\sim4\times 10^5$ dark matter particles) at infall. They both have scale radii of $\sim 2.5$ kpc, so our softening length of $80$ pc almost exactly corresponds to $0.03 r_{\rm s, \, sub}$.  Additionally, their orbital pericentres are approximately $0.1 R^{\rm host}_{\rm vir}$. That is, the aforementioned artificial disruption condition is applicable to these two galaxies. As our Supplementary Fig.~\ref{fig:figsi7} shows, Paint and Wolf became dark-matter deficient ($M_{\star}/M_{\rm dm} >1$) after spending only $\sim3$ Gyr in the host potential, before artificial stripping has time to become significant. That said, Paint and Wolf spent $\sim6$ and $\sim8$ Gyr in total in the host potential -- so during the latest couple of orbits, their associated dark-matter subhaloes likely suffered from some artificial disruption and thus only have a few particles left.

\begin{figure}
\def\figurename{Supplementary Figure}
\includegraphics[width=\columnwidth]{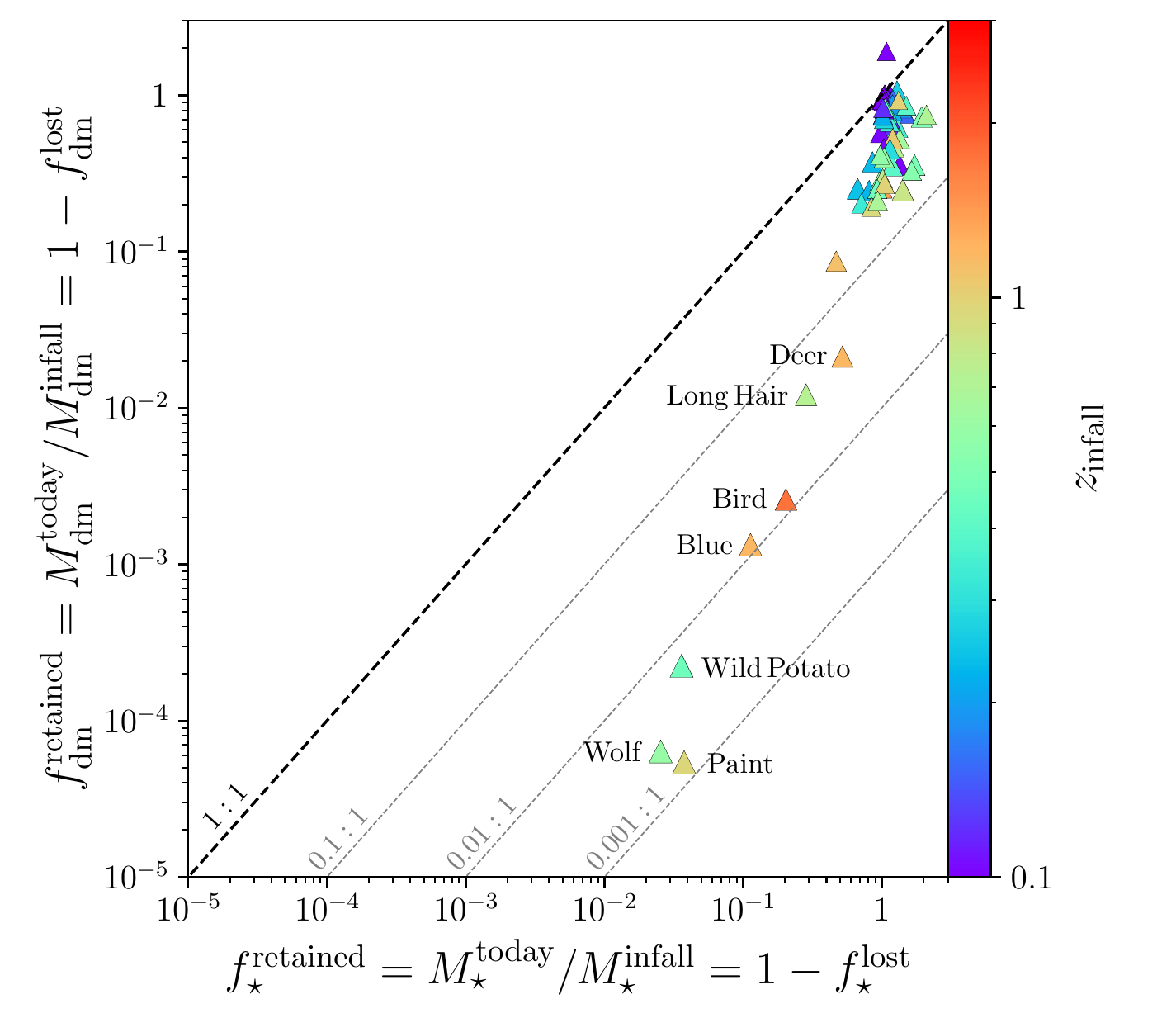}
\vspace{-2cm}
\caption{{\bf Dark-matter versus stellar mass retention.} Fraction of retained dark-matter mass versus fraction of retained stellar mass, between infall and today. The triangles represent simulated satellites with stellar masses between $10^{8-9}M_{\odot}$ (as in Figures~3 and 4), their colours represent infall redshift, and their symbol dimensions represent their dark-matter mass at infall. Our seven dark-matter deficient galaxies (labeled) experience higher $f^{\rm lost}_{\rm dm}/f^{\rm lost}_{\star}$ values than other galaxies in our parent sample with comparable $M^{\rm infall}_{\rm dm}$ and $z_{\rm infall}$. }
\label{fig:figsi8}
\end{figure}

Secondly, we also address this issue with tidal evolutionary tracks\cite{Errani2020,Errani2021}. In particular, we estimate at what stage of their structural evolution they start to deviate from their physical tidal evolutionary tracks. As Figure~A1 of Ref.\cite{Errani2021} shows, artificial deviation from the tidal track starts when the maximum-circular-velocity radius $r_{\rm max}$ approaches 8 $\times$ the softening length. In our case, this is $8 \times 80$ pc $\sim0.6$ kpc. Both Paint and Wolf started with $r_{\rm max} \sim 5$ kpc. Therefore, according to the tidal tracks in Figure~A1 of Ref.\cite{Errani2021}, the dark-matter structural evolution of these two galaxies can be trusted down to $r_{\rm max}/r^{\rm infall}_{\rm max}\sim 0.1$ --  or equivalently, when their bound mass is $1/100$ of the initial value. From our Supplementary Fig.~\ref{fig:figsi7}, we can see that both galaxies became dark-matter deficient ($M_{\star}/M_{\rm dm} >1$) before their dark-matter mass dropped to 1\% of that at infall.

We also note that the above studies adopt dark-matter-only configurations. It would be interesting to see how those numerical experiments change in the presence of an additional collision-less stellar population, plus a hydrodynamic gaseous component and feedback physics.

Lastly, we underscore the fact that all of our dark-matter deficient galaxies exhibit unusually stronger differential mass loss of dark-matter versus stars relative to other satellites in our simulation. Supplementary Figure~\ref{fig:figsi8} shows the fraction of {\it retained} dark-matter mass (defined as $M^{\rm today}_{\rm dm}/M^{\rm infall}_{\rm dm}$) versus the fraction of {\it retained} stellar mass (defined as $M^{\rm today}_{\star}/M^{\rm infall}_{\star}$). The diagonal dashed lines depict ratios between these two quantities. The triangles represent simulated satellites with $z_{\rm infall}>0$ (i.e., we exclude $z=0$ satellites that were {\it not} also satellites at the previous-to-last snapshot) and $10^8 < M_{\star}/M_{\odot} < 10^9$ (as in Figures~3 and 4). The symbol dimensions scale logarithmically with dark-matter mass at infall and the colour scheme represents infall redshift. Clearly, our seven dark-matter deficient galaxies all lose significantly more dark-matter mass than stellar-mass when compared to other satellites with similar $M^{\rm infall}_{\rm dm}$ and $z_{\rm infall}$. That is, even if artificial disruption is significant, it should affect both sets of galaxies similarly, and yet we still see a stronger differential mass loss in our sample of seven. This suggest that the environmental processes making responsible for transforming galaxies into dark-matter deficient ones are qualitatively correctly captured. 

\subsection{Simulations by other groups.}
\label{subsec:otherwork}

Since the discovery of the DF2\cite{vanDokkum2018}, several efforts to re-create analogues in simulations have been published. What is novel about our work is that, for the first time, we are able to generate a simulated low-mass galaxy (Wolf) that resembles DF2 and DF4 (in terms of stellar mass, size, 1D line-of-sight velocity dispersion and morphology) within a cosmological simulation. In other words, it is not enough to merely create dark-matter deficient galaxies (which is challenging on its own right); it is paramount for said galaxies to also exhibit internal properties that match observations. We devote this  Section to a discussion of studies on dark-matter deficient galaxies using various methods.

Works on this subject can be categorized into three classes: idealised (non-cosmological) numerical experiments, zoom-in simulations and cosmological-volume simulations. The first class\cite{Ogiya2018,Huo2020,Nusser2020,Shin2020,Yang2020,Lee2021,Maccio2021} offers unique insights at resolution levels often inaccessible to cosmological simulations. In particular, some of these controlled experiments\cite{Nusser2020,Maccio2021} have been instrumental at exposing how challenging it is to achieve velocity dispersions as {\it low} as those measured in DF2 and DF4. The majority of these efforts (but not all) also highlight the importance of satellite-host interactions, in line with our conclusions. However, a major limitation of this approach is that it relies on ad-hoc assumptions about (1) the initial structure of the progenitors, (2) the properties of their neighboring companion (all of the numerical experiments published thus far assume the presence of only one companion), and (3) the orbital geometry of their interaction. Interestingly, the idealised runs by Ref.\cite{Maccio2021} circumvent the first caveat by extracting the progenitor structure from a cosmological box, but the other two caveats remain. Nevertheless, the adoption of these assumptions does not reflect realistic cosmological conditions and provides no information on the likelihood or frequency that such configurations actually occur. Another key limitation of such works (to-date) is that they operate under the presumption that a single specific mechanism must be responsible for the creation of dark-matter deficient low-mass galaxies. For example, unlike the majority of works adhering to this approach, the idealised simulations by Ref.\cite{Shin2020} assume a priori that dark-matter deficient low-mass galaxies are created via gas-rich major mergers (i.e., the companion is not a significantly more massive central, but a galaxy of comparable mass -- see also Ref.\cite{Lee2021}). Please note that we do not find any instances of dark-matter deficient low-mass galaxies being created in this fashion. Overall, the above limitations in the idealised-simulation method underscore the importance of conducting investigations in a cosmological context.

The second class involves zoom-in simulations, where a sub-volume of a larger cosmological box is re-simulated at higher resolution. By design, this method generally only simulates a massive galaxy and a small number of galaxies in its proximity, and is thus not well suited for finding objects with rare or unusual properties. To illustrate, Ref.\cite{Applebaum2021} use a suite of Milky-Way-like zoom-in simulations and find one satellite with $M_{\star}>M_{\rm dm}$. Unlike our sample of seven (or DF2 and DF4), this galaxy has a much lower stellar mass ($\sim 10^5 M_{\odot}$) and is orbiting a Milky-Way-mass host. Unfortunately, this object is poorly resolved (its half-light radius is below their adopted force resolution). If confirmed by these authors with future simulations at higher resolution, this object will certainly become a provocative prediction, because no dark-matter deficient galaxies orbiting massive centrals at the Milky Way scale have ever been observed, and our predictions reject hosts at this scale. On a different vein, zoom-in simulations at the galaxy group scale (as suggested by our results and observations) may also be of significant interest.

The third class is cosmological-volume  simulations\cite{Yu2018,Carleton2019,Jing2019,Haslbauer2019,Sales2020,Saulder2020,Shin2020,Jackson2021}. Ref.\cite{Haslbauer2019} use an older version of a well-established cosmological simulation to challenge the viability of the $\Lambda$-Cold-Dark-Matter paradigm. These authors find zero dark-matter deficient galaxies that match DF2 simultaneously in stellar mass and size. They do find a galaxy slightly smaller than DF2 (by a factor of 0.8 in stellar-mass and size), but this object is gas-rich (unlike DF2 and DF4) and its massive companion is over one order-of-magnitude smaller in stellar mass than NGC 1052. Ref.\cite{Yu2018} use the same simulation and find a few galaxies lacking dark matter -- but most are poorly resolved, rendering their results inconclusive. These authors find one resolved dark-matter deficient galaxy. However this object has $M_{\star}\sim 9 \times 10^{10} M_{\odot}$, almost two orders of magnitude more massive than DF2 and DF4, and it is the result of a numerical artifact\cite{Saulder2020}. Ref.\cite{Jing2019} find several dark-matter deficient galaxies in two cosmological simulations, but these objects are in the $M_{\star}=10^{9-10}M_{\odot}$ regime, over an order-of-magnitude above DF2 and DF4. Using a newer simulation, Ref.\cite{Sales2020} instead investigate the tidal evolution of low-mass galaxies in massive galaxy clusters ($M_{\rm vir} \geq 10^{14} M_{\odot}$ -- see also Ref.\cite{Carleton2019}). These authors do not find any galaxies in the $M_{\star}=10^{8-9}M_{\odot}$ regime with sufficiently low values of $\sigma^{\rm 1D}_{\star}$ to be consistent with DF2 and DF4 (their Figure 5). It would be interesting if (1) this study repeated at group scale found suitable candidates; and (2) if low-mass galaxies lacking dark matter are observed in massive galaxy clusters. Ref.\cite{Shin2020} use the same simulation to test their high-velocity collision scenario, finding only one such event, which results in a gas clump that never turns into a galaxy.

The majority of the cosmological simulations we cite in this paper have resolutions much lower than ours, which could be the culprit for the lack of success in finding analogues of DF2 and DF4. Only two works\cite{Jing2019,Jackson2021} employ cosmological simulations with volumes and resolutions somewhat comparable to ours. First we compare to Ref.\cite{Jackson2021}, who employ a $10$-cMpc box (on the side) with $m_{\rm DM}=10^6 M_{\odot}$, $m_{\star}=10^4 M_{\odot}$, maximum spatial resolution of $34$ pc and star-forming gas density threshold of $10$ cm$^{-3}$ (versus our $21$-cMpc box with $m_{\rm DM}=3.3 \times 10^5 M_{\odot}$, $m_{\rm b}=6.3 \times 10^4 M_{\odot}$, maximum adaptive smoothing length resolution of $1.5$ pc and star-forming gas density threshold of $300$ cm$^{-3}$). Below we demonstrate that none of the galaxies in Ref.\cite{Jackson2021} are consistent with DF2 or DF4. We also note that their ``DM def." sample is ``too generous". Namely, they require members of this subset to be satellites with $M_{\rm dm}/M_{\star}<10$, rendering the majority of their objects dark-matter dominated by construction (i.e., with $M_{\rm dm}/M_{\star}>1$). As a result, all of their low-mass galaxies (with $M_{\star}<10^9 M_{\odot}$) are also dark-matter dominated. We note that their paper reports quantities out to $r_{\rm subhalo}$, not within $r^{\star}_{\rm 50}$, as done here and in observations.

\begin{figure}
\def\figurename{Supplementary Figure}
\includegraphics[width=\columnwidth]{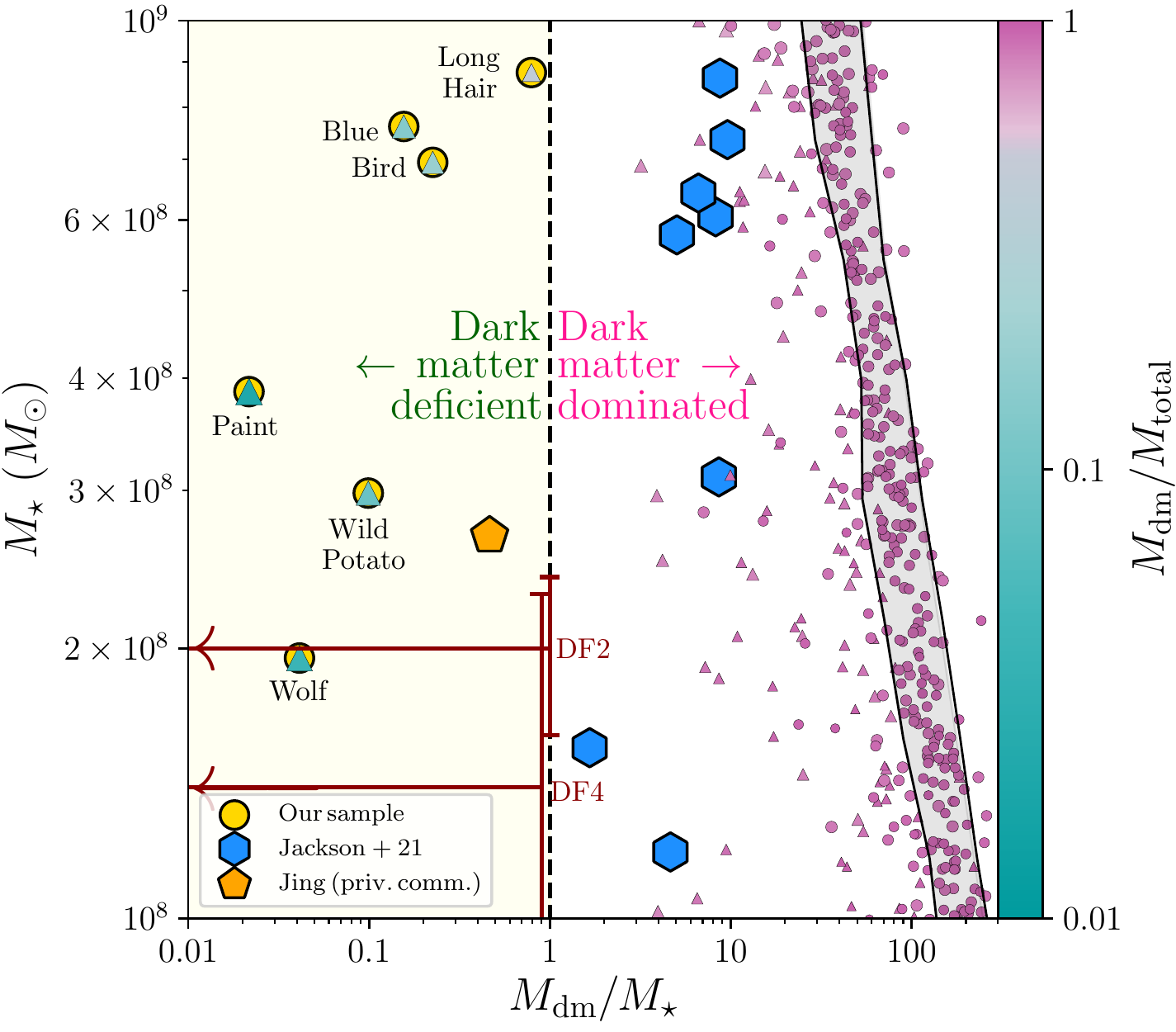}
\vspace{-1.5cm}
\caption{ {\bf Comparison with other cosmological simulations.} Stellar mass versus dark-matter-to-stars mass ratio. Quantities are measured out to either the virial radius (centrals) or the subhalo radius (satellites). The small circles (centrals) and triangles (satellites) represent our parent sample (with $M_{\star}<10^9 M_{\odot}$). The light-yellow region denotes the dark-matter deficient regime in this stellar-mass range, which contains six of our seven galaxies (large yellow circles), but none of the galaxies in Ref.\cite{Jackson2021} (large light-blue hexagons). The datum shared with us by Y. Jing (large orange pentagon, private communication) is dark-matter deficient, but its stellar mass and velocity dispersion are too high compared with observations (dark-red arrows).} 
\label{fig:figsi9}
\end{figure}

Supplementary Fig.~\ref{fig:figsi9} reproduces Figure~1 in Ref.\cite{Jackson2021} with our data, theirs, an object reported shared with us by Y. Jing (the lead author of Ref.\cite{Jing2019}, private communication) and observations. We use \texttt{WebPlotDigitizer}, a free online tool, to extract data from Ref.\cite{Jackson2021}. Quantities in this Supplementary Figure are measured out to $R_{\rm vir}$ for centrals (circles), and $r_{\rm subhalo}$ for satellites (triangles). The light-gray band represents the $M_{\star}-M_{\rm halo}/M_{\star}$ relation for centrals. The light-yellow region denotes the dark-matter deficient regime (with $M_{\rm dm}/M_{\star}<1$). The dark-red arrows with error bars represent observations. Here, the vertical location of these two arrows corresponds to $M_{\star}= M_{\star}(r<r_{\rm subhalo}) \equiv 2 \times M_{\star}(r<r^{\star}_{\rm 50})$. See Table~\ref{table:tablesi1} for the $M_{\star}(r<r^{\star}_{\rm 50})$ values measured for DF2 and DF4. We note that $M_{\rm dm}/M_{\star}$ has not been observed out to $r_{\rm subhalo}$ for these two galaxies (i.e., the horizontal extent of the two dark-red arrows). However, given that all of our seven galaxies are dark-matter deficient within both $r^{\star}_{\rm 50}$ and $r_{\rm subhalo}$, it is reasonable to assume that this is also the case for DF2 and DF4. The large yellow circles highlight our seven dark-matter deficient galaxies, which span approximately two orders-of-magnitude {\it to the left} of the $M_{\rm dm}/M_{\star}=1$ boundary (vertical dashed black line). In contrast, all low-mass galaxies in the Ref.\cite{Jackson2021} sample (large light-blue hexagons) are dark-matter dominated -- i.e., they span about one order-of-magnitude to the right of the $M_{\rm dm}/M_{\star}=1$ boundary. In particular, their least dark-matter dominated galaxy in this Figure still has $\sim66\%$ more mass in dark-matter than stars.

Note that Supplementary Fig.~\ref{fig:figsi9} only displays galaxies in the $M_{\star}=10^{8-9}M_{\odot}$ regime. Recall that our focus is on low-mass galaxies with $M_{\star}<10^9 M_{\odot}$ (within $r^{\star}_{\rm 50}$). This is because observations\cite{Tollerud2011} suggest that galaxies at higher stellar-mass regimes tend to already have lower dark-matter contributions (within $r^{\star}_{\rm 50}$), so discoveries of more-massive galaxies with low dark-matter content are not that surprising. When we expand our galactocentric extent out to $r_{\rm subhalo}$, Deer falls outside (above) the stellar-mass range considered in this Supplementary Figure. This Supplementary Figure also excludes three massive or intermediate-mass galaxies reported by Ref.\cite{Jackson2021} (with $M_{\rm dm}/M_{\star}<1$). Concretely, these three objects have approximately the following stellar masses: $1.2\times10^{9}M_{\odot}$, $1.6\times10^{10}M_{\odot}$ and $1.1\times10^{11}M_{\odot}$. The first of the three lies right above our low-mass regime cut (at $M_{\star}=10^{9} M_{\odot}$), but below theirs (at $M_{\star}=10^{9.5}M_{\odot}$). Although the line separating low- and intermediate-mass galaxies is arbitrary, this object is over one-order-of-magnitude above observations -- i.e., $12$-$\sigma$ above DF2 and $\sim13$-$\sigma$ above DF4.

Nevertheless, this ``mildly" dark-matter dominated population reported by Ref.\cite{Jackson2021} (the large light-blue hexagons in Supplementary Fig.~\ref{fig:figsi9}) is interesting on its own right. This set is situated to the left of the light-gray band. I.e., they are less dark-matter dominated than centrals at comparable stellar masses. The authors show that this effect is caused by tidal perturbations. We note that, unlike our set of seven, their objects never transit within 5\% of the host virial radius. Instead, their $d_{\rm min}/R_{\rm vir}$ values are between 20\% and 80\% (compare the left panel of their Figure 10 to our Figure~4). It is nonetheless striking to see that, even though their satellite-central interactions have larger $d_{\rm min}/R_{\rm vir}$ values than ours, they still contribute towards helping those galaxies migrate to the left of the light-gray band. But more importantly, the absence of low-mass galaxies in their sample that (1) cross to the left of the $M_{\rm dm}/M_{\star}=1$ boundary and (2) plunge within 5\% of their host virial radius further supports our scenario. Namely, that satellites are required to pierce through the innermost portions of their hosts to become dark-matter deficient galaxies (as demonstrated in Figure~4).

We now discuss Ref.\cite{Jing2019}, who found an interesting object using a third simulation (at higher resolution than the other simulations they study). This simulation employs a $25$-cMpc box (on the side) with $m_{\rm DM}=1.21 \times 10^6 M_{\odot}$, $m_{\rm b}=2.26 \times 10^5 M_{\odot}$, softening length of $350$ pc and star-forming gas density threshold of $10$ cm$^{-3}$ (versus our $21$-cMpc box with $m_{\rm DM}=3.3 \times 10^5 M_{\odot}$, $m_{\rm b}=6.3 \times 10^4 M_{\odot}$, $h_{\star}=12$ pc, $h_{\rm DM}=80$ pc, maximum adaptive smoothing length resolution of $1.5$ pc and star-forming gas density threshold of $300$ cm$^{-3}$). They report quantities within $2\times r^{\star}_{\rm 50}$, which makes it difficult to compare directly with DF2 and DF4 (which employ $r^{\star}_{\rm 50}$ instead). This object has $M_{\star}(r< 2r^{\star}_{\rm 50})=2.47 \times 10^8 M_{\odot}$, and $f_{\rm dm} \equiv M_{\rm dm}(r< 2r^{\star}_{\rm 50})/M_{\rm total}(r< 2r^{\star}_{\rm 50}) \sim 0.31$. We note that this object becomes dark-matter deficient through the same mechanism as ours, by plunging within 5\% of its host's virial radius (their Figure 6). The large orange hexagon in Supplementary Fig.~\ref{fig:figsi9} represents this object, which has $M_{\star}(r< r_{\rm subhalo})=2.67 \times 10^8 M_{\odot}$ and  $M_{\rm dm}/M_{\star}\sim 0.46$. Measurements of this object out to $r_{\rm subhalo}$ are not published in Ref.\cite{Jing2019}; rather these were shared with us by Y. Jing, the lead-author of that work (via private communication). Whilst this object has $M_{\rm dm}<M_{\star}$, its properties do not match those of DF2 and DF4. Specifically, its stellar mass is $\sim1.7$-$\sigma$ above DF2 and $\sim1.5$-$\sigma$ above DF4. Furthermore, recall that prior to our work, no other cosmological simulation has successfully created galaxies with velocity dispersions as low as those of DF2 and DF4\cite{Carleton2019,Sales2020}. This issue has also been explored by some idealised high-resolution experiments. Ref.\cite{Maccio2021} find that to achieve this, very particular satellite-central orbital configurations are required. Ref.\cite{Nusser2020}, on the other hand, invoke supernova feedback and ram-pressure stripping, aided by subsequent tidal stripping, to achieve such low $\sigma^{\rm 1D}_{\star}$ values. The velocity dispersion of this object (large orange pentagon in Supplementary Fig.~\ref{fig:figsi9}) is $\sigma_{\star} = 22.4$ km sec$^{-2}$. If we assume kinematic isotropy, this would correspond to $\sigma^{\rm 1D}_{\star} = 12.9$ km sec$^{-2}$. In other words, its 1D line-of-sight velocity dispersion is $\sim1.9$-$\sigma$ above DF2 and $\sim2.0$-$\sigma$ above DF4. (Strictly speaking, such comparisons require velocity dispersions within $r^{\star}_{\rm 50}$, but these measurements have not been published.)

There are several possibilities that could explain why galaxies that resemble the observed dark-matter deficient galaxies (such as Wolf) have not been reported in past cosmological simulations. First, as mentioned in the main text, our simulations are relatively unique in  that they explicitly track dense, molecular gas in low-mass galaxies within a volume that is large enough to contain several massive group-size haloes. The simulation in Ref.\cite{Jing2019} probes a similar volume, but uses a lower resolution of 0.35 kpc -- although it is still higher than most. However, they also employ a cruder star-formation treatment: a polytropic equation-of-state, with star-formation occurring at gas-densities $>10$ cm$^{-3}$. Ref.\cite{Jackson2021} also has similar formal spatial resolution and volume to ours, and their star-formation prescription is more sophisticated than that in Ref.\cite{Jing2019} (it is similar to ours). However, their gas component is also allowed to form stars at densities as low as $10$ cm$^{-3}$ (as in Ref.\cite{Jing2019}). What makes our simulation unique is that all of our star-forming gas reaches densities $>300$ cm$^{-3}$ -- i.e., thirty times higher than what is used in these comparably-sized simulations. Furthermore, our adaptive smoothing length reaches $1.5$ pc (versus $34$ pc for Ref.\cite{Jackson2021} and $350$ pc for Ref.\cite{Jing2019}). This allows us to track the internal baryonic structure of galaxies with higher precision. Another possibility is that some of these simulations do in fact produce such extreme objects, but their automated galaxy-recovery algorithms might pose an impediment to their query (recall that we employ a visual-recovery technique, which may not be feasible for larger runs). It would be interesting to carry out a visual search for dark-matter deficient galaxies in the simulations discussed by Ref.\cite{Jing2019} and Ref.\cite{Jackson2021}. Indeed, additional identifications of more galaxies with internal properties resembling DF2 and DF4 in other cosmological simulations will certainly put the standard paradigm on firmer ground.




\end{document}